\documentclass[aps,prc,twocolumn,superscriptaddress,showpacs,floatfix,nofootinbib]{revtex4}

\usepackage{amsmath,amssymb,amsfonts,graphicx,yhmath}
\usepackage[colorlinks=true,linktocpage=true,linkcolor=blue,citecolor=blue]{hyperref}
\usepackage[usenames,dvipsnames]{color}
\usepackage{float}

\usepackage{color}

\newcommand{\xt}{{\mathbf{x}_\perp}}

\begin{document}

\preprint{}

\title{Relativistic hydrodynamics in heavy-ion collisions: general aspects and recent developments}

\author{Amaresh Jaiswal}
\affiliation{GSI, Helmholtzzentrum f\"ur Schwerionenforschung, Planckstrasse 1, D-64291 Darmstadt, Germany}
\author{Victor Roy}
\affiliation{Institute for Theoretical Physics, Goethe University, Max-von-Laue-Strasse 1, 60438 Frankfurt am Main, Germany}

\date{\today}

\begin{abstract}

Relativistic hydrodynamics has been quite successful in explaining 
the collective behaviour of the QCD matter produced in high energy 
heavy-ion collisions at RHIC and LHC. We briefly review the latest 
developments in the hydrodynamical modeling of relativistic 
heavy-ion collisions. Essential ingredients of the model such as the 
hydrodynamic evolution equations, dissipation, initial conditions, 
equation of state, and freeze-out process are reviewed. We discuss 
observable quantities such as particle spectra and anisotropic flow 
and effect of viscosity on these observables. Recent developments 
such as event-by-event fluctuations, flow in small systems 
(proton-proton and proton-nucleus collisions), flow in ultra central 
collisions, longitudinal fluctuations and correlations and flow in 
intense magnetic field are also discussed.

\end{abstract}



\maketitle

\section{Introduction}

The existence of both confinement and asymptotic freedom in Quantum 
Chromodynamics (QCD) has led to many speculations about its 
thermodynamic and transport properties. Due to confinement, the 
nuclear matter must be made of hadrons at low energies, hence it is 
expected to behave as a weakly interacting gas of hadrons. On the 
other hand, at very high energies asymptotic freedom implies that 
quarks and gluons interact only weakly and the nuclear matter is 
expected to behave as a weakly coupled gas of quarks and gluons. In 
between these two configurations there must be a phase transition 
where the hadronic degrees of freedom disappear and a new state of 
matter, in which the quark and gluon degrees of freedom manifest 
directly over a certain volume, is formed. This new phase of matter, 
referred to as Quark-Gluon Plasma (QGP), is expected to be created 
when sufficiently high temperatures or densities are reached \cite 
{Lee, Collins:1974ky, Itoh:1970uw}.

The QGP is believed to have existed in the very early universe (a 
few microseconds after the Big Bang), or some variant of which 
possibly still exists in the inner core of a neutron star where it 
is estimated that the density can reach values ten times higher than 
those of ordinary nuclei. It was conjectured theoretically that such 
extreme conditions can also be realized on earth, in a controlled 
experimental environment, by colliding two heavy nuclei with 
ultra-relativistic energies \cite{Baumgardt:1975qv}. This may 
transform a fraction of the kinetic energies of the two colliding 
nuclei into heating the QCD vacuum within an extremely small volume 
where temperatures million times hotter than the core of the sun may 
be achieved. 

\begin{figure}[t]
\begin{center}
\includegraphics[width=\linewidth]{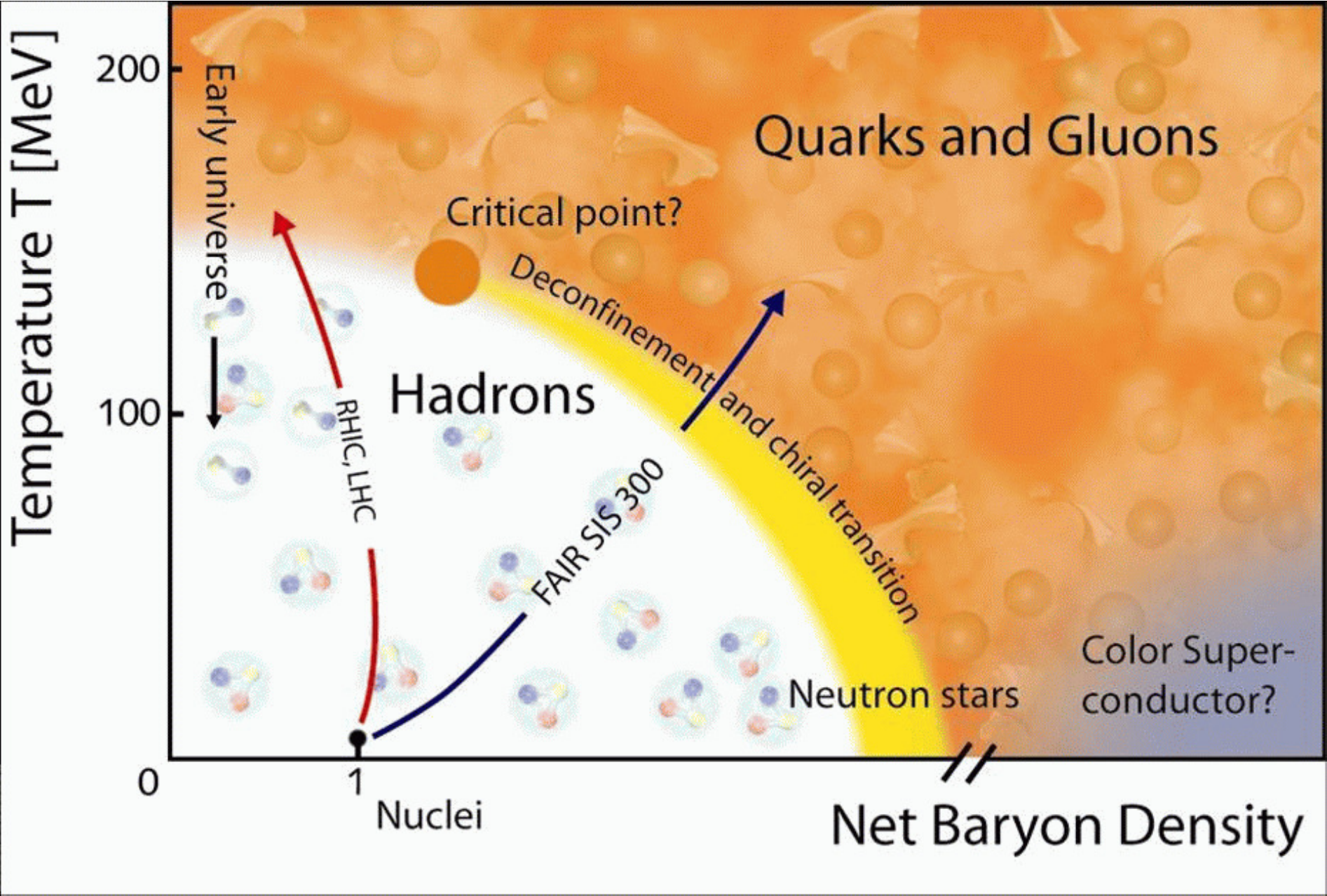}
\end{center}
\vspace{-0.4cm} 
\caption{(Color online) Schematic phase diagram of the QCD matter. 
The net baryon density on x-axis is normalized to that of the normal 
nuclear matter \cite{gsi.de}.}
\label{Phases}
\end{figure}

With the advent of modern accelerator facilities, ultra-relativistic 
heavy-ion collisions have provided an opportunity to systematically 
create and study different phases of the bulk nuclear matter. It is 
widely believed that the QGP phase is formed in heavy-ion collision 
experiments at Relativistic Heavy-Ion Collider (RHIC) located at 
Brookhaven National Laboratory, USA and Large Hadron Collider (LHC) 
at European Organization for Nuclear Research (CERN), Geneva. A 
number of indirect evidences found at the Super Proton Synchrotron 
(SPS) at CERN, strongly suggested the formation of a {\it ``new 
state of matter"} \cite{CERN}, but quantitative and clear results 
were only obtained at RHIC energies \cite{Tannenbaum:2006ch, Kolb, 
Gyulassy, Tomasik, Muller, Arsene:2004fa, Adcox:2004mh, Back:2004je, 
Adams:2005dq}, and recently at LHC energies \cite{Aamodt:2010pa, 
Aamodt:2010pb, Aamodt:2010cz, ALICE:2011ab}. The regime with 
relatively large baryon chemical potential will be probed by the 
upcoming experimental facilities like Facility for Anti-proton and 
Ion Research (FAIR) at GSI, Darmstadt. An illustration of the QCD 
phase diagram and the regions probed by these experimental 
facilities is shown in Fig.~\ref{Phases} \cite{gsi.de}.

\begin{figure}[t]
\begin{center}
\includegraphics[width=\linewidth]{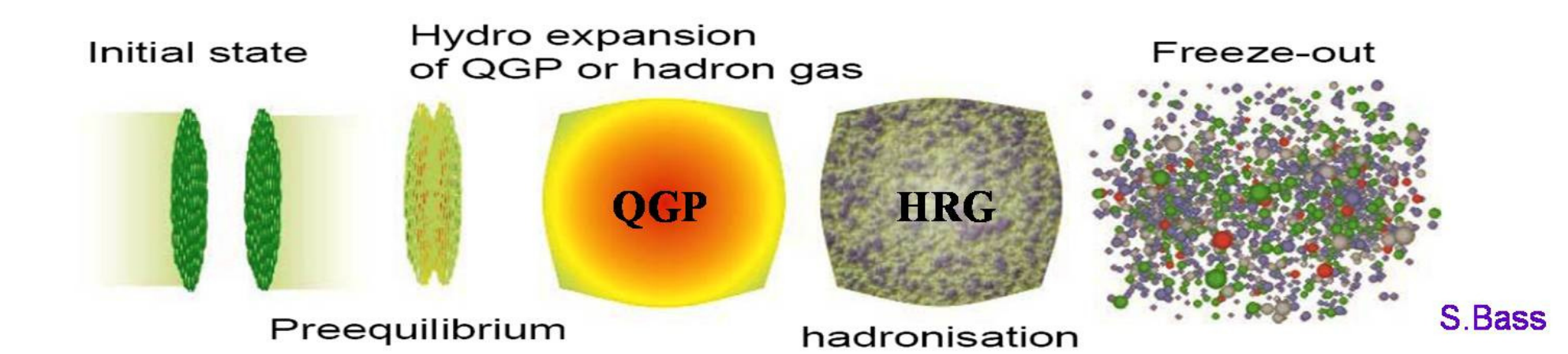}
\end{center} 
\vspace{-0.4cm}
\caption{(Color online) Various stages of ultra-relativistic 
heavy-ion collisions \cite{duke.edu}.}
\label{Stages}
\end{figure}

It is possible to create hot and dense nuclear matter with very high 
energy densities in relatively large volumes by colliding 
ultra-relativistic heavy ions. In these conditions, the nuclear 
matter created may be close to (local) thermodynamic equilibrium, 
providing the opportunity to investigate the various phases and the 
thermodynamic and transport properties of QCD. It is important to 
note that, even though it appears that a deconfined state of matter 
is formed in these colliders, investigating and extracting the 
transport properties of QGP from heavy-ion collisions is not an easy 
task since it cannot be observed directly. Experimentally, it is 
only feasible to measure energy and momenta of the particles 
produced in the final stages of the collision, when nuclear matter 
is already relatively cold and non-interacting. Hence, in order to 
study the thermodynamic and transport properties of the QGP, the 
whole heavy-ion collision process from the very beginning till the 
end has to be modelled: starting from the stage where two highly 
Lorentz contracted heavy nuclei collide with each other, the 
formation and thermalization of the QGP or de-confined phase in the 
initial stages of the collision, its subsequent space-time 
evolution, the phase transition to the hadronic or confined phase of 
matter, and eventually, the dynamics of the cold hadronic matter 
formed in the final stages of the collision. The different stages of 
ultra-relativistic heavy-ion collisions are schematically 
illustrated in Fig.~\ref{Stages} \cite{duke.edu}.

Assuming that thermalization is achieved in the early stages of 
heavy-ion collisions and that the interaction between the quarks is 
strong enough to maintain local thermodynamic equilibrium during the 
subsequent expansion, the time evolution of the QGP and hadronic 
matter can be described by the laws of fluid dynamics \cite 
{Stoecker:1986ci, Rischke:1995ir, Rischke:1995mt, Shuryak:2003xe}. 
Fluid dynamics, also loosely referred to as hydrodynamics, is an 
effective approach through which a system can be described by 
macroscopic variables, such as local energy density, pressure, 
temperature and flow velocity. The most appealing feature of fluid 
dynamics is the fact that it is simple and general. It is simple in 
the sense that all the information of the system is contained in its 
thermodynamic and transport properties, i.e., its equation of state 
and transport coefficients. Fluid dynamics is also general because 
it relies on only one assumption: the system remains close to local 
thermodynamic equilibrium throughout its evolution. Although the 
hypothesis of proximity to local equilibrium is quite strong, it 
saves us from making any further assumption regarding the 
description of the particles and fields, their interactions, the 
classical or quantum nature of the phenomena involved etc. 
Hydrodynamic analysis of the spectra and azimuthal anisotropy of 
particles produced in heavy-ion collisions at RHIC \cite 
{Romatschke:2007mq, Song:2010mg} and recently at LHC \cite 
{Luzum:2010ag, Qiu:2011hf} suggests that the matter formed in these 
collisions is strongly-coupled quark-gluon plasma (sQGP).

Application of viscous hydrodynamics to high-energy heavy-ion 
collisions has evoked widespread interest ever since a surprisingly 
small value of $\eta/s$ was estimated from the analysis of the 
elliptic flow data \cite{Romatschke:2007mq}. It is interesting to 
note that in the strong coupling limit of a large class of 
holographic theories, the value of the shear viscosity to entropy 
density ratio $\eta/s=1/4\pi$. Kovtun, Son and Starinets (KSS) 
conjectured this strong coupling result to be the absolute lower 
bound for all fluids, i.e., $\eta/s\ge1/4\pi$ \cite 
{Policastro:2001yc, Kovtun:2004de}. This specific combination of 
hydrodynamic quantities, $\eta/s$, accounts for the difference 
between momentum and charge diffusion such that even though the 
diffusion constant goes to zero in the strong coupling limit, the 
ratio $\eta/s$ remains finite \cite{Schaefer:2014awa}. Similar 
result for a lower bound also follows from the quantum mechanical 
uncertainty principle. The kinetic theory prediction for viscosity, 
$\eta=\frac{1}{3}nl_{\rm mfp}\bar p$, suggests that low viscosity 
corresponds to short mean free path. On the other hand, the 
uncertainty relation implies that the product of the mean free path 
and the average momentum cannot be arbitrarily small, i.e., $l_{\rm 
mfp}\bar p\gtrsim1$. For a weakly interacting relativistic Bose gas, 
the entropy per particle is given by $s/n=3.6$. This leads to 
$\eta/s\gtrsim0.09$ which is very close to the lower KSS bound.

Indeed the estimated $\eta/s$ of QGP was so close to the KSS bound  
that it led to the claim that the matter formed at RHIC was the most 
perfect fluid ever observed. A precise estimate of $\eta/s$ is vital 
to the understanding of the properties of the QCD matter and is 
presently a topic of intense investigation, see \cite 
{Chaudhuri:2013yna} and references therein for more details. In this 
review, we shall discuss the general aspects of the formulation of 
relativistic fluid dynamics and its application to the physics of 
high-energy heavy-ion collisions. Along with these general concepts 
we shall also discuss here some of the recent developments in the 
field. Among the recent developments the most striking feature is 
the experimental observation of flow like pattern in the particle 
azimuthal distribution of high multiplicity proton-proton (p-p) and 
proton-lead (p-Pb)collisions. We will discuss in this review the 
success of hydrodynamics model in describing these recent 
experimental measurements by assuming hydrodynamics flow of small 
systems. 

The review is organised as follows, In section II we discuss the 
general formalism of a causal theory of relativistic dissipative 
fluid dynamics. Section III deals with the initial conditions 
necessary for the modeling of relativistic heavy-ion, p-p and p-Pb 
collisions. In section IV, we discuss some models of pre-equilibrium 
dynamics employed before hydrodynamic evolution. Section V briefly 
covers various equations of state, necessary to close the 
hydrodynamic equations. In section VI, particle production mechanism 
via Cooper-Frye freeze-out and anisotropic flow generation is 
discussed. In Section VII, we discuss models of hadronic 
rescattering after freeze-out and contribution to particle spectra 
and flow from resonance decays. Section VIII deals with the 
extraction of transport coefficients from hydrodynamic analysis of 
flow data. Finally, in section IX, we discuss recent developments in 
the hydrodynamic modeling of relativistic collisions.

In this review, unless stated otherwise, all physical quantities are 
expressed in terms of natural units, where, $\hbar=c=k_{\rm B}=1$, 
with $\hbar=h/2\pi$, where $h$ is the Planck constant, $c$ the 
velocity of light in vacuum, and $k_{\rm B}$ the Boltzmann constant. 
Unless stated otherwise, the spacetime is always taken to be 
Minkowskian where the metric tensor is given by 
$g_{\mu\nu}=\mbox{diag}(+1,-1,-1,-1)$. Apart from Minkowskian 
coordinates $x^\mu=(t,x,y,z)$, we will also regularly employ Milne 
coordinate system $x^\mu=(\tau,x,y,\eta_s)$ or 
$x^\mu=(\tau,r,\varphi,\eta_s)$, with proper time 
$\tau=\sqrt{t^2-z^2}$, the radial coordinate $r=\sqrt{x^2+y^2}$, the 
azimuthal angle $\varphi=\tan^{-1}(y/x)$, and spacetime rapidity 
$\eta_s=\tanh^{-1}(z/t)$. Hence, $t=\tau\cosh\eta_s$, 
$x=r\cos\varphi$, $y=r\sin\varphi$, and $z=\tau\sinh\eta_s$. For the 
coordinate system $x^\mu=(\tau,x,y,\eta_s)$, the metric becomes 
$g_{\mu\nu}=\mbox{diag}(1,-1,-1,-\tau^2)$, whereas for 
$x^\mu=(\tau,r,\varphi,\eta_s)$, the metric is 
$g_{\mu\nu}=\mbox{diag}(1,-1,-r^2,-\tau^2)$. Roman letters are used 
to indicate indices that vary from 1-3 and Greek letters for indices 
that vary from 0-3. Covariant and contravariant four-vectors are 
denoted as $p_\mu$ and $p^\mu$, respectively. The notation $p\cdot 
q\equiv p_\mu q^\mu$ represents scalar product of a covariant and a 
contravariant four-vector. Tensors without indices shall always 
correspond to Lorentz scalars. We follow Einstein summation 
convention, which states that repeated indices in a single term are 
implicitly summed over all the values of that index.

We denote the fluid four-velocity by $u^\mu$ and the Lorentz 
contraction factor by $\gamma$. The projector onto the space 
orthogonal to $u^\mu$ is defined as: $\Delta^{\mu\nu}\equiv 
g^{\mu\nu}-u^\mu u^\nu$. Hence, $\Delta^{\mu\nu}$ satisfies the 
conditions $\Delta^{\mu\nu} u_\mu=\Delta^{\mu\nu} u_\nu=0$ with 
trace $\Delta^\mu_\mu=3$. The partial derivative $\partial^{\mu}$ 
can then be decomposed as:
\begin{equation*}
\partial^{\mu} = \nabla^\mu + u^{\mu} D, ~~\mathrm{where}~~ 
\nabla^\mu\equiv\Delta^{\mu \nu} \partial_{\nu} ~~\mathrm{and}~~ 
D\equiv u^{\mu}\partial_{\mu}.
\end{equation*}
In the fluid rest frame, $D$ reduces to the time derivative and 
$\nabla^\mu$ reduces to the spacial gradient. Hence, the notation 
$\dot{f}\equiv Df$ is also commonly used. We also frequently use the 
symmetric, anti-symmetric and angular brackets notations defined as
\begin{eqnarray*}
A_{(\mu} B_{\nu)} &\equiv&\frac{1}{2}\left(A_\mu B_\nu+A_\nu
B_\mu\right),
\\
A_{[\mu} B_{\nu]} &\equiv&\frac{1}{2}\left(A_\mu B_\nu-A_\nu
B_\mu\right),
\\
A_{\langle \mu} B_{\nu\rangle} &\equiv&
\Delta^{\alpha\beta}_{\mu\nu} A_\alpha B_\beta.
\end{eqnarray*}
where, 
\begin{equation*}
\Delta^{\alpha\beta}_{\mu\nu}\equiv\frac{1}{2}\left(\Delta^\alpha_\mu \Delta^\beta_\nu +
\Delta^\alpha_\nu \Delta^\beta_\mu-\frac{2}{3} \Delta^{\alpha \beta}
\Delta_{\mu \nu}\right)
\end{equation*}
is the traceless symmetric projection operator orthogonal to $u^\mu$ 
satisfying the conditions $\Delta^{\alpha\beta}_{\mu\nu} 
\Delta_{\alpha\beta}=\Delta^{\alpha\beta}_{\mu\nu}\Delta^{\mu\nu}=0$.


\section{Relativistic fluid dynamics}

The physical characterization of a system consisting of many degrees 
of freedom is in general a non-trivial task. For instance, the 
mathematical formulation of a theory describing the microscopic 
dynamics of a system containing a large number of interacting 
particles is one of the most challenging problems of theoretical 
physics. However, it is possible to provide an effective macroscopic 
description, over large distance and time scales, by taking into 
account only the degrees of freedom that are relevant at these 
scales. This is a consequence of the fact that on macroscopic 
distance and time scales the actual degrees of freedom of the 
microscopic theory are imperceptible. Most of the microscopic 
variables fluctuate rapidly in space and time, hence only average 
quantities resulting from the interactions at the microscopic level 
can be observed on macroscopic scales. These rapid fluctuations lead 
to very small changes of the average values, and hence are not 
expected to contribute to the macroscopic dynamics. On the other 
hand, variables that do vary slowly, such as the conserved 
quantities, are expected to play an important role in the effective 
description of the system. Fluid dynamics is one of the most common 
examples of such a situation. It is an effective theory describing 
the long-wavelength, low frequency limit of the underlying 
microscopic dynamics of a system. 

A fluid is defined as a continuous system in which every 
infinitesimal volume element is assumed to be close to thermodynamic 
equilibrium and to remain near equilibrium throughout its evolution. 
Hence, in other words, in the  neighbourhood of each point in space, 
an infinitesimal volume called fluid element is defined in which the 
matter is assumed to be homogeneous, i.e., any spatial gradients can 
be ignored, and is described by a finite number of thermodynamic 
variables. This implies that each fluid element must be large 
enough, compared to the microscopic distance scales, to guarantee 
the proximity to thermodynamic equilibrium, and, at the same time, 
must be small enough, relative to the macroscopic distance scales, 
to ensure the continuum limit. The co-existence of both continuous 
(zero volume) and thermodynamic (infinite volume) limits within a 
fluid volume might seem paradoxical at first glance. However, if the 
microscopic and the macroscopic length scales of the system are 
sufficiently far apart, it is always possible to establish the 
existence of a volume that is small enough compared to the 
macroscopic scales, and at the same time, large enough compared to 
the microscopic ones. Here, we will assume the existence of a clear 
separation between microscopic and macroscopic scales to guarantee 
the proximity to local thermodynamic equilibrium.  

Relativistic fluid dynamics has been quite successful in explaining 
the various collective phenomena observed in astrophysics, cosmology 
and the physics of high-energy heavy-ion collisions. In cosmology 
and certain areas of astrophysics, one needs a fluid dynamics 
formulation consistent with the General Theory of Relativity \cite 
{Ibanez}. On the other hand, a formulation based on the Special 
Theory of Relativity is quite adequate to treat the evolution of the 
strongly interacting matter formed in high-energy heavy-ion 
collisions when it is close to a local thermodynamic equilibrium. In 
fluid dynamical approach, although no detailed knowledge of the 
microscopic dynamics is needed, however, knowledge of the equation 
of state relating pressure, energy density and baryon density is 
required. The collective behaviour of the hot and dense quark-gluon 
plasma created in ultra-relativistic heavy-ion collisions has been 
studied quite extensively within the framework of relativistic fluid 
dynamics. In application of fluid dynamics, it is natural to first 
employ the simplest version which is ideal hydrodynamics which 
neglects the viscous effects and assumes that local equilibrium is 
always perfectly maintained during the fireball expansion. 
Microscopically, this requires that the microscopic scattering time 
be much shorter than the macroscopic expansion (evolution) time. In 
other words, ideal hydrodynamics assumes that the mean free path of 
the constituent particles is much smaller than the system size. 
However, as all fluids are dissipative in nature due to the quantum 
mechanical uncertainty principle \cite{Danielewicz:1984ww}, the 
ideal fluid results serve only as a benchmark when dissipative 
effects become important.

When discussing the application of relativistic dissipative fluid 
dynamics to heavy-ion collision, one is faced with yet another 
predicament: the theory of relativistic dissipative fluid dynamics 
is not yet conclusively established. In fact, introducing 
dissipation in relativistic fluids is not at all a trivial task and 
still remains one of the important topics of research in high-energy 
physics. Ideal hydrodynamics assumes that local thermodynamic 
equilibrium is perfectly maintained and each fluid element is 
homogeneous, i.e., spatial gradients are absent (zeroth order in 
gradient expansion). If this is not satisfied, dissipative effects 
come into play. The earliest theoretical formulations of 
relativistic dissipative hydrodynamics also known as first-order 
theories, are due to Eckart \cite{Eckart:1940zz} and Landau-Lifshitz 
\cite{Landau}. However, these formulations, collectively called 
relativistic Navier-Stokes (NS) theory, suffer from acausality and 
numerical instability. The reason for the acausality is that in the 
gradient expansion the dissipative currents are linearly 
proportional to gradients of temperature, chemical potential, and 
velocity, resulting in parabolic equations. Thus, in Navier-Stokes 
theory the gradients have an instantaneous influence on the 
dissipative currents. Such instantaneous effects tend to violate 
causality and cannot be allowed in a covariant setup, leading to the 
instabilities investigated in Refs.~\cite {Hiscock:1983zz, 
Hiscock:1985zz, Hiscock:1987zz}.

The second-order Israel-Stewart (IS) theory \cite{Israel:1979wp}, 
restores causality but may not guarantee stability \cite 
{Huovinen:2008te}. The acausality problems were solved by 
introducing a time delay in the creation of the dissipative currents 
from gradients of the fluid-dynamical variables. In this case, the 
dissipative quantities become independent dynamical variables 
obeying equations of motion that describe their relaxation towards 
their respective Navier-Stokes values. The resulting equations are 
hyperbolic in nature which preserves causality. Israel-Stewart 
theory has been widely applied to ultra-relativistic heavy-ion 
collisions in order to describe the time evolution of the QGP and 
the subsequent freeze-out process of the hadron resonance gas. 
Although IS hydrodynamics has been quite successful in modelling 
relativistic heavy-ion collisions, there are several inconsistencies 
and approximations in its formulation which prevent proper 
understanding of the thermodynamic and transport properties of the 
QGP. Moreover, the second-order IS theory can be derived in several 
ways, each leading to a different set of transport coefficients. 
Therefore, in order to quantify the transport properties of the QGP 
from experiment and confirm the claim that it is indeed the most 
perfect fluid ever created, the theoretical foundations of 
relativistic dissipative fluid dynamics must be first addressed and 
clearly understood. In this section, we review the basic aspects of 
thermodynamics and discuss the formulation of relativistic fluid 
dynamics from a phenomenological perspective. The salient features 
of kinetic theory in the context of fluid dynamics will also be 
discussed. 


\subsection{Thermodynamics}

Thermodynamics is an empirical description of the macroscopic or 
large-scale properties of matter and it makes no hypotheses about 
the small-scale or microscopic structure. It is concerned only with 
the average behaviour of a very large number of microscopic 
constituents, and its laws can be derived from statistical 
mechanics. A thermodynamic system can be described in terms of a 
small set of extensive variables, such as volume ($V$), the total 
energy ($E$), entropy ($S$), and number of particles ($N$), of the 
system. Thermodynamics is based on four phenomenological laws that 
explain how these quantities are related and how they change with 
time \cite{Fermi, Reif, Reichl}.
\begin{itemize}
\item {\bf Zeroth Law}: If two systems are both in thermal 
equilibrium with a third system then they are in thermal equilibrium 
with each other. This law helps define the notion of temperature.
\item {\bf First Law}: All the energy transfers must be accounted 
for to ensure the conservation of the total energy of a 
thermodynamic system and its surroundings. This law is the principle 
of conservation of energy.
\item {\bf Second Law}: An isolated physical system spontaneously 
evolves towards its own internal state of thermodynamic equilibrium. 
Employing the notion of entropy, this law states that the change in 
entropy of a closed thermodynamic system is always positive or zero. 
\item {\bf Third Law}: Also known an Nernst's heat theorem, states 
that the difference in entropy between systems connected by a 
reversible process is zero in the limit of vanishing temperature. In 
other words, it is impossible to reduce the temperature of a system 
to absolute zero in a finite number of operations. 
\end{itemize}

The first law of thermodynamics postulates that the changes in the 
total energy of a thermodynamic system must result from: {\bf (1)} 
heat exchange, {\bf (2)} the mechanical work done by an external 
force, and {\bf (3)} from particle exchange with an external medium. 
Hence the conservation law relating the small changes in state 
variables, $E$, $V$, and $N$ is
\begin{equation}\label{FLTI}
\delta E = \delta Q - P\delta V + \mu\,\delta N,
\end{equation}
where $P$ and $\mu$ are the pressure and chemical potential, 
respectively, and $\delta Q$ is the amount of heat exchange.

The heat exchange takes into account the energy variations due to 
changes of internal degrees of freedom that are not described by the 
state variables. The heat itself is not a state variable since it 
can depend on the past evolution of the system and may take several 
values for the same thermodynamic state. However, when dealing with 
reversible processes (in time), it becomes possible to assign a 
state variable related to heat. This variable is the entropy, $S$ , 
and is defined in terms of the heat exchange as $\delta Q = T\delta 
S$, with the temperature $T$ being the proportionality constant. 
Then, when considering variations between equilibrium states that 
are infinitesimally close to each other, it is possible to write the 
first law of thermodynamics in terms of differentials of the state 
variables,
\begin{equation}\label{FLT}
dE = TdS - PdV + \mu\, dN.
\end{equation}
Hence, using Eq.~(\ref{FLT}), the intensive quantities, $T$, $\mu$ 
and $P$, can be obtained in terms of partial derivatives of the 
entropy as
\begin{equation}\label{DSDEDNDV}
\left.{\frac{\partial S}{\partial E}}\right\vert_{N,V} = \frac{1}{T}, \qquad
\left.{\frac{\partial S}{\partial V}}\right\vert_{N,E} = \frac{P}{T}, \qquad
\left.{\frac{\partial S}{\partial N}}\right\vert_{E,V} = -\frac{\mu}{T}.
\end{equation}

The entropy is mathematically defined as an extensive and additive 
function of the state variables, which means that
\begin{equation}\label{EEFSV}
S(\lambda E, \lambda V, \lambda N) = \lambda S(E, V, N).
\end{equation}
Differentiating both sides with respect to $\lambda$, we obtain
\begin{equation}\label{DSDL}
S = E\left.{\frac{\partial S}{\partial\lambda E}}\right\vert_{\lambda N,\lambda V} 
+ V\left.{\frac{\partial S}{\partial\lambda V}}\right\vert_{\lambda N,\lambda E} 
+ N\left.{\frac{\partial S}{\partial\lambda N}}\right\vert_{\lambda E,\lambda V},
\end{equation}
which holds for any arbitrary value of $\lambda$. Setting $\lambda=1$
and using Eq.~(\ref{DSDEDNDV}), we obtain the so-called Euler's 
relation
\begin{equation}\label{TDER}
E = - PV + TS + \mu\, N.
\end{equation}
Using Euler's relation, Eq.~(\ref{TDER}), along with the first law 
of thermodynamics, Eq.~(\ref{FLT}), we arrive at the Gibbs-Duhem 
relation 
\begin{equation}\label{GDR}
VdP = SdT + Nd\mu.
\end{equation}

In terms of energy, entropy and number densities defined as 
$\epsilon\equiv E/V$, $s\equiv S/V$, and $n\equiv N/V$ respectively, 
the Euler's relation, Eq.~(\ref{TDER}) and Gibbs-Duhem relation, 
Eq.~(\ref{GDR}), reduce to 
\begin{align}
\epsilon &= - P + Ts + \mu\, n  \label{ERD} \\
dP &= s\,dT + n\,d\mu. \label{GDRD}
\end{align}
Differentiating Eq.~(\ref{ERD}) and using Eq.~(\ref{GDRD}), we 
obtain the relation analogous to first law of thermodynamics
\begin{equation}\label{ERGDR}
d\epsilon = Tds + \mu\, dn \quad\Rightarrow\quad ds = \frac{1}{T}\, d\epsilon - \frac{\mu}{T}\, dn.
\end{equation}
It is important to note that all the densities defined above 
$(\epsilon,~s,~n)$ are intensive quantities.

The equilibrium state of a system is defined as a stationary state 
where the extensive and intensive variables of the system do not 
change. We know from the second law of thermodynamics that the 
entropy of an isolated thermodynamic system must either increase or 
remain constant. Hence, if a thermodynamic system is in equilibrium, 
the entropy of the system being an extensive variable, must remain 
constant. On the other hand, for a system that is out of 
equilibrium, the entropy must always increase. This is an extremely 
powerful concept that will be extensively used in this section to 
constrain and derive the equations of motion of a dissipative fluid. 
This concludes a brief outline of the basics of thermodynamics; for 
a more detailed review, see Ref.~\cite{Reichl}. In the next section, 
we introduce and derive the equations of relativistic ideal fluid 
dynamics. 


\subsection{Relativistic ideal fluid dynamics}

An ideal fluid is defined by the assumption of local thermal 
equilibrium, i.e., all fluid elements must be exactly in 
thermodynamic equilibrium \cite{Landau, Weinberg}. This means that 
at each space-time coordinate of the fluid $x\equiv x^\mu$, there 
can be assigned a temperature $T(x)$, a chemical potential $\mu(x)$, 
and a collective four-velocity field, 
\begin{equation}\label{umu}
u^\mu(x)\equiv \frac{dx^\mu}{d\tau}. 
\end{equation}
The proper time increment $d\tau$ is given by the line element
\begin{equation}\label{dtau2}
(d\tau)^2 = g_{\mu\nu}dx^\mu dx^\nu = (dt)^2 - (d\vec x)^2 
= (dt)^2\left[ 1 - (\vec v)^2\right],
\end{equation}
where $\vec v \equiv d\vec x/dt$. This implies that
\begin{equation}\label{umuf}
u^\mu(x) = \frac{dt}{d\tau}\frac{dx^\mu}{dt} = \gamma(\vec v)\binom{1}{\vec v}
\end{equation}
where $\gamma(\vec v) = 1/\sqrt{1-\vec v^2}$. In the 
non-relativistic limit, we obtain $u^\mu(x)=(1,\vec v)$. It is 
important to note that the four-vector $u^\mu(x)$ only contains 
three independent components since it obeys the relation
\begin{equation}\label{umuumu}
u^2 \equiv u^\mu(x) g_{\mu\nu} u^\nu(x) = \gamma^2(\vec v)\left( 1 - \vec v^2 \right) = 1.
\end{equation}
The quantities $T$, $\mu$ and $u^\mu$ are often referred to as the 
primary fluid-dynamical variables.

The state of a fluid can be completely specified by the densities 
and currents associated with conserved quantities, i.e., energy, 
momentum, and (net) particle number. For a relativistic fluid, the 
state variables are the energy- momentum tensor, $T^{\mu\nu}$, and 
the (net) particle four-current, $N^\mu$. To obtain the general form 
of these currents for an ideal fluid, we first define the local rest 
frame (LRF) of the fluid. In this frame, $\vec v=0$, and the 
energy-momentum tensor, $T^{\mu\nu}_{LRF}$, the (net) particle 
four-current, $N^\mu_{LRF}$, and the entropy four-current, 
$S^\mu_{LRF}$, should have the characteristic form of a system in 
static equilibrium. In other words, in local rest frame, there is no 
flow of energy ($T^{i0}_{LRF}=0$), the force per unit surface 
element is isotropic ($T^{ij}_{LRF}=\delta^{ij} P$) and there is no 
particle and entropy flow ($\vec N=0$ and $\vec S=0$). Consequently, 
the energy-momentum tensor, particle and entropy four-currents in 
this frame take the following simple forms
\begin{align}\label{TNSLRF}
&T^{\mu\nu}_{LRF} = \left( \begin{array}{cccc}
\epsilon & 0 & 0 & 0 \\
0 & P & 0 & 0 \\
0 & 0 & P & 0 \\
0 & 0 & 0 & P \end{array} 
\right), \nonumber\\
&N^\mu_{LRF} = \left( \begin{array}{c}
n  \\
0  \\
0  \\
0  \end{array} 
\right) , \qquad
S^\mu_{LRF} = \left( \begin{array}{c}
s  \\
0  \\
0  \\
0  \end{array} 
\right).
\end{align}

For an ideal relativistic fluid, the general form of the 
energy-momentum tensor, $T^{\mu\nu}_{(0)}$, (net) particle 
four-current, $N^\mu_{(0)}$, and the entropy four-current, 
$S^\mu_{(0)}$, has to be built out of the hydrodynamic tensor 
degrees of freedom, namely the vector, $u^\mu$, and the metric 
tensor, $g_{\mu\nu}$. Since $T^{\mu\nu}_{(0)}$ should be symmetric 
and transform as a tensor, and, $N^\mu_{(0)}$ and $S^\mu_{(0)}$ 
should transform as a vector, under Lorentz transformations, the 
most general form allowed is therefore
\begin{equation}\label{TNSTD}
T^{\mu\nu}_{(0)} = c_1 u^\mu u^\nu + c_2 g^{\mu\nu} ,\quad
N^\mu_{(0)} = c_3 u^\mu,\quad
S^\mu_{(0)} = c_4 u^\mu.
\end{equation}
In the local rest frame, $\vec v=0 \Rightarrow u^\mu=(1,\vec 0)$. 
Hence in this frame, Eq.~(\ref{TNSTD}) takes the form
\begin{align}\label{TNSLRFID}
&T^{\mu\nu}_{(0)LRF} = \left( \begin{array}{cccc}
c_1+c_2 & 0 & 0 & 0 \\
0 & -c_2 & 0 & 0 \\
0 & 0 & -c_2 & 0 \\
0 & 0 & 0 & -c_2 \end{array} 
\right) , \nonumber\\
&N^\mu_{(0)LRF} = \left( \begin{array}{c}
c_3  \\
0  \\
0  \\
0  \end{array} 
\right) , \qquad
S^\mu_{(0)LRF} = \left( \begin{array}{c}
c_4  \\
0  \\
0  \\
0  \end{array} 
\right).
\end{align}
By comparing the above equation with the corresponding general 
expressions in the local rest frame, Eq.~(\ref {TNSLRF}), one 
obtains the following expressions for the coefficients
\begin{equation}\label{c1c2c3c4}
c_1 = \epsilon + P, \quad c_2 = -P, \quad c_3 = n, \quad c_4 = s.
\end{equation}  
The conserved currents of an ideal fluid can then be expressed as
\begin{equation}\label{CCIF}
T^{\mu\nu}_{(0)} = \epsilon u^\mu u^\nu - P \Delta^{\mu\nu} ,\quad
N^\mu_{(0)} = n u^\mu,\quad
S^\mu_{(0)} = s u^\mu,
\end{equation}
where $\Delta^{\mu\nu}=g^{\mu\nu}-u^\mu u^\nu$ is the projection 
operator onto the three-space orthogonal to $u^\mu$, and satisfies 
the following properties of an orthogonal projector,
\begin{equation}\label{DMUNU}
u_\mu\Delta^{\mu\nu} = \Delta^{\mu\nu}u_\nu = 0, \quad
\Delta^\mu_\rho \Delta^{\rho\nu} = \Delta^{\mu\nu}, \quad
\Delta^\mu_\mu = 3.
\end{equation} 

The dynamical description of an ideal fluid is obtained using the 
conservation laws of energy, momentum and (net) particle number. 
These conservation laws can be mathematically expressed using the 
four-divergences of energy-momentum tensor and particle four-current 
which leads to the following equations,
\begin{equation}\label{IFDCE}
\partial_\mu T^{\mu\nu}_{(0)} = 0, \quad 
\partial_\mu N^\mu_{(0)} = 0,
\end{equation}
where the partial derivative $\partial_\mu \equiv \partial/\partial 
x^\mu$ transforms as a covariant vector under Lorentz 
transformations. Using the four-velocity, $u^\mu$, and the 
projection operator, $\Delta^{\mu\nu}$, the derivative, 
$\partial_\mu$, can be projected along and orthogonal to $u^\mu$
\begin{equation}\label{PFDAOU}
D \equiv u^\mu \partial_\mu, \quad \nabla_\mu \equiv \Delta_\mu^\rho \partial_\rho,
\quad\Rightarrow\quad  \partial_\mu = u_\mu D + \nabla_\mu.
\end{equation}
Projection of energy-momentum conservation equation along and 
orthogonal to $u^\mu$ together with the conservation law for 
particle number, leads to the equations of motion of ideal fluid 
dynamics,
\begin{align}
u_\mu\partial_\nu T^{\mu\nu}_{(0)} = 0 &\quad\Rightarrow\quad 
D\epsilon + (\epsilon+P)\theta = 0,   \label{IFDE1}\\ 
\Delta^\alpha_\mu\partial_\nu T^{\mu\nu}_{(0)} = 0 &\quad\Rightarrow\quad
(\epsilon+P)Du^\alpha -\nabla^\alpha P = 0,   \label{IFDE2}\\
\partial_\mu N^\mu_{(0)} = 0 &\quad\Rightarrow\quad
Dn + n\theta = 0,   \label{IFDE3}
\end{align}
where $\theta\equiv\partial_\mu u^\mu$. It is important to note that 
an ideal fluid is described by four fields, $\epsilon$, $P$, $n$, 
and $u_\mu$, corresponding to six independent degrees of freedom. The 
conservation laws, on the other hand, provide only five equations of 
motion. The equation of state of the fluid, $P=P(n,\epsilon)$, 
relating the pressure to other thermodynamic variables has to be 
specified to close this system of equations. The existence of 
equation of state is guaranteed by the assumption of local thermal 
equilibrium and hence the equations of ideal fluid dynamics are 
always closed.


\subsection{Covariant thermodynamics}

In the following, we re-write the equilibrium thermodynamic 
relations derived in Sec. 2.1, Eqs.~(\ref{ERD}), (\ref {GDRD}), and 
(\ref{ERGDR}), in a covariant form \cite{Israel:1979wp, 
Israel:1976tn}. For this purpose, it is convenient to introduce the 
following notations
\begin{equation}\label{CTND}
\beta \equiv \frac{1}{T}, \quad \alpha \equiv \frac{\mu}{T}, \quad \beta^\mu \equiv \frac{u^\mu}{T}.
\end{equation}
In these notations, the covariant version of the Euler's relation, 
Eq.~(\ref{ERD}), and the Gibbs-Duhem relation, Eq.~(\ref{GDRD}), can 
be postulated as,
\begin{align}
S^\mu_{(0)} &= P\beta^\mu + \beta_\nu T^{\mu\nu}_{(0)} - \alpha N^\mu_{(0)}, \label{CER}\\
d\left(P\beta^\mu\right) &= N^\mu_{(0)} d\alpha - T^{\mu\nu}_{(0)} d\beta_\nu, \label{CGDR}
\end{align}
respectively. The above equations can then be used to derive a 
covariant form of the first law of thermodynamics, Eq.~(\ref{ERGDR}),
\begin{equation}\label{CFLT}
dS^\mu_{(0)} = \beta_\nu dT^{\mu\nu}_{(0)} - \alpha dN^\mu_{(0)}.
\end{equation}

The covariant thermodynamic relations were constructed in such a way 
that when Eqs.~(\ref{CER}), (\ref{CGDR}) and (\ref{CFLT}) are 
contracted with $u_\mu$,
\begin{align}
u_\mu \left[ S^\mu_{(0)} - P\beta^\mu - \beta_\nu T^{\mu\nu}_{(0)} + \alpha N^\mu_{(0)} \right] &= 0  \nonumber\\
\Rightarrow~ s + \alpha n -\beta (\epsilon+P) &= 0, \\
u_\mu \left[ d\left(P\beta^\mu\right) - N^\mu_{(0)} d\alpha + T^{\mu\nu}_{(0)} d\beta_\nu \right] &= 0 \nonumber\\
\Rightarrow~ d(\beta P) - nd\alpha + \epsilon d\beta &= 0, \\
u_\mu \left[ dS^\mu_{(0)} - \beta_\nu dT^{\mu\nu}_{(0)} + \alpha dN^\mu_{(0)} \right] &= 0 \nonumber\\
\Rightarrow~ ds - \beta d\epsilon + \alpha dn &= 0,
\end{align}
we obtain the usual thermodynamic relations, Eqs.~(\ref{ERD}), (\ref 
{GDRD}), and (\ref{ERGDR}). Here we have used the property of the 
fluid four-velocity, $u_\mu u^\mu=1\Rightarrow u_\mu du^\mu=0$. The 
projection of Eqs.~(\ref{CER}), (\ref{CGDR}) and (\ref{CFLT}) onto 
the three-space orthogonal to $u^\mu$ just leads to trivial identities,
\begin{align}
\Delta^\alpha_\mu \left[ S^\mu_{(0)} - P\beta^\mu - \beta_\nu T^{\mu\nu}_{(0)} + \alpha N^\mu_{(0)} \right]\! &=0  
~\Rightarrow~ 0 = 0, \\
\Delta^\alpha_\mu \left[ d\left(P\beta^\mu\right) - N^\mu_{(0)} d\alpha + T^{\mu\nu}_{(0)} d\beta_\nu \right]\! &=0 
~\Rightarrow~ 0 = 0, \\
\Delta^\alpha_\mu \left[ dS^\mu_{(0)} - \beta_\nu dT^{\mu\nu}_{(0)} + \alpha dN^\mu_{(0)} \right]\! &=0
~\Rightarrow~ 0 = 0.
\end{align}
From the above equations we conclude that the covariant 
thermodynamic relations do not contain more information than the 
usual thermodynamic relations.

The first law of thermodynamics, Eq.~(\ref{CFLT}), leads to the 
following expression for the entropy four-current divergence,
\begin{equation}\label{CTEFCD}
\partial_\mu S^\mu_{(0)} = \beta_\mu\partial_\nu T^{\mu\nu}_{(0)} - \alpha\partial_\mu N^\mu_{(0)}.
\end{equation}
After employing the conservation of energy-momentum and net particle 
number, Eq.~(\ref{IFDCE}), the above equation leads to the 
conservation of entropy, $\partial_\mu S^\mu_{(0)} = 0$. It is 
important to note that within equilibrium thermodynamics, the 
entropy conservation is a natural consequence of energy-momentum and 
particle number conservation, and the first law of thermodynamics. 
The equation of motion for the entropy density is then obtained as
\begin{equation}\label{CTECE}
\partial_\mu S^\mu_{(0)} = 0 \quad\Rightarrow\quad Ds + s\theta = 0.
\end{equation}
We observe that the rate equation of the entropy density in the 
above equation is identical to that of the net particle number, 
Eq.~(\ref{IFDE3}). Therefore, we conclude that for ideal 
hydrodynamics, the ratio of entropy density to number density ($s/n$) 
is a constant of motion.


\subsection{Relativistic dissipative fluid dynamics}

The derivation of relativistic ideal fluid dynamics proceeds by 
employing the properties of the Lorentz transformation, the 
conservation laws, and most importantly, by imposing local 
thermodynamic equilibrium. It is important to note that while the 
properties of Lorentz transformation and the conservation laws are 
robust, the assumption of local thermodynamic equilibrium is a 
strong restriction. The deviation from local thermodynamic 
equilibrium results in dissipative effects, and, as all fluids are 
dissipative in nature due to the uncertainty principle \cite 
{Danielewicz:1984ww}, the assumption of local thermodynamic 
equilibrium is never strictly realized in practice. In the 
following, we consider a more general theory of fluid dynamics that 
attempts to take into account the dissipative processes that must 
happen, because a fluid can never maintain exact local thermodynamic 
equilibrium throughout its dynamical evolution. 

Dissipative effects in a fluid originate from irreversible 
thermodynamic processes that occur during the motion of the fluid. 
In general, each fluid element may not be in equilibrium with the 
whole fluid, and, in order to approach equilibrium, it 
exchanges heat with its surroundings. Moreover, the fluid elements 
are in relative motion and can also dissipate energy by friction. 
All these processes must be included in order to obtain a reasonable 
description of a relativistic fluid.

The earliest covariant formulation of dissipative fluid dynamics 
were due to Eckart \cite{Eckart:1940zz}, in 1940, and, later, by 
Landau and Lifshitz \cite{Landau}, in 1959. The formulation of these 
theories, collectively known as first-order theories (order of 
gradients), was based on a covariant generalization of the 
Navier-Stokes theory. The Navier-Stokes theory, at that time, had 
already become a successful theory of dissipative fluid dynamics. It 
was employed efficiently to describe a wide variety of 
non-relativistic fluids, from weakly coupled gases such as air, to 
strongly coupled fluids such as water. Hence, a relativistic 
generalisation of Navier-Stokes theory was considered to be the most 
effective and promising way to describe relativistic dissipative 
fluids. 

The formulation of relativistic dissipative hydrodynamics turned out 
to be more subtle since the relativistic generalisation of 
Navier-Stokes theory is intrinsically unstable \cite{Hiscock:1983zz, 
Hiscock:1985zz, Hiscock:1987zz}. The source of such instability is 
attributed to the inherent acausal behaviour of this theory \cite
{Denicol:2008ha, Pu:2009fj}. A straightforward relativistic 
generalisation of Navier-Stokes theory allows signals to propagate 
with infinite speed in a medium. While in non-relativistic theories, 
this does not give rise to an intrinsic problem and can be ignored, 
in relativistic systems where causality is a physical property that 
is naturally preserved, this feature leads to intrinsically unstable 
equations of motion. Nevertheless, it is instructive to review the 
first-order theories as they are an important initial step to 
illustrate the basic features of relativistic dissipative 
fluid-dynamics.

As in the case of ideal fluids, the basic equations governing the 
motion of dissipative fluids are also obtained from the conservation 
laws of energy-momentum and (net) particle number,
\begin{equation}\label{DFDCE}
\partial_\mu T^{\mu\nu} = 0, \quad 
\partial_\mu N^\mu = 0.
\end{equation}
However, for dissipative fluids, the energy-momentum tensor is no 
longer diagonal and isotropic in the local rest frame. Moreover, due 
to diffusion, the particle flow is expected to appear in the local 
rest frame of the fluid element. To account for these effects, 
dissipative currents $\tau^{\mu\nu}$ and $n^\mu$ are added to the 
previously derived ideal currents, $T^{\mu\nu}_{(0)}$ and 
$N^\mu_{(0)}$,
\begin{align}\label{CCDF}
T^{\mu\nu} &= T^{\mu\nu}_{(0)} + \tau^{\mu\nu} = \epsilon u^\mu u^\nu - P \Delta^{\mu\nu} + \tau^{\mu\nu},\\
N^\mu &= N^\mu_{(0)} + n^\mu = n u^\mu + n^\mu,
\end{align}
where, $\tau^{\mu\nu}$ is required to be symmetric 
($\tau^{\mu\nu}=\tau^{\nu\mu}$) in order to satisfy angular momentum 
conservation. The main objective then becomes to find the dynamical 
or constitutive equations satisfied by these dissipative currents.


\subsubsection{Matching conditions}

The introduction of the dissipative currents causes the equilibrium 
variables to be ill-defined, since the fluid can no longer be 
considered to be in local thermodynamic equilibrium. Hence, in a 
dissipative fluid, the thermodynamic variables can only be defined 
in terms of an artificial equilibrium state, constructed such that 
the thermodynamic relations are valid as if the fluid were in local 
thermodynamic equilibrium. The first step to construct such an 
equilibrium state is to define $\epsilon$ and $n$ as the total 
energy and particle density in the local rest frame of the fluid, 
respectively. This is guaranteed by imposing the so-called matching 
or fitting conditions \cite{Israel:1979wp},
\begin{equation}\label{MFC}
\epsilon \equiv u_\mu u_\nu T^{\mu\nu} ,\quad
n \equiv u_\mu N^\mu .
\end{equation}
These matching conditions enforces the following constraints on the 
dissipative currents
\begin{equation}\label{RMFC}
u_\mu u_\nu \tau^{\mu\nu} = 0 ,\quad
u_\mu n^\mu = 0 .
\end{equation}
Subsequently, using $n$ and $\epsilon$, an artificial equilibrium 
state can be constructed with the help of the equation of state. It 
is however important to note that while the energy and particle 
densities are physically defined, all the other thermodynamic 
quantities ($s,~P,~T,~\mu,\cdots $) are defined only in terms of 
an artificial equilibrium state and do not necessarily retain their 
usual physical meaning.


\subsubsection{Tensor decompositions of dissipative quantities}

To proceed further, it is convenient to decompose $\tau^{\mu\nu}$ in 
terms of its irreducible components, i.e., a scalar, a four-vector, 
and a traceless and symmetric second-rank tensor. Moreover, this 
tensor decomposition must be consistent with the matching or 
orthogonality condition, Eq.~(\ref {RMFC}), satisfied by 
$\tau^{\mu\nu}$. To this end, we introduce another projection 
operator, the double symmetric, traceless projector orthogonal to 
$u^\mu$,
\begin{equation}\label{DSTPO}
\Delta^{\mu\nu}_{\alpha\beta} \equiv \frac{1}{2}\left(\Delta^{\mu}_{\alpha}\Delta^{\nu}_{\beta} 
+ \Delta^{\mu}_{\beta}\Delta^{\nu}_{\alpha} - \frac{2}{3}\Delta^{\mu\nu}\Delta_{\alpha\beta}\right),
\end{equation}
with the following properties,
\begin{align}\label{PDSTPO}
\Delta^{\mu\nu}_{~~~\alpha\beta} &= \Delta^{~~~\mu\nu}_{\alpha\beta}, \quad
\Delta^{\mu\nu}_{\rho\sigma}\Delta^{\rho\sigma}_{\alpha\beta} = \Delta^{\mu\nu}_{\alpha\beta}, \\
u_\mu\Delta^{\mu\nu}_{\alpha\beta} &= g_{\mu\nu}\Delta^{\mu\nu}_{\alpha\beta} = 0, \quad
\Delta^{\mu\nu}_{\mu\nu} = 5.
\end{align}
The parentheses in the above equation denote symmetrization of the 
Lorentz indices, i.e., $A^{(\mu\nu)}\equiv(A^{\mu\nu}+A^{\nu\mu})/2$. 
The dissipative current $\tau^{\mu\nu}$ then can be tensor 
decomposed in its irreducible form by using $u^\mu$, 
$\Delta^{\mu\nu}$ and $\Delta^{\mu\nu}_{\alpha\beta}$ as
\begin{equation}\label{TDT}
\tau^{\mu\nu} \equiv -\Pi\Delta^{\mu\nu} + 2u^{(\mu}h^{\nu)} + \pi^{\mu\nu},
\end{equation}
where we have defined
\begin{equation}\label{DDQ}
\Pi \equiv -\frac{1}{3}\Delta_{\alpha\beta}\tau^{\alpha\beta}, \quad
h^\mu \equiv \Delta^\mu_\alpha u_\beta\tau^{\alpha\beta}, \quad
\pi^{\mu\nu} \equiv \Delta^{\mu\nu}_{\alpha\beta}\tau^{\alpha\beta}.
\end{equation}
The scalar $\Pi$ is the bulk viscous pressure, the vector $h^\mu$ is 
the energy-diffusion four-current, and the second-rank tensor 
$\pi^{\mu\nu}$ is the shear-stress tensor. The properties of the 
projection operators $\Delta^\mu_\alpha$ and 
$\Delta^{\mu\nu}_{\alpha\beta}$ imply that both $h^\mu$ and 
$\pi^{\mu\nu}$ are orthogonal to $u^\mu$ and, additionally, 
$\pi^{\mu\nu}$ is traceless. Armed with these definitions, all the 
irreducible hydrodynamic fields are expressed in terms of $N^\mu$ 
and $T^{\mu\nu}$ as
\begin{align}\label{IHFD}
\epsilon &= u_\alpha u_\beta T^{\alpha\beta}, \quad
n = u_\alpha N^\alpha, \quad
\Pi = -P -\frac{1}{3}\Delta_{\alpha\beta}T^{\alpha\beta}, \nonumber\\
h^\mu &= u_\alpha T^{\langle\mu\rangle\alpha}, \quad
n^\mu = N^{\langle\mu\rangle}, \quad
\pi^{\mu\nu} = T^{\langle\mu\nu\rangle},
\end{align}
where the angular bracket notations are defined as, 
$A^{\langle\mu\rangle}\equiv\Delta^\mu_\alpha A^\alpha$ and 
$B^{\langle\mu\nu\rangle}\equiv\Delta^{\mu\nu}_{\alpha\beta} 
B^{\alpha\beta}$.

We observe that $T^{\mu\nu}$ is a symmetric second-rank tensor with 
ten independent components and $N^\mu$ is a four-vector; overall 
they have fourteen independent components. Next we count the number 
of independent components in the tensor decompositions of 
$T^{\mu\nu}$ and $N^\mu$. Since $n^\mu$ and $h^\mu$ are orthogonal 
to $u^\mu$, they can have only three independent components each. 
The shear-stress tensor $\pi^{\mu\nu}$ is symmetric, traceless and 
orthogonal to $u^\mu$, and hence, can have only five independent 
components. Together with $u^\mu$, $\epsilon$, $n$ and $\Pi$, which 
have in total six independent components ($P$ is related to 
$\epsilon$ via equation of state), we count a total of seventeen 
independent components, three more than expected. The reason being 
that so far, the velocity field $u^\mu$ was introduced as a general 
normalized four-vector and was not specified. Hence $u^\mu$ has to 
be defined to reduce the number of independent components to the 
correct value.


\subsubsection{Definition of the velocity field}

In the process of formulating the theory of dissipative fluid 
dynamics, the next important step is to fix $u^\mu$. In the case of 
ideal fluids, the local rest frame was defined as the frame in which 
there is, simultaneously, no net energy and particle flow. While the 
definition of local rest frame was unambiguous for ideal fluids, 
this definition is no longer possible in the case of dissipative 
fluids due to the presence of both energy and particle diffusion. 
From a mathematical perspective, the fluid velocity can be defined 
in numerous ways. However, from the physical perspective, there are 
two natural choices. The \emph{Eckart definition} \cite
{Eckart:1940zz}, in which the velocity is defined by the flow of 
particles
\begin{equation}\label{EDF}
N^\mu = nu^\mu \quad \Rightarrow \quad n^\mu = 0,
\end{equation}
and the \emph{Landau definition} \cite{Landau}, in which the 
velocity is specified by the flow of the total energy
\begin{equation}\label{LDF}
u_\nu T^{\mu\nu} = \epsilon u^\mu \quad \Rightarrow \quad h^\mu = 0.
\end{equation}

We note that the above two definitions of $u^\mu$ impose different 
constraints on the dissipative currents. In the Eckart definition 
the particle diffusion is always set to zero, while in the Landau 
definition, the energy diffusion is zero. In other words, the Eckart 
definition of the velocity field eliminates any diffusion of 
particles whereas the Landau definition eliminates any diffusion of 
energy. In this review, we shall always use the Landau definition, 
Eq.~(\ref{LDF}). The conserved currents in this frame take the 
following form
\begin{equation}\label{CCDLF}
T^{\mu\nu} =  \epsilon u^\mu u^\nu - (P+\Pi) \Delta^{\mu\nu} + \pi^{\mu\nu},\quad
N^\mu = n u^\mu + n^\mu.
\end{equation}

As done for ideal fluids, the energy-momentum conservation equation 
in Eq.~(\ref{DFDCE}) is decomposed parallel and orthogonal to $u^\mu$. 
Using Eq.~(\ref{CCDLF}) together with the conservation law for 
particle number in Eq.~(\ref{DFDCE}), leads to the equations of 
motion for dissipative fluids. For $u_\mu\partial_\nu T^{\mu\nu}=0$, 
$\Delta^\alpha_\mu\partial_\nu T^{\mu\nu}=0$ and $\partial_\mu 
N^\mu=0$, one obtains
\begin{align} 
\dot\epsilon + (\epsilon+P+\Pi)\theta - \pi^{\mu\nu}\sigma_{\mu\nu} &= 0,   \label{DFDE1}\\ 
(\epsilon+P+\Pi)\dot u^\alpha -\nabla^\alpha (P+\Pi) + \Delta^\alpha_\mu\partial_\nu\pi^{\mu\nu} &= 0,   \label{DFDE2}\\
\dot n + n\theta + \partial_\mu n^\mu &= 0,   \label{DFDE3}
\end{align}
respectively. Here $\dot A\equiv DA=u^\mu\partial_\mu A$, and the 
shear tensor $\sigma^{\mu\nu}\equiv\nabla^{\langle\mu}u^{\nu\rangle} 
=\Delta^{\mu\nu}_{\alpha\beta}\nabla^\alpha u^\beta$.

We observe that while there are fourteen total independent 
components of $T^{\mu\nu}$ and $N^\mu$, Eqs.~(\ref {DFDE1})-(\ref
{DFDE3}) constitute only five equations. Therefore, in order to 
derive the complete set of equations for dissipative fluid dynamics, 
one still has to obtain the additional nine equations of motion that 
will close Eqs.~(\ref{DFDE1})-(\ref {DFDE3}). Eventually, this 
corresponds to finding the closed dynamical or constitutive 
relations satisfied by the dissipative tensors $\Pi$, $n^\mu$ and 
$\pi^{\mu\nu}$.


\subsubsection{Relativistic Navier-Stokes theory}

In the presence of dissipative currents, the entropy is no longer a 
conserved quantity, i.e., $\partial_\mu S^\mu \neq 0$. Since the 
form of the entropy four-current for a dissipative fluid is not 
known \emph{a priori}, it is not trivial to obtain its equation. We 
proceed by recalling the form of the entropy four-current for ideal 
fluids, Eq.~(\ref{CER}), and extending it for dissipative fluids,
\begin{equation}\label{EFCDF}
S^\mu = P\beta^\mu + \beta_\nu T^{\mu\nu} - \alpha N^\mu.
\end{equation}
The above extension remains valid because an artificial equilibrium 
state was constructed using the matching conditions to satisfy the 
thermodynamic relations as if in equilibrium. This was the key step 
proposed by Eckart, Landau and Lifshitz in order to derive the 
relativistic Navier-Stokes theory \cite{Eckart:1940zz,Landau}. The 
next step is to calculate the entropy generation, $\partial_\mu 
S^\mu$, in dissipative fluids. To this end, we substitute the form 
of $T^{\mu\nu}$ and $N^\mu$ for dissipative fluids from Eq.~(\ref 
{CCDLF}) in Eq.~(\ref{EFCDF}). Taking the divergence and using 
Eqs.~(\ref{DFDE1})-(\ref{DFDE3}), we obtain
\begin{equation}\label{CTEFCDR1}
\partial_\mu S^\mu = -\beta\Pi\theta - n^\mu\nabla_\mu\alpha + \beta\pi^{\mu\nu}\sigma_{\mu\nu}.
\end{equation}

The relativistic Navier-Stokes theory can then be obtained by 
applying the second law of thermodynamics to each fluid element, 
i.e., by requiring that the entropy production $\partial_\mu S^\mu$ 
must always be positive,
\begin{equation}\label{DEFCDE}
-\beta\Pi\theta - n^\mu\nabla_\mu\alpha + \beta\pi^{\mu\nu}\sigma_{\mu\nu} \geq 0.
\end{equation}
The above inequality can be satisfied for all possible fluid 
configurations if one assumes that the bulk viscous pressure $\Pi$, 
the particle-diffusion four-current $n^\mu$, and the shear-stress 
tensor $\pi^{\mu\nu}$ are linearly proportional to $\theta$, 
$\nabla^\mu\alpha$, and $\sigma^{\mu\nu}$, respectively. This leads to
\begin{equation}\label{RNSE}
\Pi = -\zeta\theta, \quad
n^\mu = \kappa\nabla^\mu\alpha, \quad
\pi^{\mu\nu} = 2\eta\sigma^{\mu\nu},
\end{equation}
where the proportionality coefficients $\zeta$, $\kappa$ and $\eta$ 
refer to the bulk viscosity, the particle diffusion, and the shear 
viscosity, respectively. Substituting the above equation in Eq.~(\ref
{CTEFCDR1}), we observe that the source term for entropy production 
becomes a quadratic function of the dissipative currents
\begin{equation}\label{STEP}
\partial_\mu S^\mu = \frac{\beta}{\zeta}\, \Pi^2 - \frac{1}{\kappa}\, n_\mu n^\mu + \frac{\beta}{2\eta}\, \pi_{\mu\nu}\pi^{\mu\nu}.
\end{equation}
In the above equation, since $n^\mu$ is orthogonal to the timelike 
four-vector $u^\mu$, it is spacelike and hence $n_\mu n^\mu<0$. 
Moreover, $\pi^{\mu\nu}$ is symmetric in its Lorentz indices, and in 
the local rest frame $\pi^{0\mu}=\pi^{\mu0}=0$. Since the trace of 
the square of a symmetric matrix is always positive, therefore 
$\pi_{\mu\nu}\pi^{\mu\nu}>0$. Hence, as long as 
$\zeta,\kappa,\eta\geq 0$, the entropy production is always 
positive. Constitutive relations for the dissipative quantities, 
Eq.~(\ref{RNSE}), along with Eqs.~(\ref{DFDE1})-(\ref {DFDE3}) are 
known as the relativistic Navier-Stokes equations.

The relativistic Navier-Stokes theory in this form was obtained 
originally by Landau and Lifshitz \cite{Landau}. A similar theory 
was derived independently by Eckart, using a different definition of 
the fluid four-velocity \cite{Eckart:1940zz}. However, as already 
mentioned, the Navier-Stokes theory is acausal and, consequently, 
unstable. The source of the acausality can be understood from the 
constitutive relations satisfied by the dissipative currents, 
Eq.~(\ref{RNSE}). The linear relations between dissipative currents 
and gradients of the primary fluid-dynamical variables imply that any 
inhomogeneity of $\alpha$ and $u^\mu$, immediately results in 
dissipative currents. This instantaneous effect is not allowed in a 
relativistic theory which eventually causes the theory to be 
unstable. Several theories have been developed to incorporate 
dissipative effects in fluid dynamics without violating causality: 
Grad-Israel-Stewart theory \cite{Israel:1979wp, Grad,Israel:1976tn}, 
the divergence-type theory \cite{Muller:1967zza, Muller:1999in}, 
extended irreversible thermodynamics \cite{Jou}, Carter's theory 
\cite{Carter}, \"Ottinger-Grmela theory \cite{Grmela:1997zz}, among 
others. Israel and Stewart's formulation of causal relativistic 
dissipative fluid dynamics is the most popular and widely used; in 
the following we briefly review their approach. 


\subsubsection{Causal fluid dynamics: Israel-Stewart theory}

The main idea behind the Israel-Stewart formulation was to apply the 
second law of thermodynamics to a more general expression of the 
non-equilibrium entropy four-current \cite{Israel:1979wp, Grad, 
Israel:1976tn}. In equilibrium, the entropy four-current was 
expressed exactly in terms of the primary fluid-dynamical variables, 
Eq.~(\ref{CER}). Strictly speaking, the nonequilibrium entropy 
four-current should depend on a larger number of independent 
dynamical variables, and, a direct extension of Eq.~(\ref{CER}) to 
Eq.~(\ref{EFCDF}) is, in fact, incomplete. A more realistic 
description of the entropy four-current can be obtained by 
considering it to be a function not only of the primary 
fluid-dynamical variables, but also of the dissipative currents. The 
most general off-equilibrium entropy four-current is then given by 
\begin{equation}\label{EFCMGF}
S^\mu = P\beta^\mu + \beta_\nu T^{\mu\nu} - \alpha N^\mu - Q^\mu\left(\delta N^\mu, \delta T^{\mu\nu}\right).
\end{equation}
where $Q^\mu$ is a function of deviations from local equilibrium, 
$\delta N^\mu\equiv N^\mu-N^\mu_{(0)}$, $\delta T^{\mu\nu}\equiv 
T^{\mu\nu}-T^{\mu\nu}_{(0)}$. Using Eq.~(\ref{CCDLF}) and 
Taylor-expanding $Q^\mu$ to second order in dissipative fluxes, we 
obtain
\begin{align}\label{AEFCC2}
S^\mu =&~ su^\mu - \alpha n^\mu - \left(\beta_0\Pi^2 - \beta_1 n_\nu n^\nu 
+ \beta_2\pi_{\rho\sigma} \pi^{\rho\sigma}\right) \frac{u^\mu}{2T} \nonumber\\
&- \left(\alpha_0\Pi\Delta^{\mu\nu} + \alpha_1\pi^{\mu\nu}\right)\frac{n_\nu}{T}
+ {\cal O}(\delta^3),
\end{align}
where ${\cal O}(\delta^3)$ denotes third order terms in the 
dissipative currents and 
$\beta_0,~\beta_1,~\beta_2,~\alpha_0,~\alpha_1$ are the 
thermodynamic coefficients of the Taylor expansion and are 
complicated functions of the temperature and chemical potential.

We observe that the existence of second-order contributions to the 
entropy four-current in Eq.~(\ref{AEFCC2}) should lead to 
constitutive relations for the dissipative quantities which are 
different from relativistic Navier-Stokes theory obtained previously 
by employing the second law of thermodynamics. The relativistic 
Navier-Stokes theory can then be understood to be valid only up to 
first order in the dissipative currents (hence also called 
first-order theory). Next, we re-calculate the entropy production, 
$\partial_\mu S^\mu$, using the more general entropy four-current 
given in Eq.~(\ref{AEFCC2}),
\begin{align}\label{EFCD3C2}
\partial_\mu S^\mu = 
& - \beta\Pi\Big[ \theta + \beta_0\dot\Pi 
+ \beta_{\Pi\Pi} \Pi\theta 
+ \psi\alpha_{n\Pi } n_\mu \dot u^\mu \nonumber \\
&\qquad\quad + \alpha_0 \nabla_\mu n^{\mu}
+ \psi\alpha_{\Pi n} n_\mu \nabla^{\mu}\alpha  \Big] \nonumber \\
&- \beta n^\mu \Big[ T\nabla_\mu \alpha - \beta_1\dot n_\mu 
- \beta_{nn} n_\mu \theta  
+ \alpha_0\nabla_{\mu}\Pi \nonumber \\
&\qquad\quad~~ + \alpha_1\nabla_\nu \pi^\nu_\mu 
+ \tilde \psi\alpha_{n\Pi } \Pi\dot u_\mu 
+ \tilde \psi\alpha_{\Pi n}\Pi\nabla_{\mu}\alpha\nonumber \\
&\qquad\quad~~ + \tilde \chi\alpha_{\pi n} \pi^\nu_\mu \nabla_\nu \alpha  
+ \tilde \chi\alpha_{n\pi } \pi^\nu_\mu \dot u_\nu \Big] \nonumber \\
& + \beta\pi^{\mu\nu}\Big[ \sigma_{\mu\nu} 
- \beta_2\dot\pi_{\mu\nu} 
- \beta_{\pi\pi}\theta\pi_{\mu\nu}
- \alpha_1 \nabla_{\langle\mu}n_{\nu\rangle} \nonumber \\
&\qquad\quad~~ - \chi\alpha_{\pi n}n_{\langle\mu}\nabla_{\nu\rangle}\alpha 
- \chi\alpha_{n\pi }n_{\langle\mu}\dot u_{\nu\rangle} \Big],
\end{align}
As argued before, the only way to explicitly satisfy the second law 
of thermodynamics is to ensure that the entropy production is a 
positive definite quadratic function of the dissipative currents. 

The second law of thermodynamics, $\partial_{\mu}S^{\mu}\ge 0$, is 
guaranteed to be satisfied if we impose linear relationships between 
thermodynamical fluxes and extended thermodynamic forces, leading to 
the following evolution equations for bulk pressure, 
particle-diffusion four-current and shear stress tensor,
\begin{align}
\Pi =\,& -\zeta\Big[ \theta 
+ \beta_0 \dot \Pi 
+ \beta_{\Pi\Pi} \Pi \theta 
+ \alpha_0 \nabla_\mu n^\mu  
+ \psi\alpha_{n\Pi} n_\mu \dot u^\mu \nonumber \\
&\quad\quad + \psi\alpha_{\Pi n} n_\mu \nabla^\mu \alpha  \Big], \label{bulkC2} \\ 
n^{\mu} =\ & \lambda \Big[ T \nabla^\mu \alpha 
- \beta_1\dot n^{\langle\mu\rangle} 
- \beta_{nn} n^\mu \theta
+ \alpha_0 \nabla^\mu \Pi \nonumber \\
&\quad + \alpha_1 \Delta^\mu_\rho \nabla_\nu \pi^{\rho\nu}
+ \tilde \psi\alpha_{n\Pi} \Pi \dot u^{\langle\mu\rangle} 
+ \tilde \psi\alpha_{\Pi n} \Pi \nabla^\mu \alpha \nonumber \\
&\quad + \tilde \chi\alpha_{\pi n} \pi_\nu^\mu \nabla^\nu \alpha
+ \tilde \chi\alpha_{n\pi} \pi_\nu^\mu \dot u^\nu \Big], \label{currentC2} \\ 
\pi^{\mu\nu} =\ & 2\eta\Big[ \sigma^{\mu\nu} 
- \beta_2\dot\pi^{\langle\mu\nu\rangle} 
- \beta_{\pi\pi}\theta\pi^{\mu\nu}
- \alpha_1 \nabla^{\langle\mu}n^{\nu\rangle} \nonumber \\
&\quad~ - \chi\alpha_{\pi n} n^{\langle\mu} \nabla^{\nu\rangle} \alpha 
- \chi\alpha_{n\pi } n^{\langle\mu} \dot u^{\nu\rangle} \Big] , \label{shearC2}
\end{align}
where $\lambda\equiv\kappa/T$. This implies that the dissipative 
currents must satisfy the dynamical equations,
\begin{align}
\dot\Pi + \frac{\Pi}{\tau_\Pi} =\,& -\frac{1}{\beta_0}\Big[ \theta  
+ \beta_{\Pi\Pi} \Pi \theta   
+ \psi\alpha_{n\Pi} n_\mu \dot u^\mu \nonumber \\
&+ \alpha_0 \nabla_\mu n^\mu
+ \psi\alpha_{\Pi n} n_\mu \nabla^\mu \alpha  \Big], \label{bulkC2F} \\ 
\dot n^{\langle\mu\rangle} + \frac{n^\mu}{\tau_n} =\ & \frac{1}{\beta_1}\Big[ T \nabla^\mu \alpha  
- \beta_{nn} n^\mu \theta
+ \alpha_1 \Delta^\mu_\rho \nabla_\nu \pi^{\rho\nu} \nonumber \\
&+ \alpha_0 \nabla^\mu \Pi
+ \tilde \psi\alpha_{n\Pi} \Pi \dot u^{\langle\mu\rangle} 
\!+ \tilde \psi\alpha_{\Pi n} \Pi \nabla^\mu \alpha \nonumber \\
& + \tilde \chi\alpha_{\pi n} \pi_\nu^\mu \nabla^\nu \alpha
+ \tilde \chi\alpha_{n\pi} \pi_\nu^\mu \dot u^\nu \Big], \label{currentC2F} \\ 
\dot\pi^{\langle\mu\nu\rangle} + \frac{\pi^{\mu\nu}}{\tau_\pi} =\ & \frac{1}{\beta_2}\Big[ \sigma^{\mu\nu}  
- \beta_{\pi\pi}\theta\pi^{\mu\nu}
- \alpha_1 \nabla^{\langle\mu}n^{\nu\rangle} \nonumber \\
&- \chi\alpha_{\pi n} n^{\langle\mu} \nabla^{\nu\rangle} \alpha 
- \chi\alpha_{n\pi } n^{\langle\mu} \dot u^{\nu\rangle} \Big] . \label{shearC2F}
\end{align}
The above equations for the dissipative quantities are 
relaxation-type equations with the relaxation times defined as
\begin{equation}\label{RTC2}
\tau_{\Pi} \equiv \zeta\,\beta_0, \quad
\tau_{n} \equiv \lambda\,\beta_1 = \kappa\,\beta_1/T, \quad
\tau_{\pi} \equiv 2\,\eta\,\beta_2 ,
\end{equation}
Since the relaxation times must be positive, the Taylor expansion 
coefficients $\beta_0$, $\beta_1$ and $\beta_2$ must all be larger 
than zero. 

The most important feature of the Israel-Stewart theory is the 
presence of relaxation times corresponding to the dissipative 
currents. These relaxation times indicate the time scales within 
which the dissipative currents react to hydrodynamic gradients, in 
contrast to the relativistic Navier-Stokes theory where this process 
occurs instantaneously. The introduction of such relaxation 
processes restores causality and transforms the dissipative currents 
into independent dynamical variables that satisfy partial 
differential equations instead of constitutive relations. However, 
it is important to note that this welcome feature comes with a 
price: five new parameters, $\beta_0$, $\beta_1$, $\beta_2$, 
$\alpha_0$ and $\alpha_1$, are introduced in the theory. These 
coefficients cannot be determined within the present framework, 
i.e., within the framework of thermodynamics alone, and as a 
consequence the evolution equations remain incomplete. Microscopic 
theories, such as kinetic theory, have to be invoked in order to 
determine these coefficients. In the next section, we review the 
basics of relativistic kinetic theory and Boltzmann transport 
equation, and discuss the details of the coarse graining procedure 
to obtain dissipative hydrodynamic equations.


\subsection{Relativistic kinetic theory}

Macroscopic properties of a many-body system are governed by the 
interactions among its constituent particles and the external 
constraints on the system. Kinetic theory presents a statistical 
framework in which the macroscopic quantities are expressed in terms 
of single-particle phase-space distribution function. The various 
formulations of relativistic dissipative hydrodynamics, presented in 
this review, are obtained within the framework of relativistic 
kinetic theory. In the following, we briefly outline the salient 
features of relativistic kinetic theory and dissipative 
hydrodynamics which have been employed in the subsequent 
calculations \cite{deGroot}.

Let us consider a system of relativistic particles, each having rest 
mass $m$, momentum $\vec p$ and energy $p^0$. Therefore from 
relativity, we have, $p^0=\sqrt{(\vec p)^2+m^2}$. For a large number 
of particles, we introduce a single-particle distribution function 
$f(x,p)$ which gives the distribution of the four-momentum 
$p=p^\mu=(p^0,\vec p)$ at each space-time point such that 
$f(x,p)\Delta^3x\Delta^3p$ gives the average number of particles at 
a given time $t$ in the volume element $\Delta^3x$ at point $\vec x$ 
with momenta in the range $(\vec p,\vec p+\Delta\vec p)$. However, 
this definition of the single-particle phase-space distribution 
function $f(x,p)$ assumes that, while on one hand, the number of 
particles contained in $\Delta^3x$ is large, on the other hand, 
$\Delta^3x$ is small compared to macroscopic point of view. 

The particle density $n(x)$ is introduced to describe, in general, a 
non-uniform system, such that $n(x)\Delta^3x$ is the average number 
of particles in volume $\Delta^3x$ at $(\vec x,~t)$. Similarly, 
particle flow $\vec j(x)$ is defined as the particle current. With 
the help of the distribution function, the particle density and 
particle flow are given by
\begin{equation}\label{PDPFC2}
n(x) = \int d^3p~f(x,p), \quad 
\vec j(x) = \int d^3p~\vec v\,f(x,p),
\end{equation}
where $\vec v=\vec p/p^0$ is the particle velocity. These two local 
quantities, particle density and particle flow constitute a 
four-vector field $N^\mu=(n,\vec j)$, called particle four-flow, and 
can be written in a unified way as
\begin{equation}\label{PFFC2}
N^\mu(x) = \int \frac{d^3p}{p^0}\,p^\mu\,f(x,p).
\end{equation}
Note that since $d^3p/p^0$ is a Lorentz invariant quantity, $f(x,p)$ 
should be a scalar in order that $N^\mu$ transforms as a four-vector.

Since the energy per particle is $p^0$, the average energy density 
and the energy flow can be written in terms of the distribution 
function as
\begin{equation}\label{AEEFC2}
T^{00}(x) = \int d^3p~p^0\,f(x,p), \quad T^{0i}(x) = \int d^3p~p^0\,v^i\,f(x,p).
\end{equation}
The momentum density is defined as the average value of particle 
momenta $p^i$, and, the momentum flow or pressure tensor is defined 
as the flow in direction $j$ of momentum in direction $i$. For these 
two quantities, we have
\begin{equation}\label{MDMFC2}
T^{i0}(x) = \int d^3p~p^i\,f(x,p), \quad T^{ij}(x) = \int d^3p~p^i\,v^j\,f(x,p).
\end{equation}
Combining all these in a compact covariant form using $v^i=p^i/p^0$, 
we obtain the energy-momentum tensor of a macroscopic system
\begin{equation}\label{EMTC2}
T^{\mu\nu}(x) = \int \frac{d^3p}{p^0}\,p^\mu\,p^\nu\,f(x,p).
\end{equation}
Observe that the above definition of the energy momentum tensor 
corresponds to second moment of the distribution function, and 
hence, it is a symmetric quantity.

The H-function introduced by Boltzmann implies that the 
nonequilibrium local entropy density of a system can be written as
\begin{equation}\label{EDC2}
s(x) = -\int d^3p~f(x,p)\left[ \ln f(x,p) - 1\right].
\end{equation}
The entropy flow corresponding to the above entropy density is
\begin{equation}\label{EFC2}
\vec S(x) = -\int d^3p~\vec v\,f(x,p)\left[ \ln f(x,p) - 1 \right].
\end{equation}
These two local quantities, entropy density and entropy flow 
constitute a four-vector field $S^\mu=(s,\vec S)$, called entropy 
four-flow, and can be written in a unified way as
\begin{equation}\label{EFCC2}
S^\mu(x) = -\int \frac{d^3p}{p^0}\, p^\mu\,f(x,p)\left[ \ln f(x,p) - 1 \right].
\end{equation}
The above definition of entropy four-current is valid for a system 
comprised of Maxwell-Boltzmann gas. This expression can also be 
extended to a system consisting of particles obeying Fermi-Dirac 
statistics ($r=1$), or Bose-Einstein statistics ($r=-1$) as
\begin{equation}\label{EFCQC2}
S^\mu(x) = -\!\!\int\!\! \frac{d^3p}{p^0}\, p^\mu\! \left[\! f(x,p)\!\ln f(x,p) + r\tilde f(x,p)\!\ln\tilde f(x,p) \!\right]\!, 
\end{equation}
where $\tilde f \equiv 1 - rf$. The expressions for the entropy 
four-current given in Eqs.~(\ref{EFCC2}) and (\ref{EFCQC2}) can be 
used to formulate the generalized second law of thermodynamics 
(entropy law), and, define thermodynamic equilibrium.

For small departures from equilibrium, $f(x,p)$ can be written as 
$f=f_0+\delta f$. The equilibrium distribution function $f_0$ is 
defined as
\begin{equation}\label{EDFC2}
f_0(x,p) = \frac{1}{\exp(\beta u\cdot p -\alpha) + r}, 
\end{equation}
where the scalar product is defined as $u\cdot p\equiv u_\mu p^\mu$ 
and $r=0$ for Maxwell-Boltzmann statistics. Note that in 
equilibrium, i.e., for $f(x,p)=f_0(x,p)$, the particle four-flow and 
energy momentum tensor given in Eqs.~(\ref{PFFC2}) and (\ref 
{EMTC2}) reduce to that of ideal hydrodynamics $N^\mu_{(0)}$ and 
$T^{\mu\nu}_{(0)}$. Therefore using Eq.~(\ref{CCDLF}), the 
dissipative quantities, {\it viz.}, the bulk viscous pressure $\Pi$, 
the particle diffusion current $n^\mu$, and the shear stress tensor 
$\pi^{\mu\nu}$ can be written as
\begin{align}
\Pi &= -\frac{1}{3}\,\Delta_{\alpha\beta}\int \frac{d^3p}{p^0}\, p^\alpha p^\beta\, \delta f, \label{bulk_def}\\
n^\mu &=  \Delta^{\mu\nu} \int \frac{d^3p}{p^0}\, p_\nu\, \delta f, \label{ch_curr_def}\\
\pi^{\mu\nu} &= \Delta^{\mu\nu}_{\alpha\beta} \int \frac{d^3p}{p^0}\, p^\alpha p^\beta \delta f. \label{shear_def}
\end{align}

The evolution equations for the dissipative quantities expressed in 
terms of the non-equilibrium distribution function, Eqs.~(\ref 
{bulk_def})-(\ref{shear_def}), can be obtained provided the 
evolution of distribution function is specified from some 
microscopic considerations. Boltzmann equation governs the evolution 
of the phase-space distribution function which provides a reliably 
accurate description of the microscopic dynamics. Relativistic 
Boltzmann equation can be written as
\begin{equation}\label{rel_boltz_eqn}
p^\mu \partial_\mu f = C[f],
\end{equation}
where $dp\equiv d^3p/p^0$ and $C[f]$ is the collision functional. 
For microscopic interactions restricted to $2 \leftrightarrow 2$ 
elastic collisions, the form of the collision functional is given by
\begin{equation}\label{coll_term}
C[f] = \frac{1}{2}\!\int\! dp' dk \ dk' \ W_{pp' \to kk'}(f_k f_{k'} 
\tilde f_p \tilde f_{p'} - f_p f_{p'} \tilde f_k \tilde f_{k'}),
\end{equation}
where $W_{pp' \to kk'}$ is the collisional transition rate. The 
first and second terms within the integral of Eq.~(\ref{coll_term}) 
refer to the processes $kk'\to pp'$ and $pp'\to kk'$, respectively. 
In the relaxation-time approximation, where it is assumed that the 
effect of the collisions is to restore the distribution function to 
its local equilibrium value exponentially, the collision integral 
reduces to \cite{Anderson_Witting}
\begin{equation}\label{rel_time_approx}
C[f] = -(u\cdot p)\frac{\delta f}{\tau_R},
\end{equation}
where $\tau_R$ is the relaxation time.


\subsection{Dissipative fluid dynamics from kinetic theory}

The derivation of a causal theory of relativistic dissipative 
hydrodynamics by Israel and Stewart \cite{Israel:1979wp} proceeds by 
invoking the second law of thermodynamics, viz., $\partial_\mu S^\mu 
\geq 0$, from the algebraic form of the entropy four-current given 
in Eq.~(\ref{AEFCC2}). As noted earlier, the new parameters, 
$\beta_0$, $\beta_1$, $\beta_2$, $\alpha_0$ and $\alpha_1$, cannot 
be determined within the framework of thermodynamics alone and 
microscopic theories, such as kinetic theory, have to be invoked in 
order to determine these coefficients. On the other hand, one may 
demand the second law of thermodynamics from the definition of the 
entropy four-current, given in Eqs.~(\ref{EFCC2}) and (\ref 
{EFCQC2}), in order to obtain the dissipative equations \cite 
{Jaiswal:2013fc}. This essentially ensures that the non-equilibrium 
corrections to the distribution function, $\delta f$, does not 
violate the second law of thermodynamics. In Ref.~\cite 
{Jaiswal:2013fc}, the generalized method of moments developed by 
Denicol {\it et al.} \cite{Denicol:2012cn} was used to quantify the 
dissipative corrections to the distribution function. The form of 
the resultant dissipative equations, obtained in Ref.~\cite 
{Jaiswal:2013fc}, are identical to Eqs.~(\ref{bulkC2F})-(\ref 
{shearC2F}), with the welcome exception that all the transport 
coefficients are now determined in terms of the thermodynamical 
quantities.

The moment method, originally proposed by Grad \cite{Grad}, has been 
used quite extensively to quantify the dissipative corrections to 
the distribution function \cite{Jaiswal:2013fc, Denicol:2012cn, 
Denicol:2010xn, Jaiswal:2012qm, Jaiswal:2012dd, Bhalerao:2013aha, 
Betz:2008me, El:2008yy, Muronga:2003ta}. In this method, the 
distribution function is Taylor expanded in powers of four-momenta 
around its local equilibrium value. Truncating the Taylor expansion 
at second-order in momenta results in 14 unknowns that have to be 
determined to describe the distribution function. This expansion 
implicitly assumes a converging series in powers of momenta. An 
alternative derivation of causal dissipative equations, which do not 
make use of the moment method, was proposed in Ref.~\cite 
{Jaiswal:2013npa}. In this method, which is based on a 
Chapman-Enskog like expansion, the Boltzmann equation in the 
relaxation time approximation 
\begin{equation}\label{RBEC5}
p^\mu\partial_\mu f =  -\frac{u\cdot p}{\tau_R}(f-f_0)~,
\end{equation}
is solved iteratively to obtain $\delta f$ up to any arbitrary order 
in derivatives. To first and second-order in gradients, one obtains
\begin{align}
\delta f^{(1)} &= -\frac{\tau_R}{u\cdot p} \, p^\mu \partial_\mu f_0, \label{FOCC5} \\
\delta f^{(2)} &= \frac{\tau_R}{u\cdot p}p^\mu p^\nu\partial_\mu\Big(\frac{\tau_R}{u\cdot p} \partial_\nu f_0\Big). \label{SOCC5}
\end{align}
This method of obtaining the form of the nonequilibrium distribution 
function is consistent with dissipative hydrodynamics, which is also 
formulated as a gradient expansion. 

The second-order evolution equations for the dissipative quantities 
are then obtained by substituting $\delta f=\delta f^{(1)}+\delta 
f^{(2)}$ from Eqs.~(\ref{FOCC5}) and (\ref{SOCC5}) in Eqs.~(\ref 
{bulk_def})-(\ref{shear_def}),
\begin{align}
\frac{\Pi}{\tau_\Pi} =& -\dot{\Pi}
-\beta_{\Pi}\theta 
-\delta_{\Pi\Pi}\Pi\theta
+\lambda_{\Pi\pi}\pi^{\mu\nu}\sigma_{\mu \nu }  \nonumber \\
&-\tau_{\Pi n}n\cdot\dot{u}
-\lambda_{\Pi n}n\cdot\nabla\alpha
-\ell_{\Pi n}\partial\cdot n ~, \label{BULK}\\
\frac{n^{\mu}}{\tau_n} =& -\dot{n}^{\langle\mu\rangle}
+\beta_{n}\nabla^{\mu}\alpha
-n_{\nu}\omega^{\nu\mu}
-\lambda_{nn}n^{\nu}\sigma_{\nu}^{\mu}
-\delta_{nn}n^{\mu}\theta   \nonumber \\
&+\lambda_{n\Pi}\Pi\nabla^{\mu}\alpha
-\lambda_{n\pi}\pi^{\mu\nu}\nabla_{\nu}\alpha 
-\tau_{n\pi}\pi_{\nu}^{\mu}\dot{u}^{\nu}  \nonumber \\
&+\tau_{n\Pi}\Pi\dot{u}^{\mu}
+\ell_{n\pi}\Delta^{\mu\nu}\partial_{\gamma}\pi_{\nu}^{\gamma}
-\ell_{n\Pi}\nabla^{\mu}\Pi~,  \label{HEAT} \\
\frac{\pi^{\mu\nu}}{\tau_\pi} =& -\dot{\pi}^{\langle\mu\nu\rangle}
+2\beta_{\pi}\sigma^{\mu\nu}
+2\pi_{\gamma}^{\langle\mu}\omega^{\nu\rangle\gamma}
-\tau_{\pi\pi}\pi_{\gamma}^{\langle\mu}\sigma^{\nu\rangle\gamma}  \nonumber \\
&-\delta_{\pi\pi}\pi^{\mu\nu}\theta 
+\lambda_{\pi\Pi}\Pi\sigma^{\mu\nu}
-\tau_{\pi n}n^{\langle\mu}\dot{u}^{\nu\rangle }  \nonumber \\
&+\lambda_{\pi n}n^{\langle\mu}\nabla ^{\nu\rangle}\alpha
+\ell_{\pi n}\nabla^{\langle\mu}n^{\nu\rangle } ~, \label{SHEAR}
\end{align}
where $\omega^{\mu\nu}=(\nabla^\mu u^\nu-\nabla^\nu u^\mu)/2$ is the 
vorticity tensor. It is interesting to note that although the form 
of the evolution equations for dissipative quantities in Eqs.~(\ref 
{BULK})-(\ref{SHEAR}), are identical to those obtained in Ref.~\cite
{Denicol:2010xn} using the moment method, the transport 
coefficients are, in general, different \cite{Jaiswal:2014isa, 
Florkowski:2015lra}. Moreover, it was shown that the above described 
method, based on iterative solution of Boltzmann equation, leads to 
phenomenologically consistent corrections to the distribution 
function \cite{Bhalerao:2013pza} and the transport coefficients 
exhibits intriguing similarities with strongly coupled conformal 
field theory \cite{Jaiswal:2015mxa, Jaiswal:2016pmi}.

Proceeding in a similar way, a third-order dissipative evolution 
equation can also be obtained \cite{Jaiswal:2013vta, 
Jaiswal:2014raa, Chattopadhyay:2014lya}
\begin{align}\label{TOSHEAR}
\dot{\pi}^{\langle\mu\nu\rangle} =& -\frac{\pi^{\mu\nu}}{\tau_\pi}
+2\beta_\pi\sigma^{\mu\nu}
+2\pi_{\gamma}^{\langle\mu}\omega^{\nu\rangle\gamma}
-\frac{10}{7}\pi_\gamma^{\langle\mu}\sigma^{\nu\rangle\gamma}  \nonumber \\
&-\frac{4}{3}\pi^{\mu\nu}\theta
+\frac{25}{7\beta_\pi}\pi^{\rho\langle\mu}\omega^{\nu\rangle\gamma}\pi_{\rho\gamma}
-\frac{1}{3\beta_\pi}\pi_\gamma^{\langle\mu}\pi^{\nu\rangle\gamma}\theta \nonumber \\
&-\frac{38}{245\beta_\pi}\pi^{\mu\nu}\pi^{\rho\gamma}\sigma_{\rho\gamma}
-\frac{22}{49\beta_\pi}\pi^{\rho\langle\mu}\pi^{\nu\rangle\gamma}\sigma_{\rho\gamma} \nonumber \\
&-\frac{24}{35}\nabla^{\langle\mu}\left(\pi^{\nu\rangle\gamma}\dot u_\gamma\tau_\pi\right)
+\frac{4}{35}\nabla^{\langle\mu}\left(\tau_\pi\nabla_\gamma\pi^{\nu\rangle\gamma}\right) \nonumber \\
&-\frac{2}{7}\nabla_{\gamma}\left(\tau_\pi\nabla^{\langle\mu}\pi^{\nu\rangle\gamma}\right)
+\frac{12}{7}\nabla_{\gamma}\left(\tau_\pi\dot u^{\langle\mu}\pi^{\nu\rangle\gamma}\right) \nonumber \\
&-\frac{1}{7}\nabla_{\gamma}\left(\tau_\pi\nabla^{\gamma}\pi^{\langle\mu\nu\rangle}\right)
+\frac{6}{7}\nabla_{\gamma}\left(\tau_\pi\dot u^{\gamma}\pi^{\langle\mu\nu\rangle}\right) \nonumber \\
&-\frac{2}{7}\tau_\pi\omega^{\rho\langle\mu}\omega^{\nu\rangle\gamma}\pi_{\rho\gamma}
-\frac{2}{7}\tau_\pi\pi^{\rho\langle\mu}\omega^{\nu\rangle\gamma}\omega_{\rho\gamma} \nonumber \\
&-\frac{10}{63}\tau_\pi\pi^{\mu\nu}\theta^2
+\frac{26}{21}\tau_\pi\pi_\gamma^{\langle\mu}\omega^{\nu\rangle\gamma}\theta.
\end{align}
It is reassuring that the results obtained using third-order 
evolution equation indicates convergence of the gradient expansion 
and shows improvement over second-order, when compared to the direct 
solutions of the Boltzmann equation \cite{Jaiswal:2013vta, 
Jaiswal:2014raa, Chattopadhyay:2014lya, El:2009vj}.

Apart from these standard formulations, there are several other 
formulations of relativistic dissipative hydrodynamics from kinetic 
theory. Among them, the ones which have gained widespread interest 
are anisotropic hydrodynamics and derivations based on 
renormalization group method. Anisotropic hydrodynamics is a 
non-perturbative reorganization of the standard relativistic 
hydrodynamics which takes into account the large momentum-space 
anisotropies generated in ultrarelativistic heavy-ion collisions 
\cite{Bazow:2013ifa, Florkowski:2013lza, Tinti:2013vba, 
Florkowski:2010cf, Martinez:2010sc, Romatschke:2003ms}. On the other 
hand, the derivation based on renormalization group method attempts 
to solve the Boltzmann equation, as faithfully as possible, in an 
organized perturbation scheme and resum away the possible secular 
terms by a suitable setting of the initial value of the distribution 
function \cite{Kikuchi:2015swa, Tsumura:2015fxa, Tsumura:2012kp, 
Tsumura:2012gq}. Since it is widely accepted that the QGP is 
momentum-space anisotropic, application of anisotropic hydrodynamics 
to high energy heavy-ion collisions has phenomenological 
implications. Nevertheless, the dissipative hydrodynamic formulation 
based on renormalization group method is important in order to 
accurately determine the higher-order transport coefficients. 

Since it is well established that QGP formed in high energy 
heavy-ion collisions is strongly coupled, it is of interest to 
compare the transport coefficients obtained from kinetic theory with 
that of a strongly coupled system \cite{York:2008rr}. In contrast to 
kinetic theory, strongly coupled quantum systems, in general, does 
not allow for a quasiparticle interpretation. This can be attributed 
to the fact that the quasiparticle notion hinges on the presence of 
a well-defined peak in the spectral density, which may not exist at 
strong coupling. Therefore it is interesting to study the 
hydrodynamic limit of an infinitely strongly coupled system, which 
are different than systems described by kinetic theory. In the 
following we discuss the evolution equation for shear stress tensor 
for a strongly coupled conformal system which is equivalent to a 
system of massless particles in kinetic theory. 

\begin{table}[t!]
\centering
\begin{tabular}{ |c|c|c|c|c| } 
 \hline
 ~& $\tau_\pi T/(\eta/s)$ & $\lambda_1 T/(\eta/s)$ & $\lambda_2 T/(\eta/s)$ & $\lambda_3$\\ 
 \hline
 ADS/CFT & $2(2-\ln 2)$ & $2\eta$ & $4\eta\ln 2$ & 0\\ 
 \hline
 KT & $5$ & $(25/7)\eta$ & $10\eta$ & 0\\ 
 \hline
\end{tabular}
\caption{Comparison of transport coefficients}
\label{comp_tr_coeff}
\end{table}

For such a system, the evolution of shear stress tensor is governed 
by the equation
\begin{align}\label{shear}
\pi^{\mu\nu} =&~ 2\eta\sigma^{\mu\nu} 
- \tau_\pi\left( \dot\pi^{\langle\mu\nu\rangle} + \frac{4}{3}\pi^{\mu\nu}\theta \right)
- \frac{\lambda_1}{\eta^2}\pi_\gamma^{\langle\mu}\pi^{\nu\rangle\gamma}\nonumber \\
&+ \frac{\lambda_2}{\eta}\pi_\gamma^{\langle\mu}\omega^{\nu\rangle\gamma} 
+\lambda_3\omega_\gamma^{~\langle\mu}\omega^{\nu\rangle\gamma}.
\end{align}
For a system of massless particles, the kinetic theory results for 
second-order transport coefficients agree in the case of both moment 
method \cite{Denicol:2012cn} as well as the the Chapman-Enskog like 
iterative solution of the Boltzmann equation \cite{Jaiswal:2013npa}. 
In Table~\ref{comp_tr_coeff}, we compare the transport coefficients 
obtained from kinetic theory and from calculations employing ADS/CFT 
correspondence of strongly coupled $\mathcal{N}=4$ SYM and its 
supergravity dual \cite{Baier:2007ix, Bhattacharyya:2008jc}. We see 
that the second-order transport coefficients obtained from Kinetic 
theory are, in general, larger than those obtained from the ADS/CFT 
calculations. 


\section{Initial conditions}

In order to apply hydrodynamics to study the collective phenomena 
observed at relativistic heavy-ion collisions, one needs to first 
characterize the system. To this end, we shall discuss here, and in 
the next few sections about initial conditions, Equation of State 
(EoS), and freeze-out procedure as used in state of the art 
relativistic hydrodynamics simulations. However, we note that the 
following discussions are in no way complete but we will try to 
provide appropriate references wherever possible. Most of the 
following discussions can be also found in more details in Ref.~\cite
{Chaudhuri:2013yna, Chaudhuri:2012yt, Romatschke:2009im, 
Heinz:2009xj, Ollitrault:2008zz, Teaney:2009qa, deSouza:2015ena}.

In high energy heavy-ion collisions, bunch of nucleus of heavy 
elements are accelerated inside the beam pipes and in the final 
state (after the collisions) we have hundred or thousands of newly 
created particles coming out from the collision point in all 
directions. The underlying processes of the collisions between the 
constituent partons of the colliding nucleus and the conversion of 
initial momentum along the beam direction to the (almost) isotropic 
particle production is still not very well understood. Particularly 
the state just after the collisions when the longitudinal momentum 
distribution of the partons started to become isotropic and 
subsequently achieve the local thermal equilibrium state is poorly 
understood. But the precise knowledge of this so called 
pre-equilibrium stage is essential input in the viscous 
hydrodynamics models. The knowledge of distribution of 
energy/entropy density and the thermalisation time is one of the 
uncertainty present in the current hydrodynamics model studies. 
Below we discuss four most popular initial condition models used in 
hydrodynamics simulation of heavy-ion collisions.


\subsection{Glauber model}

The Glauber model of nuclear collisions is based on the original 
idea of Roy J. Glauber to describe the quantum mechanical scattering 
of proton-nucleus and nucleus-nucleus collisions at low energies. 
The original idea of Glauber was further modified by Bialas {\it et. 
al.} \cite{Bialas:1976ed} to explain inelastic nuclear collisions. 
For a nice and more complete review of Glauber model see Ref.~\cite 
{Miller:2007ri} and references therein for more details. We discuss 
below the very essential part of this model as used in heavy-ion 
collisions, particularly in the context of relativistic 
hydrodynamics. 

At present, there are two main variation of Glauber model in use. 
One of them is based on the optical limit approximation for nuclear 
scattering, where the nuclear scattering amplitude can be described 
by an eikonal approach. In this limit each of the colliding nucleons 
see a smooth density of nucleon distribution in the other nucleus. 
This variation of Glauber model, also known as optical Glauber 
model, uses the Wood-Saxon nuclear density distribution for a 
nucleus with mass number $A$ as
\begin{equation}
\rho_A(x,y,z)=\frac{\rho_0}{1+\exp[(r-R_0)/a_0]},
\end{equation}
where $R_0,a_0$ are the nuclear radius and skin thickness parameter 
of the nucleus, and $\rho_0$ is an overall constant that is 
determined by requiring $\int d^3 x\rho_A(x,y,z)=A$. One 
additionally defines the ``thickness function'' \cite{Miller:2007ri} 
\begin{equation}\label{thick_fn}
T_A(x,y)=\int_{-\infty}^{\infty} dz \rho_A(x,y,z),
\end{equation}
which indicates the Lorentz contraction in the laboratory frame. The 
Wood-Saxon nuclear density distribution is used along with the 
experimentally measured inelastic nuclear cross section to calculate 
the number of participating nucleons ($N_{part})$ and number of 
binary collisions ($N_{coll}$) for the two colliding nucleus.

In order to calculate the $N_{coll}(x,y,\vec{b})$ and 
$N_{part}(x,y,\vec{b})$ from Glauber model, one can choose the $X$ 
axis along the impact parameter vector $\vec{b}$. The $N_{part}$ and 
$N_{coll}$ distribution are functions of impact parameter, the 
inelastic Nucleon-Nucleon cross-section and the nuclear density 
distribution function. For a collision of two spherical nuclei with 
different mass number `A' and `B', the transverse density of binary 
collision and wounded nucleon profile is given by \cite
{Miller:2007ri} 
\begin{align}
N_{coll}(x,y; b) =&~ \sigma_{in}\, T_{A}\!\left(x{+}\frac{b}{2},y\right)T_{B}\!\left(x{-}\frac{b}{2},y\right), \label{eq:diff_ncoll}\\ 
N_{part}(x,y; b) =&~ T_{A}\!\left(x{+}{\textstyle\frac{b}{2}},y\right) 
F\left[ T_{B}\!\left(x{-}{\textstyle\frac{b}{2}},y\right)\!,\,B \right] \nonumber\\
+&~T_{B}\!\left(x{-}{\textstyle\frac{b}{2}},y\right) 
F\left[ T_{A}\!\left(x{+}{\textstyle\frac{b}{2}},y\right)\!,\,A \right], \label{eq:diff_npart} 
\end{align}
where
\begin{equation}
F\left[ T_{A}\!\left(x,y\right)\!,\,A \right] \equiv 1-\left(1-\frac{\sigma_{in} T_{A}\!\left(x,y\right)}{A}\right)^A.
\end{equation}
Here $\sigma_{in}$ is the inelastic nucleon-nucleon cross section 
whose value depends on the $\sqrt{s_{NN}}$ and is obtained from the 
experimental data. 

The distribution of $N_{part}$ and $N_{coll}$ in the transverse 
plane as obtained from Glauber model is used to calculate the 
initial energy/entropy density for the hydrodynamics simulation. The 
exact form for calculation of the energy density in the transverse 
plane using optical Glauber model is given by
\begin{equation}
\epsilon\left(x,y\right)= \epsilon_0 \left[\alpha N_{coll}\left(x,y\right)+(1-\alpha)
\frac{N_{part}\left(x,y\right)}{2}\right],
\end{equation}
where $\epsilon_0$ is a multiplicative constant used to fix the 
charged hadron multiplicity, $\alpha$ is the fraction of hard 
scattering \cite{Roy:2010zd}. The energy density corresponds to the 
MC-Glauber model is obtained with similar contribution from number 
of binary collisions and number of participants. 

In the second variation, the distribution of nucleons inside the 
colliding nucleus are sampled according to the nuclear density 
distribution by using statistical Monte-Carlo~(MC) method. The 
collisions between two nucleons occurs when the distance between 
them becomes equal or smaller than the radius obtained from the 
inelastic nucleon-nucleon cross section. This is also known as 
MC-Glauber model.

In MC-Glauber model the positions of binary collisions and 
participating nucleons are random and they are delta function in 
configuration space. These delta functions cannot be used in the 
numerical simulation of hydrodynamics. The usual practice is to use 
two-dimensional Gaussian profile to make a smooth profile of initial 
energy density as given by \cite{Holopainen:2012id}
\begin{equation}
\epsilon \left(x,y\right) = K\sum_{WN,BC}^{ }{\frac{1}{2\pi{\sigma}^{2}}} 
\exp\left[ -\frac{{\left( x-{x}_{i} \right)}^{2} + {\left( y-{y}_{i} \right)}^{2}}{2{\sigma}^{2}} \right],
\end{equation}
where $\sigma$ is a free parameter controlling the width of the 
Gaussian. Typical values for this fluctuation size parameter are of 
the order of $0.5$ fm, WN is the abbreviation for wounded nucleons 
which is same as number of participant $N_{part}$ and BC represent 
number of binary collisions. With this short discussion we now move 
on to the next topic.


\subsection{Color-Glass-Condensate}

The Color-Glass-Condensate (CGC) model takes into account the 
non-linear nature of the QCD interactions. Due to Lorentz 
contraction at relativistic energies, the nucleus in the laboratory 
frame is contracted into a sheet and therefore one only needs to 
consider the transverse plane. The density of partons inside such a 
highly Lorentz contracted nucleus is dominated by gluons. According 
to the uncertainty principle, the radius, $r_{gl}$, of a gluon is 
related to its momentum, $Q$, via $|r_{gl}|\times |Q|\sim\hbar=1$. 
Therefore the cross-section of gluon-gluon interactions is 
\begin{equation}
\sigma \sim \alpha_s(Q^2)\pi r_{gl}^2 \sim \alpha_s(Q^2)\frac{\pi}{Q^2},
\end{equation}
where $\alpha_s$ is the strong coupling constant. The total number 
of gluons in a nucleus can be considered to be approximately 
proportional to the number of partons, and therefore also to its 
mass number $A$. The density of gluons in the transverse plane is 
then given by $A/(\pi R_0^2)$, where $R_0$ is the radius of the 
nucleus. Gluons starts interacting with each other when the 
scattering probability becomes of the order unity,
\begin{equation}
\frac{A}{\pi R_0^2}\sigma = \alpha_s(Q^2)\frac{A}{R_0^2 Q^2} \sim 1.
\end{equation}
This indicates that there exists a typical momentum scale 
$Q_s^2=\alpha_sA/R_0^2$ separating perturbative ($Q^2\gg Q_s^2$) and 
non-perturbative ($Q^2\ll Q_s^2$) regimes. Classical chromodynamics 
is a good approximation at low momenta due to the high occupation 
number (``saturation"). The CGC model was developed to 
incorporate the saturation physics at low momenta $Q^2$ in 
relativistic heavy-ion collisions \cite{McLerran:1993ni, 
McLerran:1993ka}.

The presence of non-abelian plasma instabilities \cite 
{Romatschke:2005pm, Fukushima:2006ax, Attems:2012js} makes it 
difficult to determine the energy density distribution in the 
transverse plane. As a result, one has to resort to phenomenological 
models for the transverse energy density distribution in the CGC 
model \cite{Dumitru:2007qr}
\begin{equation}\label{en_den_CGC}
\epsilon({\bf x_\perp},b) = {\rm const}\times\left[ \frac{dN_g}{d^2{\bf x}_{T}dY}({\bf x}_T,b) \right]^{4/3}.
\end{equation}
Here $N_g$ is the number of gluons produced in the collision whose 
momentum distribution is given by
\begin{align}
\frac{dN_g}{d^2 {\bf x}_{T}dY} \sim& \int \frac{d^2{\bf p}_T}{p^2_T} \int^{p_T} d^2 {\bf k}_T \;\alpha_s(k_T) \nonumber\\
&\phi_+\!\!\left(\! \frac{({\bf p}_T+{\bf k}_T)^2}{4};{\bf x}_T \!\right)
\phi_-\!\!\left(\! \frac{({\bf p}_T - {\bf k}_T)^2}{4};{\bf x}_T \!\right), \\
\phi_{\pm} (k^2_{T}; {\bf x}_{T}) =&~
\frac{Q^2_s ~ (1-x)^4}{\alpha_s (Q^2_s)\textrm{max}(Q^2_s,k^2_{T})} 
\left(\frac{n_{\rm part}^A({\bf x_\perp},\pm b)}{T_A(x\pm b/2,y)}\right), \\
Q_s^2(x,{\bf x_\perp})=&~
\frac{2\,T_A^2(x\pm b/2,y)\,{\rm GeV}^2}{n_{\rm part}^A({\bf x_\perp},\pm b)}\!\!\left(\!\frac{{\rm fm}^2}{1.53}\!\right)\!
\!\left(\!\frac{0.01}{x}\!\right)^{\!0.288}, 
\end{align}
where $x=p_T/\sqrt{s}$. It is important to note that the Glauber and 
CGC models lead to different values of spatial eccentricity, defined 
by
\begin{equation}\label{eccentricity}
e_x(b)=\frac{\langle y^2-x^2\rangle_{\epsilon}}{\langle y^2+x^2\rangle_{\epsilon}},
\end{equation}
where $\langle\rangle_{\epsilon}$ represents averaging over the 
transverse plane with weight $\epsilon({\bf x_\perp},b)$. It has 
been observed that the CGC model typically has a larger eccentricity 
than the Glauber model which means that the anisotropy in fluid 
velocities is larger for the CGC model.

In a variant of the CGC model, also known as the 
Kharzeev-Levin-Nardi (KLN) model \cite{Kharzeev:2000ph, 
Kharzeev:2001gp, Kharzeev:2001yq}, the entropy production is 
determined by the initial gluon multiplicity. A Monte-Carlo version 
of KLN model (MC-KLN) has also been proposed to incorporate 
event-by-event fluctuations in the nucleon positions \cite 
{Drescher:2006ca, Drescher:2007ax}. In these models, the initial 
gluon production is calculated using the perturbative merging of two 
gluons from the target and projectile nuclei. The gluon structure 
functions are parametrized by a position-dependent gluon saturation 
momentum, $Q_s$, which is computed from the longitudinally projected 
density of wounded nucleons. The positions of the wounded nucleons 
are sampled according to Eq.~(\ref{eq:diff_npart}) using the 
MC-Glauber model. However, one should keep in mind that the 
MC-Glauber and MC-KLN models are unable to account for fluctuations 
of the gluon fields inside the colliding nucleons.


\section{Pre-equilibrium dynamics}

The initial condition models described in the previous section are 
static models because after the collisions the energy/entropy 
remains constant in space-time until the initial time $\tau_0$ when 
the hydrodynamics evolution starts. More realistic condition should 
include dynamical evolution of the constituent partons in the 
pre-equilibrium phase. The simplest choice for the dynamical 
evolution is the free-streaming of the produced partons in the 
pre-equilibrium phase, but this is in contrary to the assumption of 
local thermal equilibrium which needs multiple collisions among the 
constituent to achieve the local thermal equilibrium. In the 
following, we describe a few state-of-the-art models which takes 
into account the pre-equilibrium dynamics, until hydrodynamics sets 
in.


\subsection{IP-Glasma}

In the IP-Glasma model, the initial conditions is determined within 
the CGC framework by combining the impact parameter dependent 
saturation model (IP-Sat). In addition to fluctuations of nucleon 
positions within a nucleus, the IP-Glasma description also 
incorporates quantum fluctuations of color charges on the 
length-scale determined by the inverse saturation scale, $1/Q_s$. 
The initial Glasma fields are then evolved using the classical 
Yang-Mills (CYM) equation. One of the most important feature of this 
model is that long-range rapidity correlations from the initial 
state wavefunctions are efficiently converted into hydrodynamic flow 
of the final state quark-gluon matter \cite{Voloshin:2003ud, 
Shuryak:2007fu}. Moreover, initial energy fluctuations produced 
within this model naturally follows a negative binomial distribution. 

The color charges, $\rho^a(x^-,\xt)$, in the IP-Sat model behaves as 
local sources for small-$x$ classical gluon Glasma fields. The 
classical gluon fields are then determined by solving the classical 
Yang-Mills equations,
\begin{equation}\label{eq:YM1}
[D_{\mu},F^{\mu\nu}] = J^\nu.
\end{equation}
The color current in the above equation, generated by a nucleus $A$ 
($B$) moving along the $x^+$ ($x^-$) direction, is given by
\begin{equation}\label{eq:current}
J^\nu = \delta^{\nu \pm}\rho_{A (B)}(x^\mp,\xt),
\end{equation}
where the upper indices are for nucleus $A$.

It is easy to solve Eq.~(\ref{eq:YM1}) in Lorentz gauge, 
$\partial_\mu A^\mu = 0$, where the equation transforms into a 
two-dimensional Poisson equation
\begin{equation}
-\boldsymbol{\nabla}_\perp^2 A_{A(B)}^\pm = \rho_{A (B)}(x^\mp,\xt)\,.
\end{equation}
The solution of the above equation can be written as
\begin{equation}
A_{A(B)}^\pm = -\rho_{A (B)}(x^\mp,\xt)/\boldsymbol{\nabla}_\perp^2\,.
\end{equation}
Using the path-ordered exponential
\begin{equation}\label{eq:wilson}
V_{A (B)} (\xt) = P \exp\left({-ig\int dx^{-} \frac{\rho^{A (B)}(x^-,\xt)}{\boldsymbol{\nabla}_T^2+m^2} }\right)\,,
\end{equation}
one can gauge transform the results of Lorentz gauge to light-cone 
gauge, $A^+ (A^-) = 0$. The pure gauge fields are then given by \cite 
{McLerran:1994vd, JalilianMarian:1996xn, Kovchegov:1996ty}
\begin{align}\label{eq:sol}
A^i_{A (B)}(\xt) &= \theta(x^-(x^+))\frac{i}{g}V_{A (B)}(\xt)\partial_i V^\dag_{A (B)}(\xt)\,,\\
A^- (A^+) &= 0\,.\label{eq:sol2}
\end{align}
The discontinuity in the fields on the light-cone corresponds to the 
localized valence charge source \cite{Kovner:1995ja}.

The initial condition for a heavy-ion collision, at time $\tau=0$, 
is determined by the solution of the CYM equations in 
Fock--Schwinger gauge $A^\tau=(x^+ A^- + x^- A^+)/\tau=0$, where the 
$\tau,\eta$ coordinates are defined as $\tau = \sqrt{2 x^+ x^-}$ and 
$\eta = 0.5\ln(x^+/x^-)$. The Fock--Schwinger gauge is a natural 
gauge choice because it interpolates between the light-cone gauge 
conditions of the incoming nuclei. In terms of the gauge fields of 
the colliding nuclei, one obtains \cite{Kovner:1995ja, Kovner:1995ts}:
\begin{align}
A^i &= A^i_{(A)} + A^i_{(B)}\,,\label{eq:init1}\\
A^\eta &= \frac{ig}{2}\left[A^i_{(A)},A^i_{(B)}\right]\,,\label{eq:init2}\\
\partial_\tau A^i &= 0\,,\\
\partial_\tau A^\eta &= 0.
\end{align}
In the limit $\tau\rightarrow 0$, $A^\eta=-E_\eta/2$, where $E_\eta$ 
is the longitudinal component of the electric field. At $\tau=0$, 
one can non-perturbatively calculate the longitudinal magnetic and 
electric fields, which are the only non-vanishing components of the 
field strength tensor. These fields determine the energy density of 
the Glasma at each transverse position in a single event \cite 
{Krasnitz:1999wc, Krasnitz:2000gz, Lappi:2003bi}.

The Glasma fields are then evolved in time numerically according to 
Eq.~(\ref {eq:YM1}), up to a proper time $\tau_{\rm switch}$, which 
is the switching time from classical Yang-Mills dynamics to 
hydrodynamics \cite{Gale:2012rq}. At the switching time, one can 
construct the fluid's initial energy momentum tensor 
$T^{\mu\nu}_{\rm fluid} = (\epsilon + {\cal P})u^\mu u^\nu - {\cal 
P}g^{\mu\nu} + \Pi^{\mu\nu}$ from the energy density in the fluid's 
rest frame $\varepsilon$ and the flow velocity $u^\mu$. The local 
pressure ${\cal P}$ at each transverse position is obtained using an 
equation of state. The hydrodynamic quantities $\varepsilon$ and 
$u^\mu$ are obtained by solving the Landau frame condition, $u_\mu 
T^{\mu\nu}_{\rm CYM} = \varepsilon u^\nu$.  


\subsection{Transport: AMPT and UrQMD}

In Refs.~\cite{Steinheimer:2007iy, Pang:2012he, Bhalerao:2015iya} a 
different approach was taken in order to incorporate the 
pre-equilibrium dynamics for obtaining the initial condition of 
hydrodynamics evolution. While the authors of Ref.~\cite 
{Steinheimer:2007iy} employ ultrarelativistic quantum molecular 
dynamics (UrQMD) string dynamics model, A Multi Phase Transport 
Model (AMPT) was used in Refs.~\cite{Pang:2012he, Bhalerao:2015iya} 
to simulate the pre-equilibrium dynamics. In these studies the 
partons produced in the collisions were evolved until the initial 
time $\tau_0$ according to a simplified version of Boltzmann 
transport equation. We shall  discuss here the particular procedure 
used in Ref.~\cite {Pang:2012he} for calculating initial conditions 
for a (3+1)D hydrodynamics evolution with the parton transport in 
the pre-equilibrium phase. Additional benefit for choosing this type 
of initial condition is that one naturally incorporate the 
fluctuating energy density in the longitudinal direction due to the 
discrete nature of partons, details of which will be discussed in a 
later section.

In Ref.~\cite{Pang:2012he}, A Multi Phase Transport Model (AMPT) 
\cite{Zhang:1999bd, Lin:2004en} was used to obtain the local initial 
energy-momentum tensor in each computational cell. The AMPT model 
uses the Heavy-Ion Jet INteraction Generator (HIJING) model \cite
{Wang:1991hta, Gyulassy:1994ew} to generate initial partons from 
hard and semi-hard scatterings and excited strings from soft 
interactions. The number of excited strings in each event is equal 
to that of participant nucleons. The number of mini-jets per binary 
nucleon-nucleon collision follows a Poisson distribution with the 
average number given by the mini-jet cross section, which depends on 
both the colliding energy and the impact parameter via an 
impact-parameter dependent parton shadowing in a nucleus. The total 
energy-momentum density of parton depends on the number of 
participants, number of binary collisions, multiplicity of mini jets 
in each nucleon-nucleon collisions and the fragmentation of excited 
strings. HIJING uses MC-Glauber model to calculate number of 
participant and binary collisions with the Wood-Saxon nuclear 
density distribution function.

After the production of partons from hard collisions and from the 
melting strings, they are evolved within a parton cascade model, 
where only two parton collisions are considered. The positions and 
momentum of each partons are then recorded and used to calculate the 
initial energy-momentum tensor using a Gaussian smearing at time 
$\tau_0$ as 
\begin{align}\label{Tmunu_hydro}
T^{\mu\nu}&\!\left( \tau_0,x,y,\eta_s \right) = K \sum_{i=1}^N \, \frac{ p_i^\mu p_i^\nu }{ p_i^\tau }\,
\frac{ 1 }{ 2 \pi\, \tau_0 \sigma_r^2 \sqrt{ 2\pi \sigma_{\eta_s}^2 }} \nonumber \\
&\times\! \exp\!\left[ -\frac{ \left( x-x_i \right)^2 \!+\! \left( y-y_i \right)^2 }{ 2\sigma_r^2 } 
- \frac{ \left( \eta_s \!-\eta _{is} \right)^2 }{ 2\sigma_{\eta_s}^2 } \right]\!,
\end{align}
where $p_i^\tau=m_{iT}\cosh\left(Y_i-\eta_{is} \right)$, and 
$p_i^{x,y}=p_{x,yi}$, $p_i^\eta=m_{iT}\sinh\left(Y_i-\eta_{is} 
\right)/\tau_0$ are the four-momenta of the $i$th parton and $Y_i$, 
$\eta_{is}$, and $m_{iT}$ are the momentum rapidity, the spatial 
rapidity, and the transverse mass of the $i$th parton, respectively. 
Unless otherwise stated, the smearing parameters are taken as: 
$\sigma_r=0.6$ fm and $\sigma_{\eta_s}=0.6$ from Refs.~\cite 
{Pang:2012he} where the soft hadron spectra, rapidity distribution 
and elliptic flow can be well described. The sum index $i$ runs over 
all produced partons~($N$) in a given nucleus-nucleus collision. The 
scale factor $K$ and the initial proper time $\tau_0$ are the two 
free parameters that we adjust to reproduce the experimental 
measurements of hadron spectra for central Pb+Pb collisions at 
mid-rapidity \cite{Pang:2012he}. The initial energy density and the 
local fluid velocity in each cell is obtained from the calculated 
$T^{\mu\nu}$ via a root finding method which is used as an input to 
the subsequent hydrodynamics evolution, see Ref.~\cite{Pang:2012he} 
for further details. 


\subsection{Numerical relativity: AdS/CFT}

Another method to simulate the pre-equilibrium stage is via 
numerical relativity solutions to AdS/CFT \cite 
{vanderSchee:2013pia}. In this method, one employs the dynamics of 
the energy-momentum tensor of the strongly coupled Conformal Field 
Theory (CFT) on the boundary using the gravitational field in the 
bulk of $AdS_5$. Therefore a relativistic nucleus may be described 
using a gravitational shockwave in AdS, whereby the energy-momentum 
tensor of a nucleus can be exactly matched \cite{deHaro:2000vlm}. 
For a central collision, the dynamics of the colliding shockwaves 
has been solved near the boundary of AdS in Ref.~\cite 
{Romatschke:2013re}, resulting in the energy-momentum tensor at 
early times. The starting point of this simulation is the energy 
density of a highly boosted and Lorentz contracted nucleus, 
$T^{tt}=\delta(t+z) T_A(x,\,y)$. Here the thickness function, 
$T_A(x,y)$, is the same as defined in Eq.~(\ref{thick_fn}) but with 
an extra normalization, $\epsilon_0$, which is used to match the 
experimentally observed particle multiplicity, $dN/dY$.

In terms of the polar Milne coordinates $\tau,\xi,\rho,\theta$ with 
$t=\tau \cosh \xi$, $z=\tau \sinh \xi$, $\rho^2=x^2+y^2$, 
$\tan\theta=y/x$, the energy density, fluid velocity and pressure 
anisotropy was found up to leading order in $t$ \cite
{Romatschke:2013re}
\begin{eqnarray}\label{earlytimeTmunu}
\epsilon= 2 T_A^2(\rho)\tau^2\,,\ u^{\rho}=-\frac{T_A^\prime(\rho)}{3 T_A(\rho)} \tau\,,\ \frac{P_L}{P_T}=-\frac{3}{2},\quad
\end{eqnarray}
where in the local rest frame $T^\mu_\nu={\rm 
diag}(-\epsilon,P_T,P_T,P_L)$ \cite{Vredevoogd:2008id, 
Casalderrey-Solana:2013aba, Grumiller:2008va, Taliotis:2010pi}. One 
finds that the corresponding line-element $ds^2$ turns out to be 
$\xi$-independent (boost-invariant), up to leading order in $\tau$, 
and can be written as
\begin{eqnarray}
ds^2&=&-A d\tau^2+\Sigma^2 \left(e^{-B-C} d\xi^2+e^{B}d\rho^2+e^{C} d\theta^2\right)
\nonumber\\
&&+2 dr d\tau+2 F d\rho d\tau.
\end{eqnarray}
Here all functions depend on $\tau$, $\rho$ and the fifth AdS space 
dimension $r$ only. In this scenario, the space boundary is located 
at $r\rightarrow \infty$ where the induced metric is given by 
$g_{\mu\nu}={\rm diag} (g_{\tau\tau}, g_{\rho\rho}, 
g_{\theta\theta}, g_{\xi\xi})={\rm diag} (-1,1,\rho^2,\tau^2)$.

The metric is then expanded near the boundary, 
\begin{equation}\label{sigma}
B(r,\tau,\rho)\rightarrow B_0(r,\tau,\rho)+\sum_{i=0}^6  \frac{b_i(\tau,\rho)r^{-i}}{1+\sigma^7 r^{-7}},
\end{equation}
where $B_0$ is given by the vacuum value. In order to have a stable 
time evolution, a function with one bulk parameter $\sigma$ has been 
introduced to extend the metric functions to arbitrary $r$. An 
analogous expansion is also made for $C$. Using Eq.~(\ref 
{earlytimeTmunu}) to fix the near-boundary coefficients at a time 
$\tau_{\rm init}$, and choosing a value for $\sigma$, the time 
evolution of the metric can be determined by solving the Einstein 
equations. This is done numerically by adopting a pseudo-spectral 
method based on Refs.~\cite{Chesler:2010bi, vanderSchee:2012qj, 
Chesler:2013lia}. At a proper time $\tau_{\rm hydro}$, which is the 
switching time from AdS/CFT to hydrodynamics, the evolution using 
Einstein equations is stopped and hydrodynamic quantities such as 
$\epsilon, u^\mu, \pi^{\mu\nu}$ are extracted from the metric using 
Eq.~(\ref{earlytimeTmunu}). These quantities are then used to create 
the energy-momentum tensor which provides the initial conditions for 
the subsequent relativistic viscous hydrodynamic evolution. The 
initial conditions for hydrodynamic evolution is therefore 
determined using an early-time, far-from-equilibrium dynamics, 
modeled as a strongly coupled CFT described by gravity in AdS. 
Recently, a non-conformal extension has also been studied in order 
to incorporate bulk viscosity \cite{Attems:2016tby}.


\section{Equation of state}

Equation of State (EoS) is the functional relationship between 
thermodynamic variables pressure (P) and number density (n) to the 
energy density ($\epsilon$). The conservation equations, 
$\partial_{\mu}T^{\nu\mu} = 0$, contains one additional variable 
than the number of equations. EoS closes the system of equations by 
providing another functional relationship and it is one of the 
important input to hydrodynamics. For a relativistic simple fluid 
the acceleration under a given pressure gradient $\nabla P$ is 
governed by the following relationship
\begin{equation}
Du^{\mu}=-\frac{1}{\epsilon+P}\nabla P,
\end{equation}
where $D=u^{\mu}\partial_{\mu}$ is the covariant derivative, 
$\epsilon$ and $P$ are the energy density and pressure 
respectively. Clearly the fluid expansion is governed by the 
gradient of pressure as well as the combined value of pressure and 
energy density. The pressure for a given energy density is defined 
via the EoS and hence the EoS governs the rate of change of fluid 
expansion.

At present, the most reliable calculation of EoS for nuclear matter 
at high temperature ($>100$ MeV) is obtained from lattice QCD (lQCD) 
calculations. However, at present the lQCD calculations are not 
reliable at lower temperatures (because of the large grid size 
needed at lower temperatures) and at higher baryon densities (due to 
the so called sign problem for finite chemical potential). The usual 
practice in the heavy-ion community is to use lQCD calculation at 
high temperature and a hadron resonance gas (HRG) model at lower 
temperature to construct the equation of state for vanishing baryon 
chemical potential ($\mu_b$). The EoS for finite $\mu_b$ is usually 
obtained by employing some approximation such as Taylor series 
expansion around $\mu_b=0$. For more details about the nuclear EoS 
relevant to the heavy-ion collisions see Ref.~\cite{Huovinen:2009yb} 
and references therein. Here we briefly outline the procedure used 
to calculate the lQCD+HRG equation of state for vanishing baryon 
chemical potential. 

\begin{figure}[t!]
{\includegraphics[width=7cm]{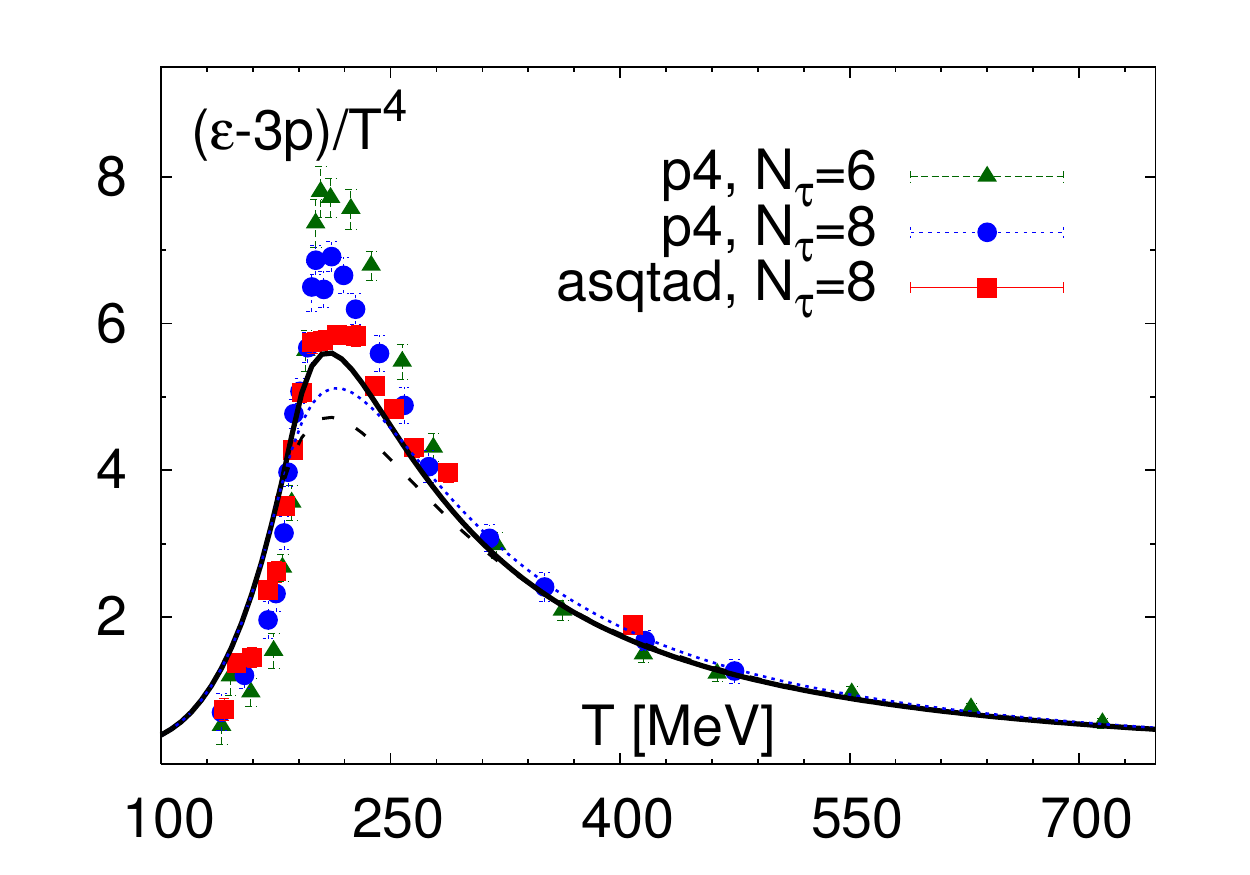}}
{\includegraphics[width=7cm]{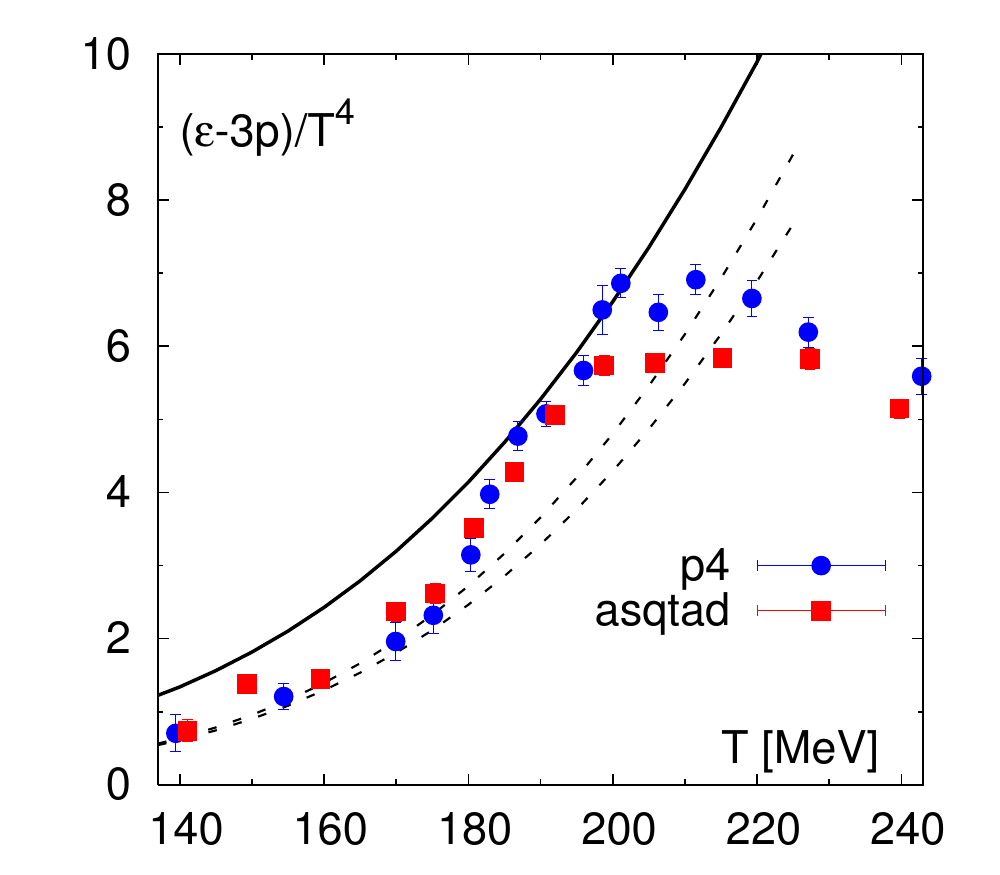}}
\caption{(Color online) The trace anomaly calculated in lattice QCD 
with p4 and asqtad actions on $N_\tau=6$ and 8 lattices compared 
with various parametrizations given by the solid, dotted and dashed 
lines (top) and the trace anomaly calculated in lattice QCD compared 
with the HRG model given by the solid and dashed lines (bottom). The 
figures are taken from Ref.~\cite{Huovinen:2009yb}.}
\label{EoS_param}
\end{figure}

Usual lQCD calculations for the thermodynamical variables assume 
that the system has infinite extent (volume $V \rightarrow \infty$) 
and it is homogeneous \cite{Boyd:1996bx}. All thermodynamic 
quantities can be derived from the partition function Z($T,V$). The 
energy density and pressure are derivatives of the partition 
function with respect to $T$ and $V$ respectively
\begin{eqnarray}
\epsilon &=& \frac{T^2}{V}\frac{\partial}{\partial T} lnZ\left(T,V\right) ,\\
P &=& T\frac{\partial}{\partial V} lnZ\left(T,V\right). 
\end{eqnarray}
The pressure for a homogeneous system of infinite extent can be simply expressed 
in terms of $f$ as 
\begin{equation}
P = \frac{T}{V}lnZ\left(T,V\right). 
\end{equation}  
Using the above relations one can arrive at the following relationships
\begin{align}
\epsilon &= T\frac{\partial P}{\partial T} - P, \label{eq:endens_therm} \\
\Theta(T) &= T\frac{\partial}{\partial T} \left(\frac{P}{T^4}\right), \label{eq:press_therm}
\end{align}
where the trace anomaly, $\Theta(T)=(\epsilon-3P)/T^4$.

In the high temperature, $\Theta(T)$ can be reliably calculated from 
lQCD. On the other hand, at lower temperature the lQCD results are 
affected by possibly large discretisation effect. Therefore the 
usual practice to construct realistic EoS is to use the lattice data 
for the trace anomaly in the high temperature region ($T>250 MeV$), 
and use HRG model in the low temperature region ($T<180 MeV$). In 
the intermediate temperature $\Theta(T)$ is obtained by joining the 
parametrised high temperature and low temperature values smoothly 
(continuous first and second derivative). Once $\Theta(T)$ is known, 
pressure can be calculated by using Eq.~(\ref{eq:press_therm}). The 
energy density then can be readily obtained from Eq.~(\ref 
{eq:endens_therm}). The top panel of Fig.~\ref{EoS_param} shows the 
trace anomaly calculated in lattice QCD with p4 and asqtad actions 
on $N_\tau=6$ and 8 lattices \cite{Bazavov:2009zn} compared with 
various parametrizations given by the solid, dotted and dashed lines 
\cite{Huovinen:2009yb}. The bottom panel shows the trace anomaly 
calculated in lattice QCD compared with the HRG model given by the 
solid and dashed lines \cite{Huovinen:2009yb}.


\section{Freeze-out: Spectra and Flow}

In the late stage of hydrodynamics evolution of hot and dense 
nuclear matter created in high energy heavy-ion collisions, the 
density and the temperature reaches a critical value when the 
constituents no longer collides among themselves and thereafter they 
move in a straight trajectory towards the detectors. This phenomenon 
is known as freeze-out, more precisely the kinetic freeze-out. There 
is another chemical freeze-out when the particle number changing 
processes ceases. The chemical freeze-out temperature so far known 
to be higher than the kinetic freeze-out temperature. 

In relativistic hydrodynamics simulations one also needs to stop the 
hydrodynamics evolution when the system reaches the kinetic 
freeze-out criterion. For this one needs to impose some physical 
constraints to calculate the freeze-out hyper-surface. This can be 
done by more than one way, we discuss here only the most popular 
choices used to calculate the freeze-out hyper-surface namely (i) 
the constant temperature freeze-out (ii) the constant energy density 
freeze-out (iii) the dynamical freeze-out. Among the three choices, 
the first two are based on the general idea that the pion (or other 
hadron) cross section is very sensitive to the  temperature/energy 
density of the system, thus within short interval of 
temperature/energy density the condition for kinetic freeze-out is 
achieved. For the computational purpose this is realised by choosing 
a constant temperature/energy density surface. 
 
The dynamical freeze-out is based on the idea that the ratio of 
expansion rate ($\theta$) to the  collision rate ($\Gamma$)  should 
be much much less than unity ($\frac{\theta}{\Gamma}\ll$ 1) in order 
to maintain the local thermal equilibrium essential for the 
applicability of the hydrodynamics evolution. One can then define 
the freeze-out criterion based on some predefined value of 
$\frac{\theta}{\Gamma}$ smaller than 1. Though the idea of dynamical 
freeze-out sounds more realistic it is not easy to implement  in 
numerical calculation see Ref.~\cite{Huovinen:2009yb} for details.

The thermodynamical quantities of the fluid such as energy density, 
pressure and the fluid velocity obtained from the hydrodynamics 
simulation on the freeze-out surface are used to evaluate momentum 
distributions of the identified hadrons. The conversion of fluid to 
hadrons is done by using the Cooper-Frye procedure, see Ref.~\cite 
{Cooper:1974mv} for details. In Cooper-Frye procedure the momentum 
distribution (or invariant yield) of hadrons are calculated as \cite 
{Cooper:1974mv}
\begin{equation}
E\frac{d^3N}{d^3p} = \frac{d^3N}{d^2p_{T}dy} = \int_{\Sigma}{f\left(x,p\right){p}^{\mu}d{\Sigma}_{\mu}} ,
\end{equation}
where E, N, and $p^{\mu}$ are energy, number and four-momentum of 
hadrons, $d\Sigma_{\mu}$ is the differential freeze-out 
hyper-surface element. The distribution function, $f(x,p)$, consists 
of an equilibrium part, $f_0(x,p)$, and dissipative corrections, 
$\delta f(x,p)$. While the equilibrium distribution corresponding to 
the local thermodynamic quantities, as given in Eq.~(\ref{EDFC2}), 
is taken to be either Bose-Einstein or Fermi-Dirac distribution 
depending on the spin of the hadronic species, the dissipative 
correction is not unique, and will be explained in the following.

In the simple case when the dissipation is only due to the shear 
viscosity, leading-order moment method, also known as the Grad's 
14-moment approximation, leads to the well-known form of the viscous 
correction \cite{Grad, Israel:1979wp}
\begin{equation}\label{deltaf_grad}
\delta f(x,p) = \frac{f_0 \tilde f_0}{2(\epsilon+P)T^2}\, p^\alpha p^\beta \pi_{\alpha\beta},
\end{equation}
where $\tilde f_0 \equiv 1-r f_0$, with $r=1,-1,0$ for Fermi, Bose, 
and Boltzmann gases, respectively. Note that the viscous correction 
in this case increases with quadratic power of momenta. On the other 
hand, the Chapman-Enskog like iterative solution of the Boltzmann 
equation, Eq.~(\ref{FOCC5}), leads to a viscous correction which is 
effectively linear in momenta \cite{Bhalerao:2013pza},
\begin{equation}\label{deltaf_CE}
\delta f(x,p) = \frac{5 f_0 \tilde f_0}{2 (\epsilon+P) T}\frac{1}{(u \cdot p)}\, p^\alpha p^\beta \pi_{\alpha\beta}.
\end{equation}
It has been shown that, in contrast to Eq.~(\ref{deltaf_grad}) 
obtained using moment method, Eq.~(\ref{deltaf_CE}) leads to 
phenomenologically consistent corrections to the equilibrium 
distribution function, and is therefore a better alternative for 
hydrodynamic modeling of relativistic heavy-ion collisions \cite
{Bhalerao:2013pza}. 

We note here that the calculation of four-dimensional freeze-out 
hyper-surface and the numerical evaluation of it is not trivial, for 
example see Ref.~\cite{Huovinen:2012is} for more details. Once we 
know the invariant momentum distribution the ``n-th" order Fourier 
coefficient the flow harmonics $v_n$ can be readily obtained as
\begin{equation}\label{vn_def}
v_{n} = \frac{\displaystyle{\int_{y}dy\int_{p_T}d^2p_{T}\frac{d^3N}{d^2p_{T}dy}
\cos\left[n\left(\phi_n-\psi_n\right)\right]}}{\displaystyle{\int_{y}dy\int_{p_T}d^2p_{T}\frac{d^3N}{d^2p_{T}dy}}}\,.
\end{equation}
These above mentioned quantities are directly compared to the 
corresponding experimental data in order to obtain information about 
the transport coefficients such as shear and bulk viscosity of the 
QGP.


\section{Resonance decay and hadronic rescattering}

In high energy nuclear collisions various hadronic resonances are 
formed. The life time of most of the resonance particles are of the 
order of the expansion life time of the nuclear matter. The end 
product for the most of the decay channels involve pions. The decay 
of hadron resonances to pion enhances the pion yield specially at 
low transverse momentum, $p_T$. One can use the formalism given in 
\cite{Sollfrank:1990qz} to calculate the relative contribution of 
the resonance decay to thermal pion spectra. The relative 
contribution of the resonance decay to pion spectra is a function of 
both the freeze-out temperature, $T_{\rm fo}$, and $p_T$. Thus the 
final $p_T$ spectra of $\pi$ are obtained by adding the contribution 
from resonance decay to the thermal $p_T$ spectra calculated from 
Cooper-Frye formula. The most dominant hadronic decay channels 
contributing to pion yield are: $\rho^\pm \rightarrow \pi^\pm \pi^0, 
~ \rho^0 \rightarrow \pi^- \pi^+, ~ K^{*\pm} \rightarrow \pi^\pm 
K^0, ~ K^{*0} \rightarrow \pi^- K^+, ~ \Delta \rightarrow 
\pi^{\pm,0} N, ~ \omega \rightarrow \pi^+ \pi^- \pi^0, ~ \eta 
\rightarrow \pi^+ \pi^- \pi^0$, which should be considered with 
their corresponding branching ratios \cite{Sollfrank:1990qz}.

According to the formalism given in \cite{Sollfrank:1990qz}, to 
calculate the pion contribution from resonances, one need to provide 
the source temperature. The parametric fit to the ratio of the total 
pion to the thermal pion for the calculation at two different 
freeze-out temperature $T_{\rm fo}=130$ MeV and $150$ MeV are 
approximately given by \cite{Victor:thesis}
\begin{align}
\left.{\frac{\pi^\pm_{\rm total}}{\pi^\pm_{\rm thermal}}}\right\vert_{\rm T_{fo}=130 MeV} & = 1.0121 +
\frac{1.4028}{1+\left(\displaystyle{\frac{\frac{p_T}{m_\pi}-0.0964}{3.666}}\right)^{\!2}}\,, \label{eq:ratio_130}\\
\left.{\frac{\pi^\pm_{\rm total}}{\pi^\pm_{\rm thermal}}}\right\vert_{ \rm T_{fo}=150 MeV} & = 1.0252 +
\frac{3.0495}{1+\left(\displaystyle{\frac{\frac{p_T}{m_\pi}-0.2302}{2.792}}\right)^{\!2}}\,, \label{eq:ratio_150} 
\end{align}
where $m_\pi=139$ MeV is the pion mass. Note that about $\sim50\%$ 
of the total pion yield come from resonance decay at LHC energy 
($\sqrt{s_{NN}} = 2.76$ TeV), whereas for RHIC energy ($\sqrt{s_{NN}} 
= 200$ GeV) the resonance contribution to total pion yield is 
$\sim30\%$ for $T_{\rm fo}=130$ MeV.

The sudden conversion of fluid to non-interacting hadrons at the 
freeze-out hyper-surface in the fluid dynamical evolution is hard to 
happen in practice. In reality the hydrodynamical picture should 
work fine for the early hot and dense phase of the QGP evolution 
when the scattering rate is comparatively large compared to the 
expansion rate. As the system grows in size and cools down with time 
the scattering rate goes down compared to the expansion rate. At 
some point of space-time, particularly in the late hadronic phase  
it is expected that the dynamical evolution most probably be 
governed by the microscopic Boltzmann equations considering multiple 
hadronic species and their collisions rather than the simplified 
macroscopic hydrodynamics evolution. Thus a complete dynamical 
evolution of high energy heavy-ion collisions contains simpler 
hydrodynamics evolution in the early time and a much computational 
expensive hadronic transport evolution in the late stage with the 
additional complexity of transforming fluid variables to position 
and momentum of hadrons. 

For the hadronic rescattering phase several microscopic algorithms 
that solve coupled Boltzmann equations for a hadronic gas were 
developed in the 1980s and 1990s \cite{Aichelin:1986wa, 
Stoecker:1986ci, Sorge:1989dy, Ehehalt:1996uq, Zhang:1999bd, 
Bass:1998ca, Bleicher:1999xi}. Hybrid codes that coupled an ideal 
fluid dynamical description of an expanding QGP to hadronic 
rescattering codes and compared the results with purely fluid 
dynamical calculations began to appear around 2000 \cite 
{Bass:2000ib, Teaney:2001gc, Hirano:2005xf, Nonaka:2006yn}. One of 
the first numerical code VISHNU that couples (2+1)D viscous hydro 
with a late hadronic Boltzmann cascade appeared in 2011 \cite 
{Song:2010aq}. The use of these more sophisticated hybrid models are 
believed to reduce the uncertainty in the extracted value of $\eta/s$
of QGP, since the late hadronic stage is known to have larger shear 
viscosity which in usual viscous hydrodynamics simulations is not 
taken into account properly. We shall not go into the details of the 
hadronic transport model nor to the technical details of various 
techniques and uncertainties arising due to the matching of viscous 
hydrodynamics to the hadronic transport, details of which can be 
found in Ref.~\cite{Song:2010aq} and references therein. Before 
finishing this section we should point out one of the major findings 
of Ref.~\cite{Song:2010aq}, the $\eta/s$ of hadronic matter is found 
to be quite sensitive to the details of preceding hydrodynamics 
phase and on the switching temperature when the viscous 
hydrodynamics is switched to the hadronic transport evolution. The 
effort to better constraint the $\eta/s$ of QGP by using such 
sophisticated numerical models is a topic of current ongoing research.


\section{Transport coefficients}

Determination of transport coefficients of the hot and dense QCD 
matter is one of the primary goal of theoretical simulations of 
relativistic heavy-ion collisions; see \cite{Heinz:2013th} for a 
recent review. Ideal hydrodynamics has been proved to be quite 
successfully in the past to describe the spectra of produced 
particles in relativistic heavy-ion collisions. The presence of 
dissipation leads to dissipative entropy generation via Eq.~(\ref 
{EFCD3C2}), which results in the increase of total particle 
multiplicity for a fixed initial entropy. Shear viscosity, in 
particular, also leads to stronger radial flow leading to an 
increase in the mean transverse momentum of particles. However, the 
most important effect of shear viscosity is to suppress the elliptic 
flow coefficient, $v_2$, defined in Eq.~(\ref{vn_def}) strongly. 
Therefore, in order to estimate the viscosity of the QCD matter 
within a hydrodynamic simulation, one has to tune the value of the 
specific shear viscosity, $\eta/s$, in order to fit the experimental 
data for $v_2$. One of the first estimates of $\eta/s$ was made 
within a hydrodynamics inspired blast-wave model \cite 
{Teaney:2003kp}. Since then there has been a lot of activity in this 
field, which is briefly reviewed in the following.

\begin{figure}[t!]
{\includegraphics[width=\linewidth]{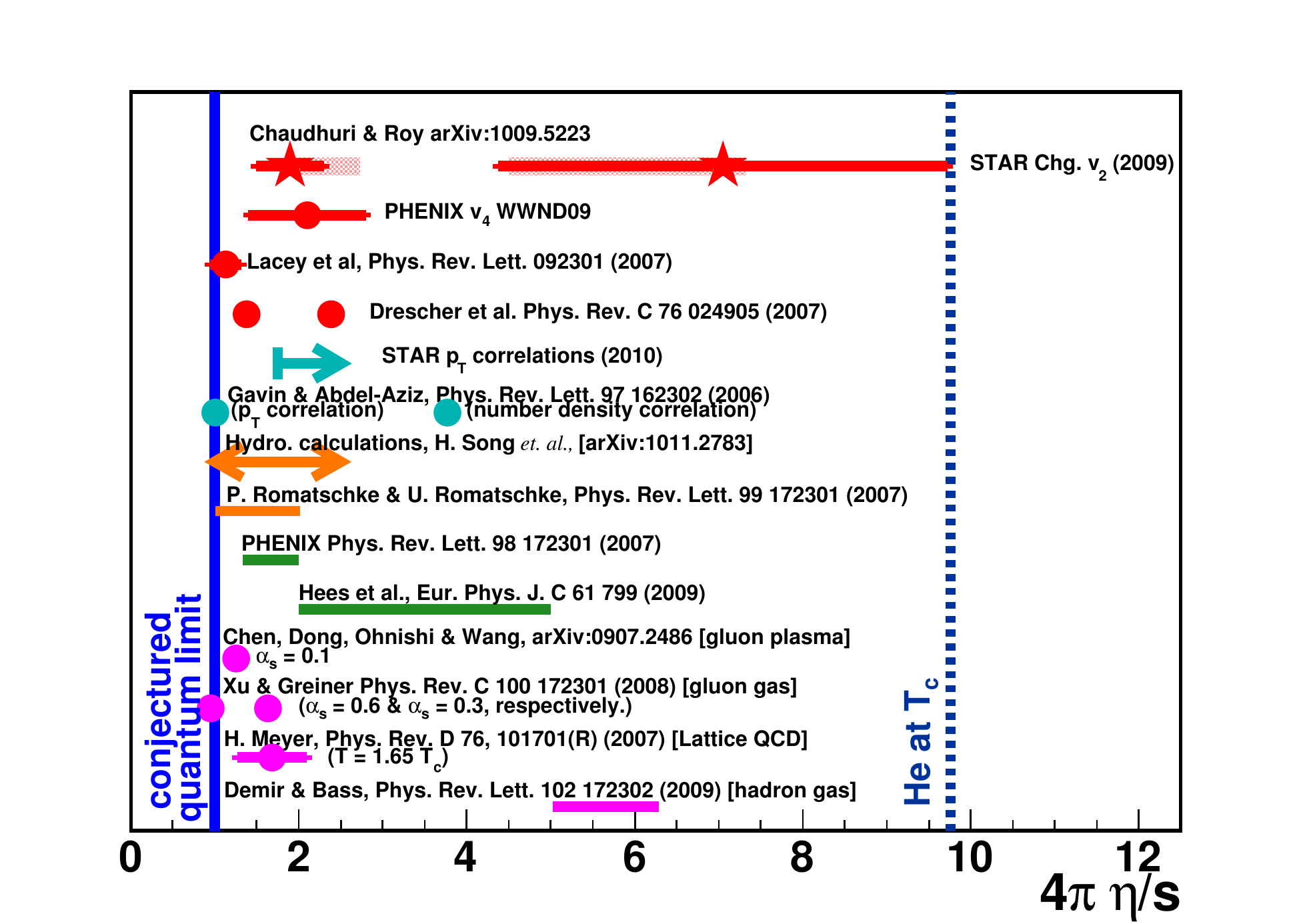}}
{\includegraphics[width=7cm]{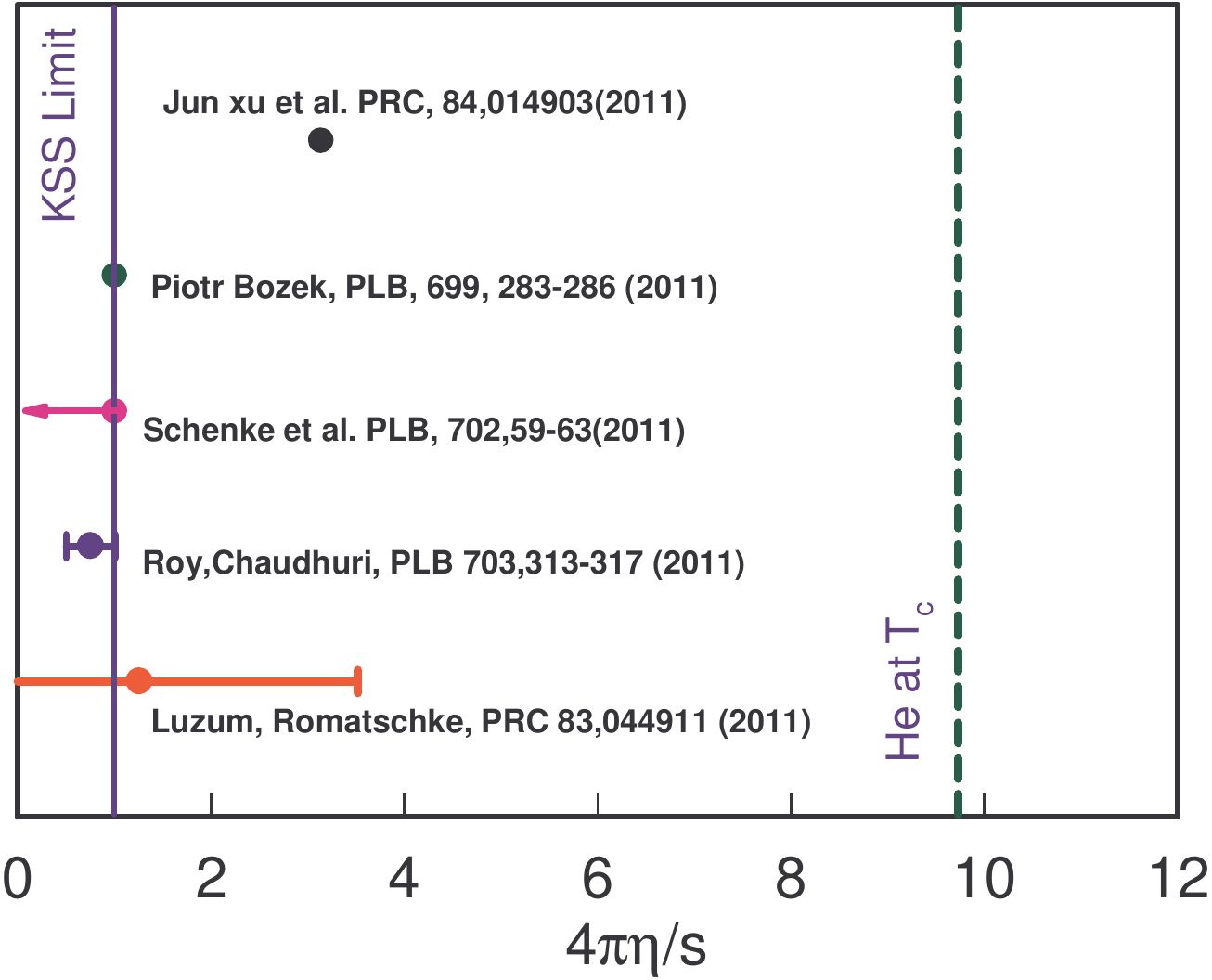}}
\caption{(Color online) Extracted values of $\eta/s$ for Au-Au 
collision at $\sqrt{s_{NN}}=200$ GeV (top) and for Pb-Pb collision 
at $\sqrt{s_{NN}}=2.76$ TeV (bottom) by different model calculations 
using different experimental observables. The solid vertical line at 
the left shows the lower limit of $\eta/s$ in unit of $1/(4\pi)$ 
\cite{Kovtun:2004de}. For comparison we have also shown the $\eta/s$ 
of He at $T_{c}$ (dashed vertical line on the right).}
\label{etas}
\end{figure}

Figure~\ref{etas} (top panel) shows the extracted values of $\eta/s$ 
in different model calculations for Au-Au collisions at 
$\sqrt{s_{NN}}=200$ GeV \cite{Romatschke:2007mq, Song:2010mg, 
Chaudhuri:2010hr, Lacey:2006bc, Drescher:2007cd, Gavin:2006xd, 
vanHees:2008gj, Chen:2009sm, Xu:2007ns, Meyer:2007ic, Demir:2008tr}. 
Most of the estimates are obtained by comparing experimental data 
for elliptic flow with model calculations. Some of the estimates 
used $p_{T}$ correlations and heavy meson $v_{2}$ data. The 
theoretical calculations include simulations with transport based 
approach as well as (2+1)D and (3+1)D viscous hydrodynamics with 
various initial conditions. Also shown are the results from lattice 
QCD calculation. All these results indicate that the $\eta/s$ value 
of the QGP fluid produced at top RHIC energies lies within $1-5\times
1/4\pi$ and is below the $\eta/s$ value of helium (blue dashed 
line) at $T_{c}$. The spread in the estimated values of $\eta/s$ 
reflects the current uncertainties associated with the theoretical 
calculations.

Figure~\ref{etas} (bottom panel) shows $\eta/s$ estimated in various 
model calculations for Pb-Pb collision at $\sqrt{s_{NN}}=2.76$ TeV 
\cite{Xu:2011fe, Bozek:2011wa, Schenke:2011tv, Luzum:2008cw}. All 
the model calculations indicates that the values of $\eta/s$ of the 
QCD matter formed in heavy-ion collision at LHC lies between 1-4 
$\times(1/4\pi)$. The specific shear viscosity was obtained in 
reference \cite{Xu:2011fe} by using A Multi Phase Transport model 
(AMPT). Bozek \cite{Bozek:2011wa} has estimated the specific shear 
viscosity of the fluid for LHC energy by using a (2+1)D viscous 
hydrodynamics model. In addition to shear viscosity, bulk viscosity 
($\zeta/s=0.04$) in the hadronic phase was considered. Freeze-out 
and resonance decay was based on THERMINATOR event generator \cite 
{Kisiel:2005hn}. Experimental data are best fitted with $\eta/s\sim 
0.08$. A (3+1)D viscous hydrodynamics calculation with fluctuating 
initial conditions was done by Schenke {\it et al.} \cite 
{Schenke:2011tv}. They explain the $v_{2}(p_{T})$ and $p_{T}$ 
integrated $v_{2}$ for different centralities. Their calculation 
shows that the experimental data measured at LHC by the ALICE 
collaboration are best described for $\eta/s$ value 0.08 or smaller. 
Luzum {\it et al.} \cite{Luzum:2008cw} have estimated $\eta/s$ by 
using a (2+1)D viscous hydrodynamics simulation with smooth initial 
conditions for LHC energy to be same as at RHIC, $\eta/s=0.1\pm 0.1 
({\rm theory}) \pm 0.08 ({\rm experiment})$. Comparison of 
experimentally measured integrated and differential $v_2$, the 
charged hadron $p_T$ spectra and multiplicity in the mid-rapidity 
and their global fit by minimising $\chi^2$ was done in Ref.~\cite 
{Roy:2011xt} by using a (2+1)D viscous hydrodynamics simulation, the 
extracted value of $\eta/s$ is $\sim0.07\pm 0.01$.

The effects of bulk viscosity in hydrodynamic simulations of 
relativistic heavy-ion collisions have not been investigated as 
thoroughly as that of shear viscosity. In principle, the bulk 
viscosity of the QCD matter should not be zero for the range of 
temperatures achieved at the RHIC and the LHC, and it may become 
large enough to significantly affect the evolution of the medium 
\cite{Meyer:2007dy, Paech:2006st}. There has been several 
simulations of heavy-ion collisions that include the effect of bulk 
viscosity where it has been demonstrated that bulk viscosity can 
have a non-negligible effect on heavy-ion observables \cite 
{Ryu:2015vwa, Noronha-Hostler:2013gga, Song:2009rh, Roy:2011pk, 
Noronha-Hostler:2014dqa, Denicol:2009am}. However, there are various 
uncertainties in the extraction of bulk viscosity from the 
anisotropic flow data of heavy-ion collisions. For example, the 
theoretical uncertainties arising due to the ambiguities in the form 
of the specific bulk viscosity, $\zeta/s$, its relaxation time and 
the bulk viscous corrections to the freeze-out process, makes it 
difficult to study the effect of bulk viscosity on the evolution of 
QCD matter. Unlike shear viscosity, the extraction of bulk viscosity 
from hydrodynamic simulations is still unresolved and is currently 
an active research area.


\section{Recent developments}


\subsection{Flow in small systems: proton-proton and proton-nucleus collisions}

As mentioned in the introduction, among the recent developments in 
the field of high energy heavy-ion collisions the most striking 
observation is the existence of radial flow like pattern in high 
multiplicity proton-proton (p-p) and proton-lead (p-Pb)collisions, 
for example see Ref.~\cite{Loizides:2016tew} for a summary of recent 
experimental results. At this point it needs some explanation why 
the observation of flow in small system is remarkable. One of the 
fundamental assumption when applying hydrodynamics to high energy 
nuclear collisions is that the system reaches a state of "local 
thermal equilibrium" very quickly because of the strong interactions 
among the quarks and gluons.  In heavy-ion collisions such as Pb-Pb 
or Au-Au  the number of participating nucleons are large, for 
example for a head-on Au-Au or Pb-Pb collisions there are 197+197 
and 208+208 participating nucleons. Each of these colliding nucleons 
on average produce more than one particles in each collisions, thus 
the total number of degrees of freedom in the system created just 
after the collisions are large. They collide among themselves 
through strong interaction and subsequently reaches local thermal 
equilibrium (we note here that the full mechanism by which the 
system reaches local thermal equilibrium within such a short period 
of time is not fully understood yet).

\begin{figure}[t!]
{\includegraphics[width=7cm]{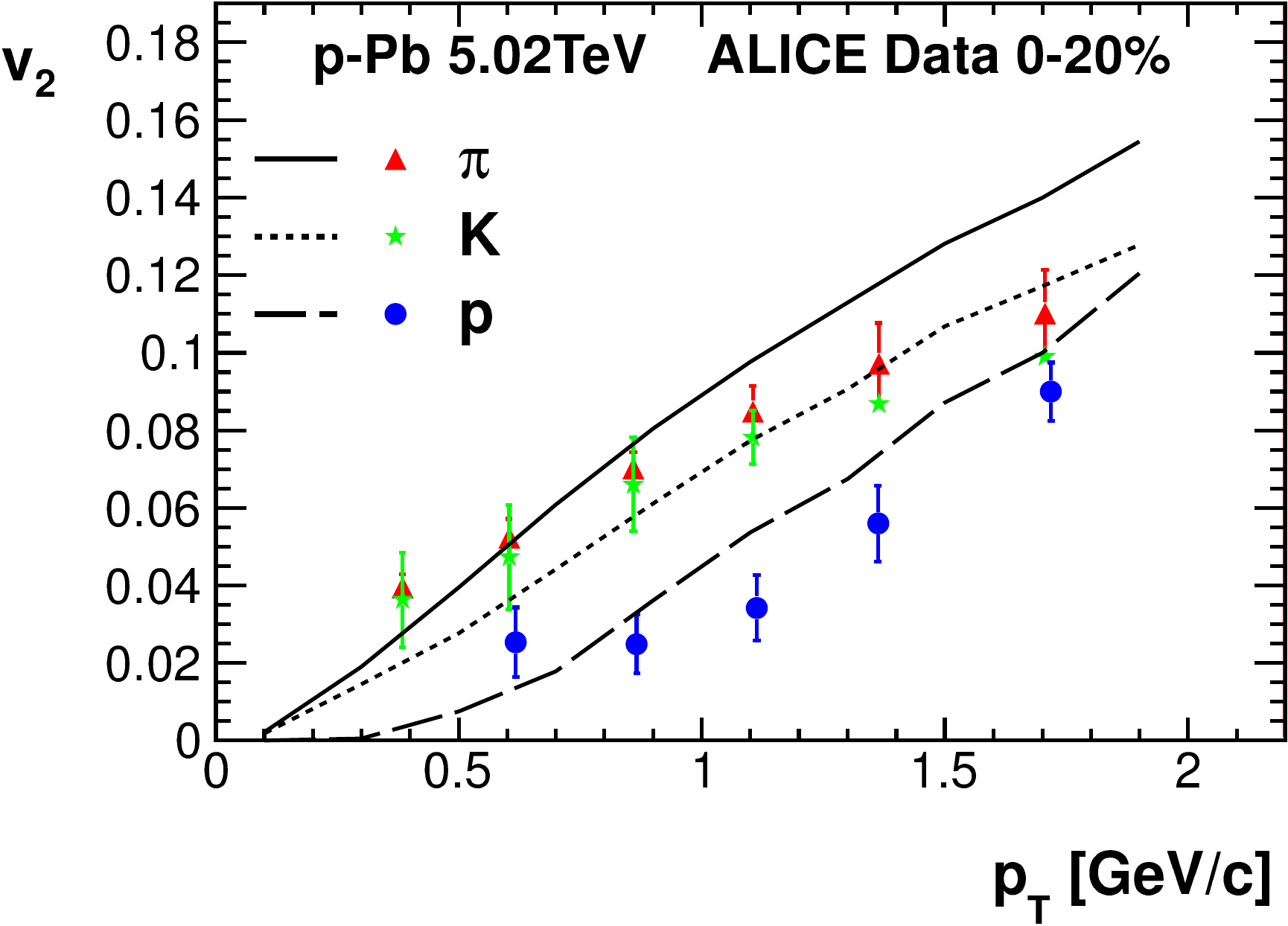}}
\caption{(Color online) Elliptic flow, $v_2$, of identified 
particles from the hydrodynamic model compared to the ALICE data 
\cite {ABELEV:2013wsa}. The figure is taken from Ref.~\cite 
{Bozek:2014wpa}.}
\label{fig_pPb_v2id}
\end{figure}

The situation is very different in smaller colliding systems such as 
p-p or p-Pb where the number of participating nucleons and the 
numbers of produced particles are comparatively smaller in numbers. 
It is very counterintuitive that with such few number of particles 
the system reaches local thermal equilibrium within a very short 
time-period. On the other hand, event generators based on 
perturbative QCD such as PYTHIA and HIJING have successfully 
described various observables associated with particle production in 
p-p collisions. Thus p-p as well as p-nucleus collisions have been 
for long time considered qualitatively different from heavy-ion 
collisions, for which the hydrodynamic description became a 
mainstream since its successful explanation of RHIC data. It is 
interesting to note that the p-p collisions were used as a benchmark 
for studying the existence of QGP in larger systems where a 
thermalised medium is believed to be created.

This situation changed recently as the CMS and ATLAS collaboration 
observed a ``ridge" like correlation in the azimuthal distribution 
of charged hadrons produced in high multiplicity p-p or p-Pb 
collisions. In those experiments a mass dependence of the slope of 
identified hadron's $m_T$ spectrum in high multiplicity p-Pb and p-p 
collisions were also observed. All these phenomenon are known to be 
the most significant indication for existence of hydrodynamic flow 
in larger colliding systems such as Au-Au or Pb-Pb. Like in 
heavy-ion collisions, the PYTHIA model failed to describe these 
observed experimental measurement for high multiplicity p-p or p-Pb 
events unless it employs some special mechanism like Color 
Reconnection~(CR) and Multi Parton Interaction~(MPI) with an 
additional free parameter to explain the experimental data \cite 
{Ortiz:2013yxa}. On the other hand, the relativistic hydrodynamic 
models with large radial velocity have been proved to be quite 
successful in describing the same experimental data. It is also 
worthwhile to mention that there are some other theoretical 
conjectures about these recent observation which does not 
incorporate this hydrodynamics like flow, but till now those studies 
lack detailed numerical calculation in order to compare it with the 
experimental data Ref.~\cite{Gyulassy:2014cfa}.

\begin{figure}[t!]
{\includegraphics[width=7cm]{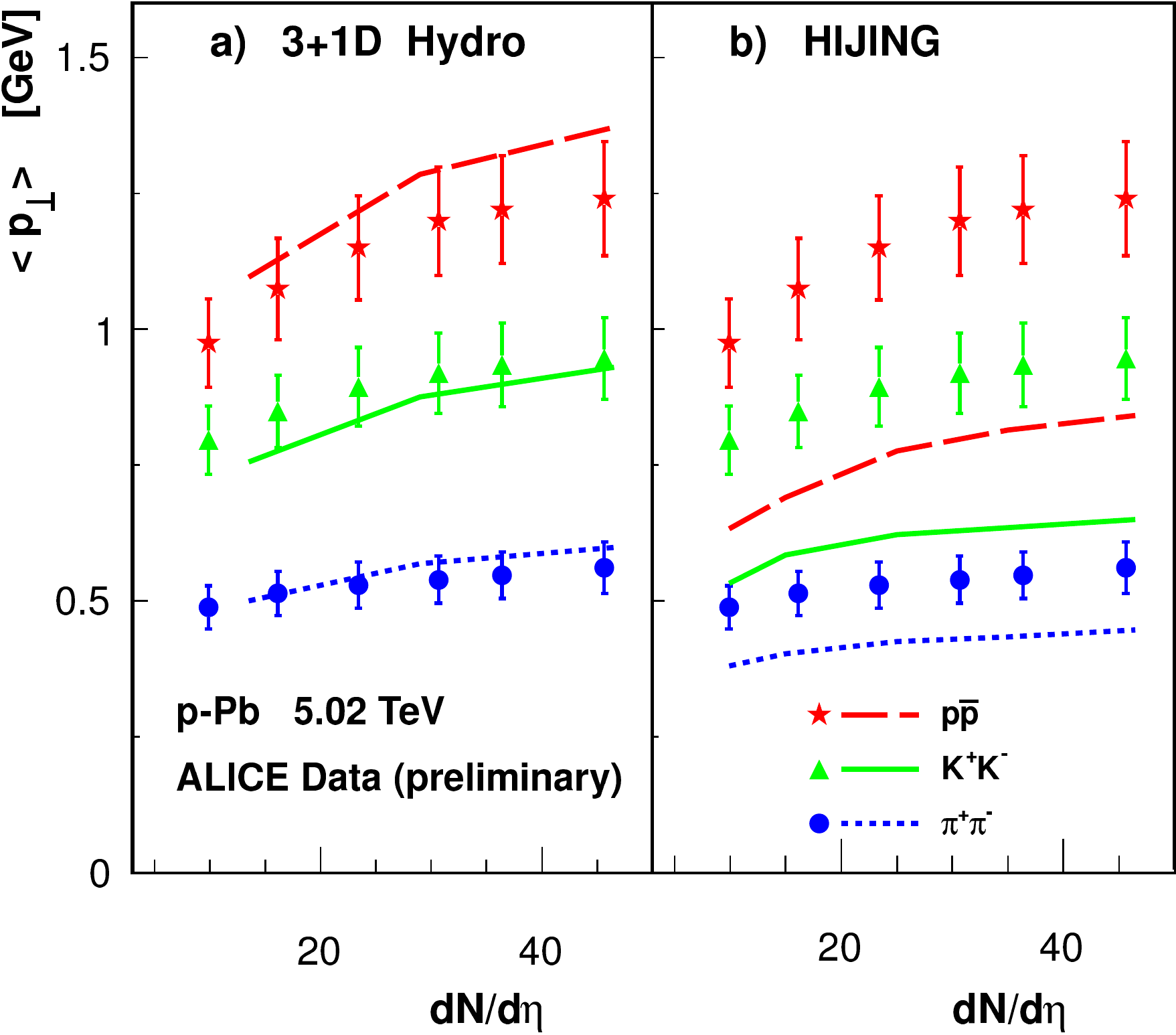}}
\caption{(Color online) Average transverse momentum of identified 
particles in p-Pb collisions at $\sqrt{s_{NN}}$=5.02 TeV, the 
experimental data~(various symbols) from ALICE Collaboration \cite
{ABELEV:2013wsa}, compared to the results of the HIJING model and of 
the viscous hydrodynamics. The figure is taken from Ref.~\cite
{Bozek:2014wpa}.}
\label{fig_pPb_avpt}
\end{figure}

\begin{itemize}
\item p-Pb collisions: Recent experimental measurement shows that 
the number of charged particle produced in p-Pb collisions at 
$\sqrt{s_{NN}}=5.02$ TeV is similar to those in peripheral Pb-Pb 
collisions Ref.~\cite{Chatrchyan:2013nka}. Considering the fact that 
final charged multiplicity is proportional to the initial 
energy/entropy density it is clear that the initial energy density 
in most violent p-Pb collisions is similar as in heavy-ion 
collisions. In fact the collision zone in p-Pb collisions is 
expected to be smaller than the peripheral Pb-Pb collisions, 
consequently the energy density is higher in high multiplicity p-Pb 
events than the peripheral Pb-Pb events. The initial high energy 
density within small volume in p-Pb collisions creates favourable 
condition for the subsequent hydrodynamics evolution. Another strong 
evidence of hydrodynamical flow in p-Pb collisions came from the 
observation of mass dependence of slope of identified hadrons $p_T$ 
spectra, measured in experiment Ref.~\cite{Chatrchyan:2013eya}. The 
experimentally measured $v_n$ and $v_2(p_T)$ data for identified 
hadrons in p-Pb collisions by CMS~\cite{Chatrchyan:2013nka} and 
ALICE~\cite{ABELEV:2013wsa} collaboration is nicely explained by a 
(3+1)D viscous hydrodynamics model study by Bozek {\it et al.} in 
Ref.~\cite{Bozek:2014wpa} (see Fig.~\ref{fig_pPb_v2id}). In addition 
to that the mean transverse momentum of identified hadrons is also 
explained within the same (3+1)D hydrodynamic model, whereas, the 
Monte-Carlo event generator model HIJING which is based on the 
perturbative QCD processes relevant to the collisions fails to 
explain the same experimental data as can be seen in Fig.~\ref
{fig_pPb_avpt}. This already gives the indication that for the high 
multiplicity p-Pb collisions QGP is produced and it flows like fluid 
before freezing out to hadrons.
\end{itemize}

\begin{figure}[t!]
{\includegraphics[width=\linewidth]{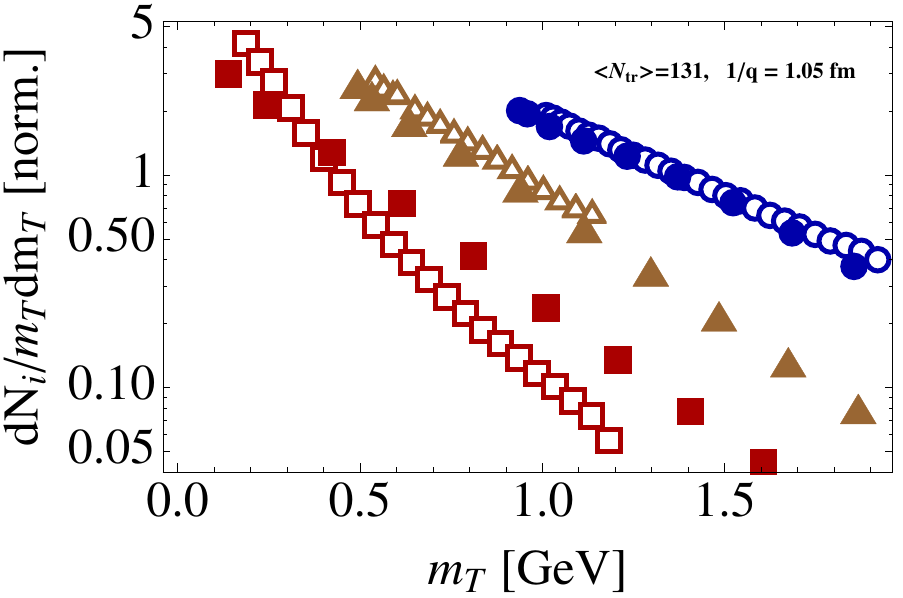}}
\caption{(Color online) Normalized spectra of pions (squares), kaons 
(triangles) and protons (discs) for p-p collisions. Open symbols 
correspond to the CMS data \cite{Chatrchyan:2012qb} for $|\eta|<2.4$ 
and $\sqrt{s}=7$ TeV, while the solid ones are obtained from the 
best one-parameter fit of the Gubser's flow. The figure is taken 
from Ref.~\cite{Kalaydzhyan:2015xba}.}
\label{fig_PPspectra}
\end{figure}

\begin{itemize}
\vspace{-0.2cm}
\item p-p collisions: Qualitatively similar signature of collective 
behaviour is also observed in high multiplicity  p-p collisions like 
high multiplicity p-Pb collisions. However, the initial measurement 
shows that the $p_T$ integrated $v_2$ and $v_3$ is 30\% and 50\% 
smaller than in p-Pb at similar multiplicity. Like heavy-ion and 
p-Pb collisions a simple hydrodynamic inspired model with large 
radial velocity has successfully explained the experimental 
observation of mass dependence of slope in p-p collisions; see 
Fig.~\ref{fig_PPspectra} which is taken from Ref.~\cite
{Kalaydzhyan:2015xba}. In a similar effort a blast wave model fit 
was shown to be inconsistent with the experimental data, see 
Ref.~\cite{Ghosh:2014eqa} for details. There are some studies of 
viscous hydrodynamics for p-p collisions for example see Ref.~\cite
{Bozek:2010pb, Prasad:2009bx, Bzdak:2013zma}, more extensive study 
is needed for the comparison of all experimentally available data. 
We note here that it is still an open question whether the small 
system created in p-p collisions are big enough or live long enough 
for hydrodynamics to be applicable, detailed discussion of which is 
out of the scope of the present review. We refer to see the Ref.~
\cite{Shuryak:2013ke} for a detailed discussion about the 
applicability of hydrodynamics in small systems.
\end{itemize}


\subsection{Flow in ultra central collisions}

\begin{figure}[t!]
{\includegraphics[width=7cm]{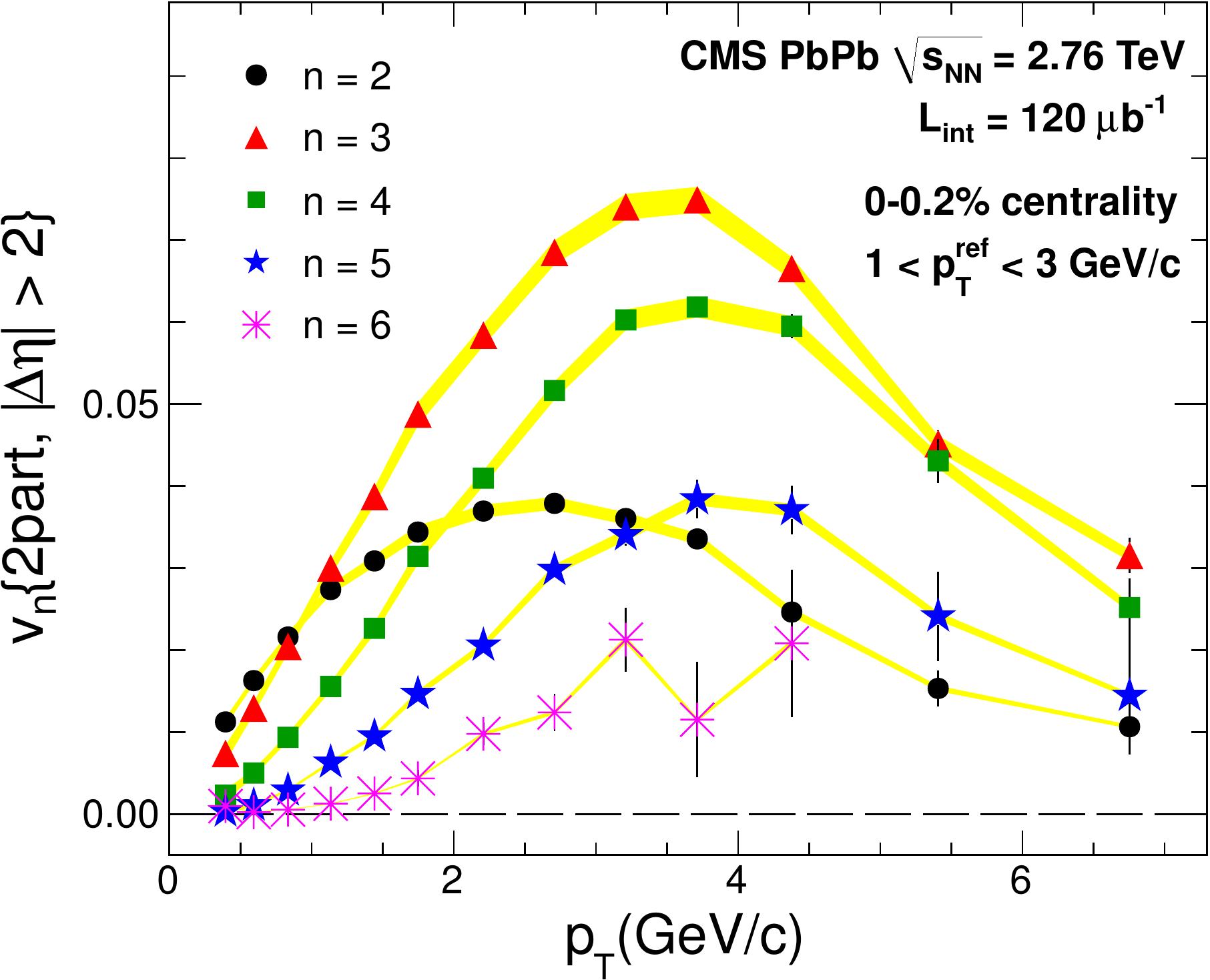}}
{\includegraphics[width=7cm]{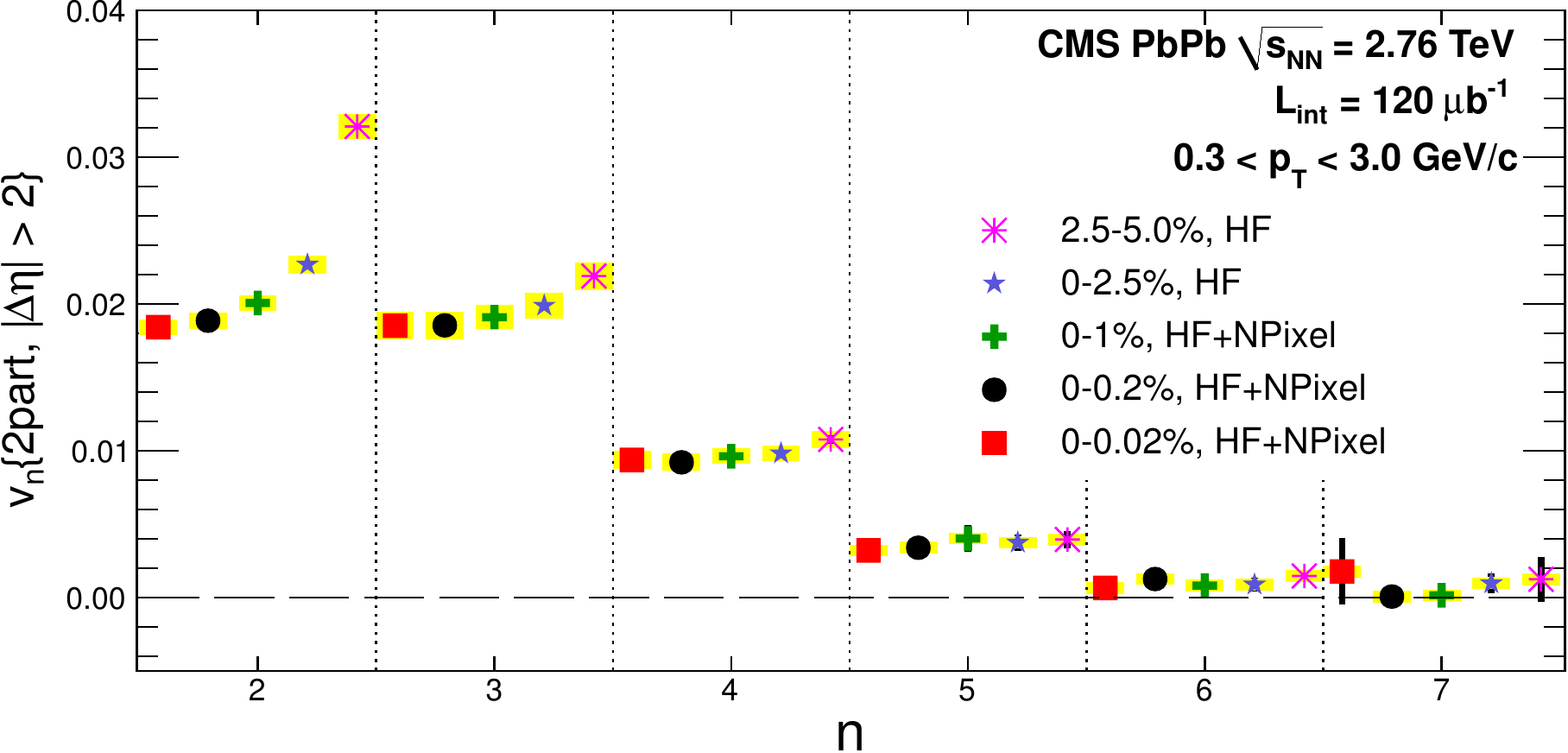}}
\caption{(Color online) Top panel: The $v_n$ (n=2-6) values as a 
function of $p_T$ in 0.0-0.2\% central Pb+Pb collisions at 
$\sqrt{s_{NN}}= 2.76$ TeV. Bottom panel: Experimental measurement of 
$p_T$-averaged (0.3-3.0 GeV) $v_n$  as a function of ``n" in five 
centrality classes (2.5-5.0\%, 0-2.5\%, 0-1\%, 0-0.2\% and 0-0.02\%) 
for Pb+Pb $\sqrt{s_{NN}}=2.76$ TeV collisions. Error bars denote the 
statistical uncertainties, while the shaded color boxes correspond 
to the systematic uncertainties. The data was measured by the CMS 
collaboration and the figures are taken from  Ref.~\cite 
{CMS:2013bza}.}
\label{fig:vn_ultraCentralExpt}
\end{figure}

As mentioned earlier, the hydrodynamic response of the anisotropy in 
the initial overlap geometry in the configuration space transforms 
to the final momentum space anisotropy giving rise to non-zero 
values of flow harmonics $v_n$. The most prominent flow harmonics 
$v_2$ originates as a hydrodynamic expansion of the initial elliptic 
shape of the fireball. The conversion efficiency of the spatial 
deformation into the momentum space anisotropy is very sensitive to 
the shear viscosity over entropy density ($\eta/s$) and the initial 
configuration of the system. The extraction of $\eta/s$ of QGP by 
comparing hydrodynamic simulation results to the corresponding 
experimental data is riddled with large uncertainties in our 
understanding of the initial-state conditions of heavy-ion 
collisions. For example, viscous hydrodynamic simulation with 
MC-Glauber initial condition gives very different values of $\eta/s$ 
compared to the same simulation with different initial condition 
such as MC-KLN. This uncertainty due to the poorly known initial 
condition can be minimised in case of ultra central collisions. In 
ultra-central collisions $v_2$ and other higher flow harmonics 
solely originate from the initial fluctuating energy density since 
the overlap zone in ultra central collisions are almost circular.

\begin{figure}[t!]
{\includegraphics[width=7cm]{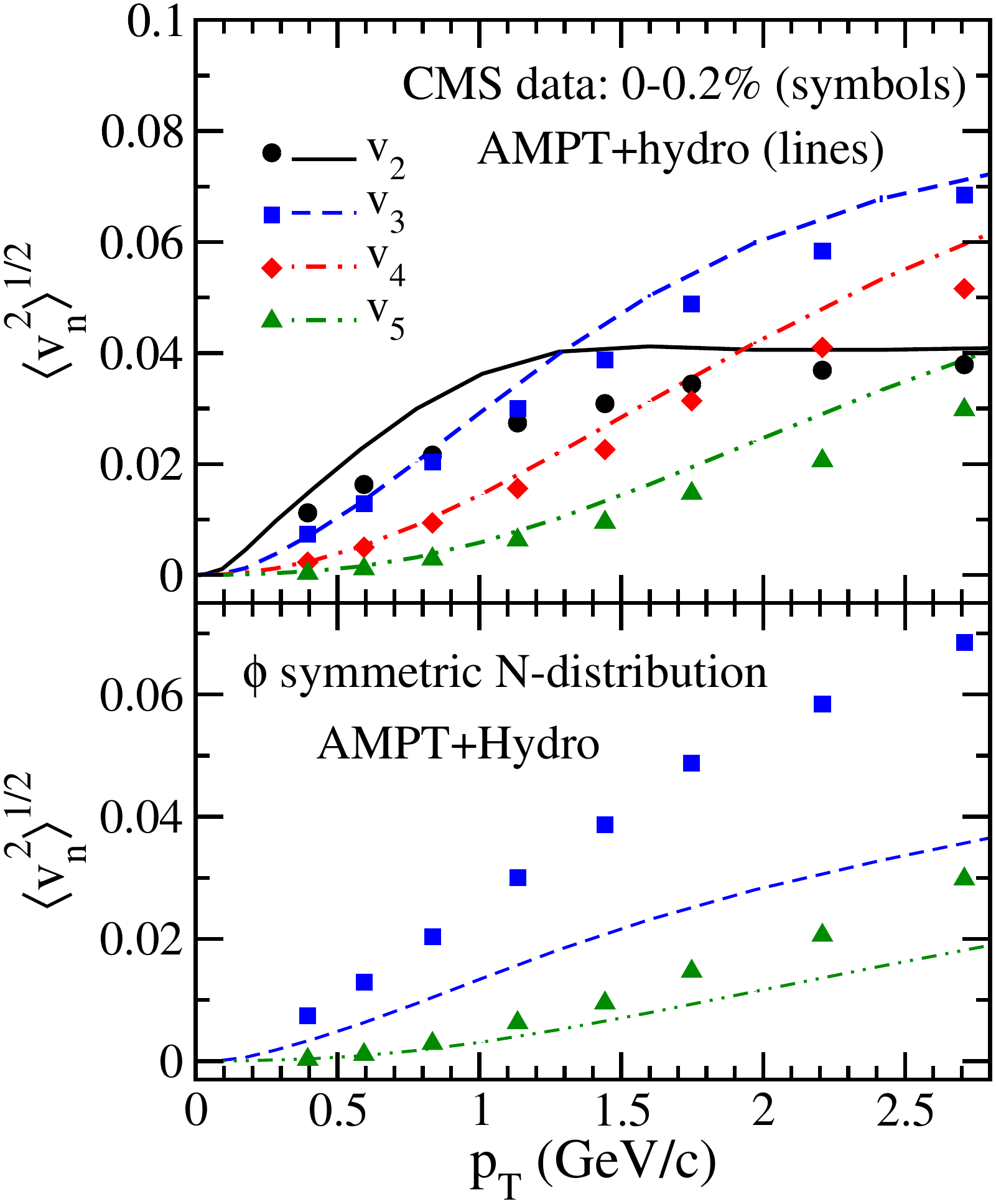}}
\caption{(Color online) Comparison of a (2+1)D viscous hydrodynamics 
simulation with the initial condition from AMPT model to 
corresponding experimental data for root mean squared values of $v_n$
for $n=2-5$; figure is from Ref.~\cite{Bhalerao:2015iya}.}
\label{fig:vn_ultraCentralHydroPT_Pal}
\end{figure}

For ultra central collisions the initial collision geometry is 
predominantly generated by fluctuations such that various orders of 
eccentricities predicted by different models tend to converge. 
Therefore, studies of $v_n$ in ultra-central heavy-ion collisions 
can help to reduce the systematic uncertainties of initial-state 
modelling in extracting the $\eta/s$ value of the system. Let us 
first discuss the recent experimental results for ultra central 
Pb-Pb collisions at LHC, after that we shall also discuss the 
corresponding results from viscous hydrodynamics simulations. Top 
panel of Fig.~\ref{fig:vn_ultraCentralExpt} shows the experimentally 
measured differential flow coefficients $v_n$ as a function of $p_T$ 
for 0-0.2\% centrality Pb-Pb collisions at $\sqrt{s_{NN}}=2.76$ TeV. 
The $v_n$'s were calculated using 2 particle correlation method with 
large pseudo-rapidity gap $|\Delta\eta|>2$ between the two hadrons. 
The bottom panel of Fig.~\ref{fig:vn_ultraCentralExpt} shows the 
$p_T$ integrated $v_n$ (n=2-7) in ultra central Pb-Pb collisions for 
five different collisions centrality. The experimental data and the 
figure are taken from Ref.~\cite{CMS:2013bza}. Before we proceed any 
further we note the following experimental observation from the CMS 
paper.
\begin{itemize}
\item At higher transverse momentum ($p_T \ge\ 2$ GeV), $v_2$ 
becomes even smaller than the higher-order $v_3, v_4$, and at much 
higher values of $p_T$ it becomes smaller than other higher order 
$v_n$.
\item The $p_T$ averaged $v_2$ and $v_3$ are found to be equal 
within 2\%, while other higher-order $v_n$ decrease as n increases.
\end{itemize}

\begin{figure}[t!]
{\includegraphics[width=\linewidth]{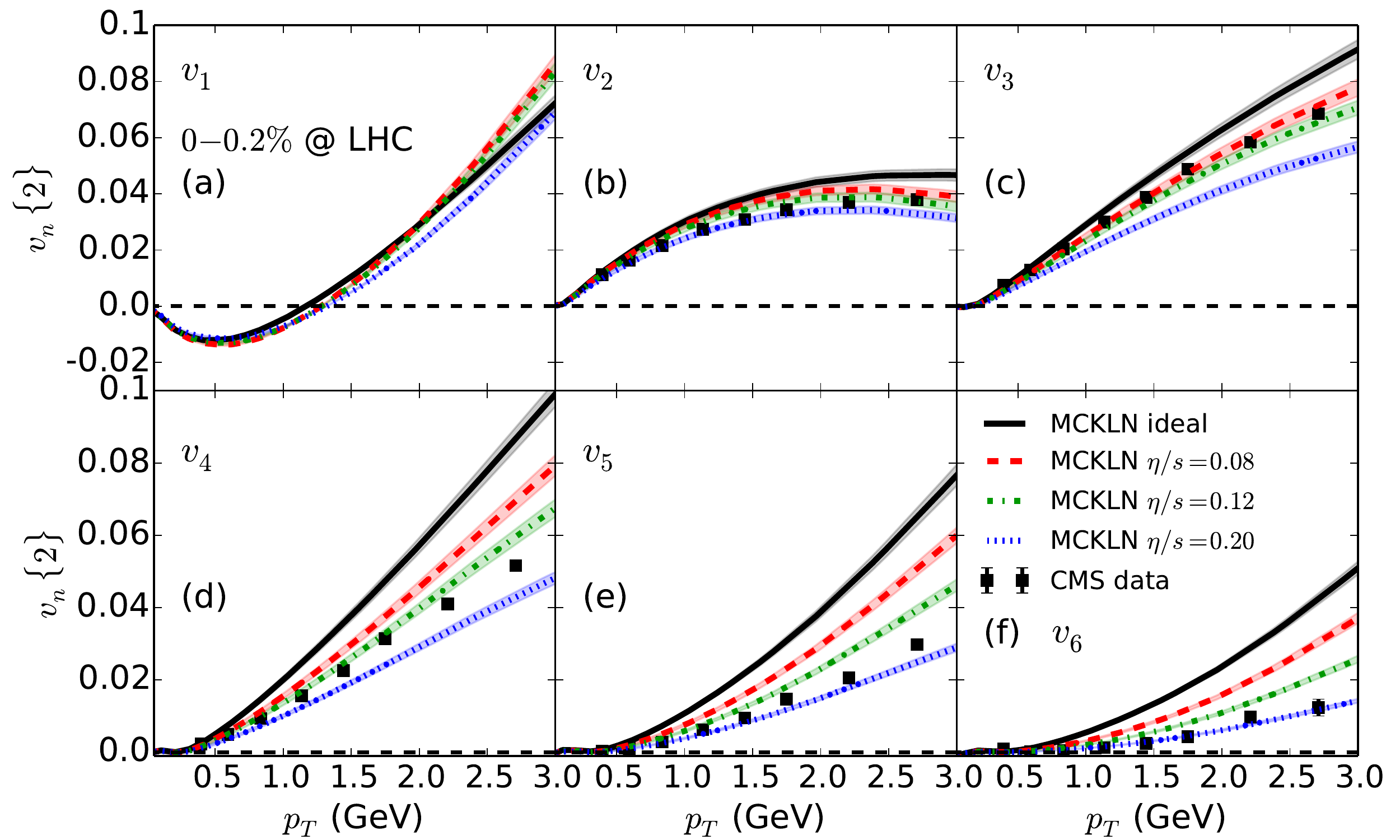}}
\caption{(Color online) Comparison of $p_T$ dependent 
$v_{n}\left\{2\right\}$ of charged hadrons in $2.76$ A TeV Pb+Pb 
collisions at 0-0.2\% centrality for viscous hydrodynamics 
simulations (various lines) with the corresponding experimental 
results (solid squares); the figure is from Ref.~\cite 
{Shen:2015qta}.}
\label{fig:vn_ultraCentralHydroPT_Shen}
\end{figure}

\begin{figure}[t!]
{\includegraphics[width=\linewidth]{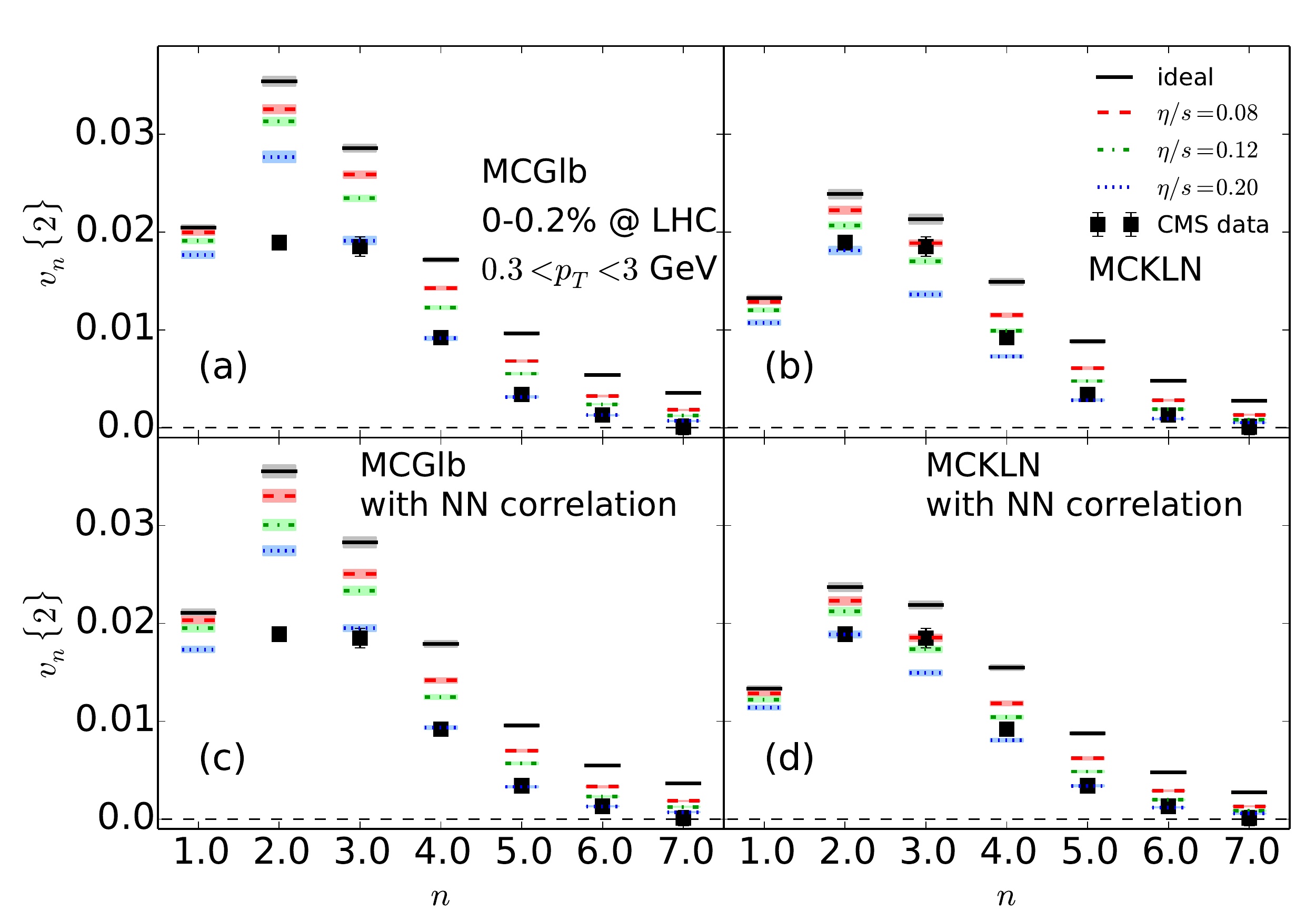}}
\caption{(Color online) Comparison of $p_T$ integrated $v_{n}\left\{ 
2\right\}$ of charged hadrons in $2.76$ A TeV Pb+Pb collisions at 
0-0.2\% centrality for viscous hydrodynamics simulations (various 
lines) with the corresponding experimental results (solid squares). 
Plot (a) and (c) are are for MC-Glauber initial conditions, and 
(b),(d) corresponds to MC-KLN initial conditions. Results in the top 
two panels (a) and (b) was obtained by considering nucleons with a 
repulsive hard core, whereas the results in the bottom panel (c) and 
(d) are obtained for the initial conditions with finite 
nucleon-nucleon correlations. The figure is taken from Ref.~\cite 
{Shen:2015qta}.}
\label{fig:vn_ultraCentralHydro}
\end{figure}

The evolution of the QGP according to relativistic hydrodynamics 
simulations have been able to consistently explain experimentally 
measured $v_n$'s for different centralities and for different 
colliding energies, it is natural to expect that it should also 
explain the measured $v_n$ in the ultra central collisions. Before 
we discuss the results of hydrodynamic simulations, we note that one 
need to carefully select events into centrality classes since the 
integrated $v_n$'s are quite sensitive on the selection of 
centrality class as can be seen from the bottom panel of Fig.~\ref 
{fig:vn_ultraCentralExpt}. We also note that it is computationally 
expensive to simulate such ultra-central collisions since the number 
of events within the given centrality class is significantly small 
compared to the total number of minimum bias events. Although the 
essential $p_T$ dependence of charged hadrons $v_n$ and their 
observed ordering for ultra-central Pb-Pb collisions was nicely 
explained by a viscous hydrodynamic simulation using initial 
conditions from AMPT model \cite{Bhalerao:2015iya}; see Fig.~\ref 
{fig:vn_ultraCentralHydroPT_Pal}. However on careful observation we 
notice that at low $p_T<1.5$ GeV, the splitting from hydrodynamics 
simulation is larger than the corresponding experimental 
measurement. Similar disagreements are also evident for $p_T>1.5$ 
GeV in Fig.~\ref {fig:vn_ultraCentralHydroPT_Shen}, which is taken 
from Ref.~\cite{Shen:2015qta}. This can be seen more clearly from 
the $p_T$ integrated $v_2$ and $v_3$ in Fig.~\ref
{fig:vn_ultraCentralHydro} which is also taken from Ref.~\cite 
{Shen:2015qta}. In Ref.~\cite{Shen:2015qta}, the $p_T$ integrated 
$v_n$ was studied using (2+1)D viscous hydrodynamics model with 
MC-Glauber and MC-KLN initial conditions.

The nucleon-nucleon correlations in the colliding nucleus were also 
considered as a potential cause behind the experimentally measured 
$v_2\sim v_3$. However, none of the initial condition model has so 
far been able to simultaneously explain the experimentally measured 
$v_n$'s, as can be seen in Figs.~\ref{fig:vn_ultraCentralHydroPT_Pal}, 
\ref{fig:vn_ultraCentralHydroPT_Shen} and \ref
{fig:vn_ultraCentralHydro}. In this regard we note that Denicol {\it 
et al.}, Ref.~\cite{Denicol:talk_ECT}, have considered bulk 
viscosity along with the shear viscosity and the nucleon-nucleon 
correlations in order to explain this apparent discrepancy between 
the experimental data and corresponding theoretical results, 
although there was some improvement but so far the effort remains 
unfruitful.


\subsection{Longitudinal fluctuations and correlations}

In relativistic heavy-ion collision experiments, a fraction of the 
incoming kinetic energy is converted into new matter deposited in 
the collision zone. The distribution of this matter in the plane 
transverse to the colliding beams is inhomogeneous and fluctuates 
from collision to collision. The lumpy initial energy density 
distribution and its event-by-event fluctuations lead to anisotropic 
flows of final hadrons through collective expansion in high energy 
heavy-ion collisions. The first numerical demonstration of the role 
of lumpy initial energy density (or event-by-event fluctuation) in 
the transverse plane (plane defined by the impact parameter vector 
and one of the perpendicular axis to the beam direction) to the 
experimentally observed non-zero odd flow harmonics (particularly 
third harmonics $v_3$) in heavy-ion collision was made by Alver and 
Roland \cite{Alver:2010gr}. From then on experimentally measured 
flow harmonics for all order (even and odd) has been successfully 
explained by viscous hydrodynamics model studies with fluctuating 
initial conditions such as Monte-Carlo (from now on we denote it by 
MC) Glauber\cite{Broniowski:2007ft, Hirano:2009ah}, MC-CGC \cite 
{Roy:2012pn}, URQMD \cite{Petersen:2008dd}, EPOS \cite 
{Werner:2010aa}, AMPT \cite{Pang:2012he}, and IP-Glasma \cite 
{Schenke:2012wb}. Fluctuations in the transverse plane not only give 
rise to odd flow harmonics but also significant even and odd $v_n$ 
in ultra central collisions \cite{Ma:2010dv}. They also result in 
$p_T$ dependent event planes, which break down the flow 
factorization $v_{n,n}(p_{T1},p_{T2})=v_n(p_{T1})v_n(p_{T2})$ \cite 
{Aamodt:2011by}. Like the lumpy initial energy density in the 
transverse plane, it is also expected (the reason for which will be 
discussed shortly) that the energy density is lumpy in the 
longitudinal (space rapidity) direction.

\begin{figure}[t!]
{\includegraphics[width=6cm]{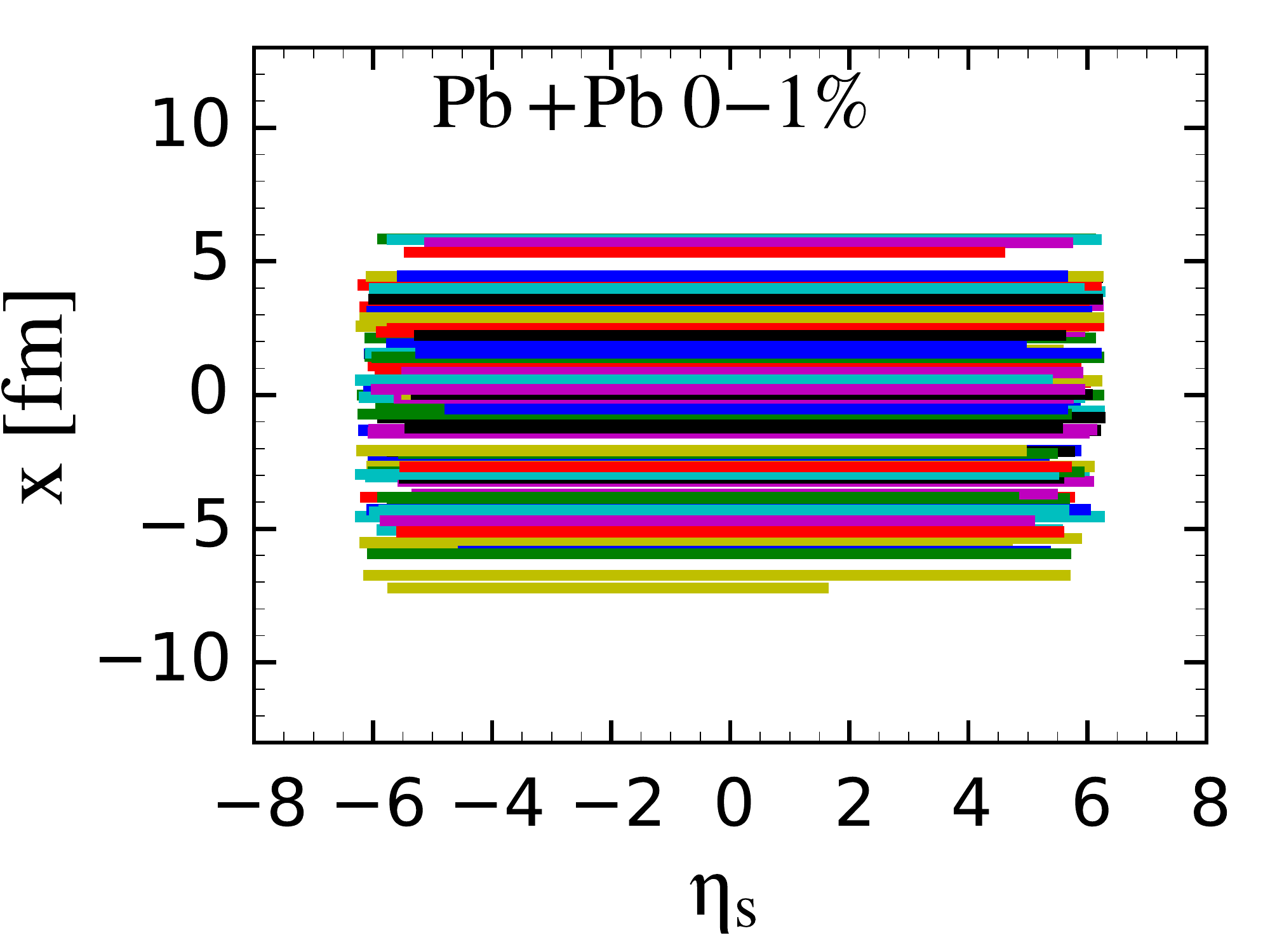}}
{\includegraphics[width=6cm]{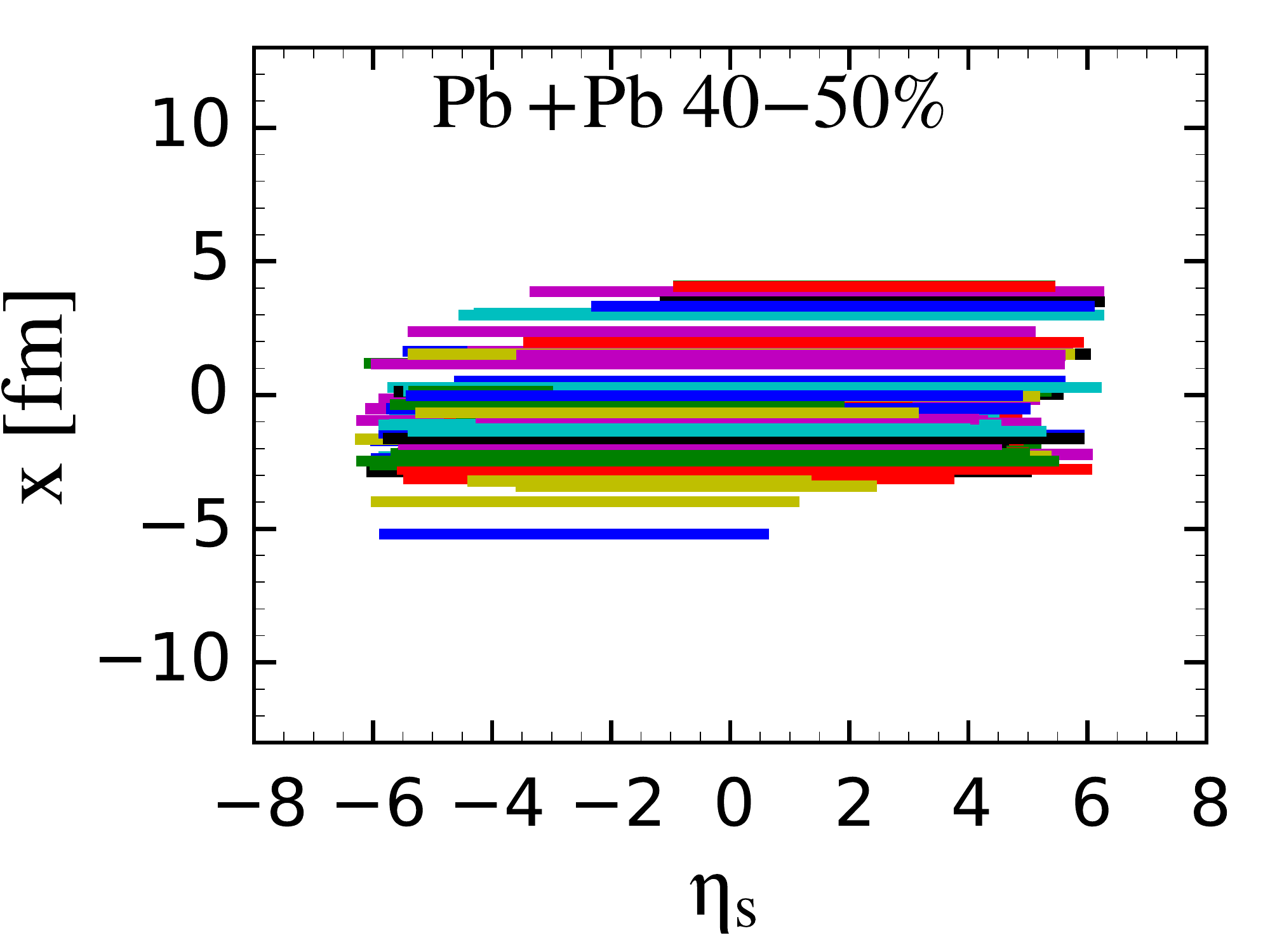}}
\caption{(Color online) Distribution of strings created between the 
partons of two colliding Pb nucleus as a function of space-time 
rapidity ($\eta_s$) at $\sqrt{s_{NN}}=2.76$ TeV for $0-1$\% (top 
panel) and 40-50\% (bottom panel) collision centrality. The figure 
is taken from Ref.~\cite{Pang:2015zrq}.}
\label{fig_long_string}
\end{figure}

Recent measurement of decorrelation of anisotropic flow along 
longitudinal direction by CMS collaboration has corroborated the 
above expectation. Studies of fluctuations along the longitudinal 
direction and their effects on anisotropic flows of final charged 
hadrons have only recently been started. At present the current 
understanding of longitudinal correlation (or decorrelation) of flow 
harmonics is as follows
\begin{itemize}
\item The fluctuations of energy density along the longitudinal 
direction due to the fragmentation and different lengths of the 
coloured string produced in the scattering of nucleons \cite
{Pang:2015zrq, Pang:2014pxa, Broniowski:2015oif}.
\item A gradual twist of the fireball (or more specifically the 
event plane) along the longitudinal direction Ref.~\cite
{Bozek:2010vz, Jia:2014ysa}.
\end{itemize}
Let us discuss each of them separately. Regarding the contribution 
of colour string we shall particularly discuss here a recent study 
Ref.~\cite{Pang:2015zrq} where AMPT transport model is used to 
evaluate the initial conditions for (3+1)D hydrodynamic model. 

\begin{figure}[t!]
{\includegraphics[width=\linewidth]{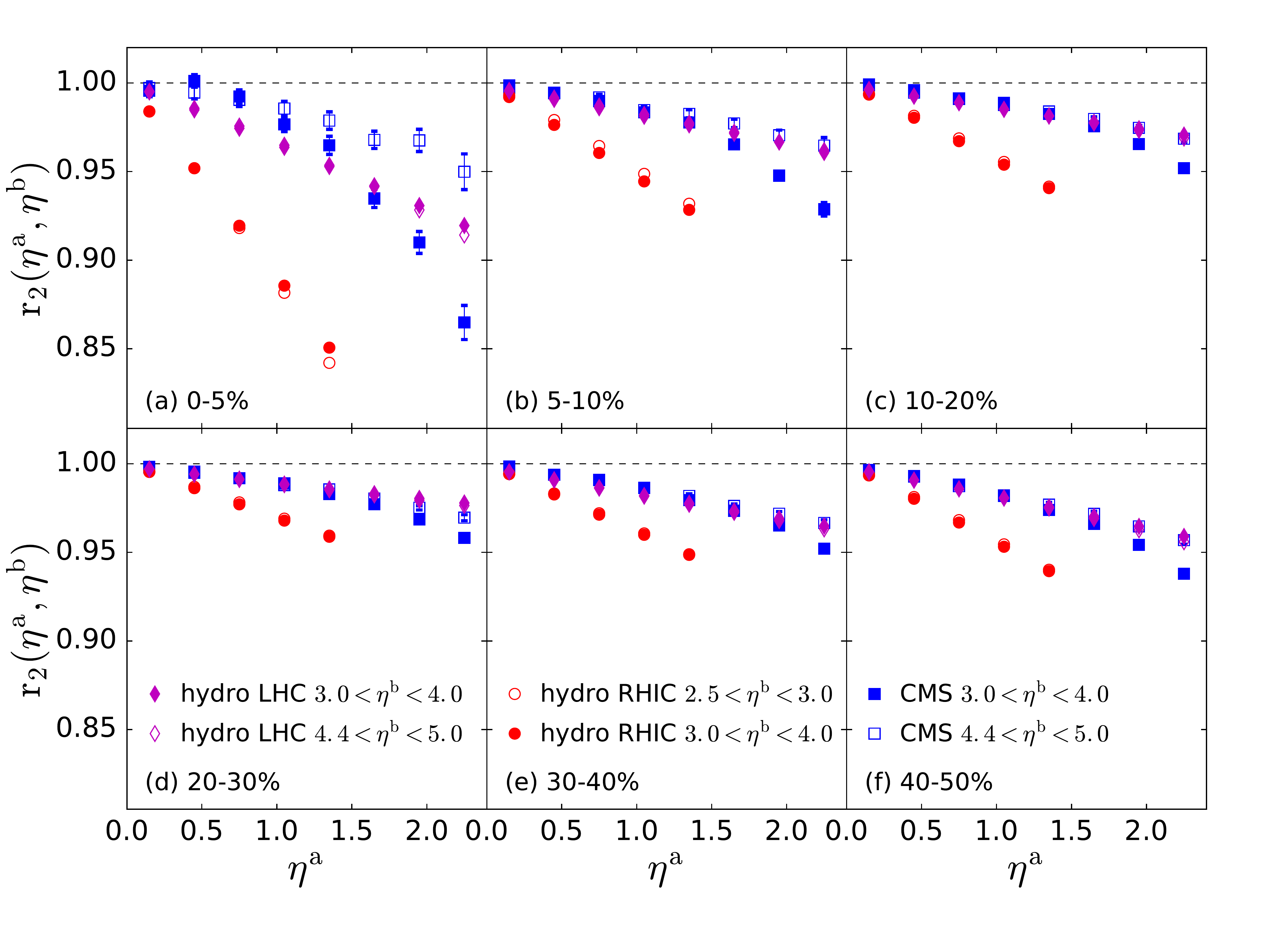}}
{\includegraphics[width=\linewidth]{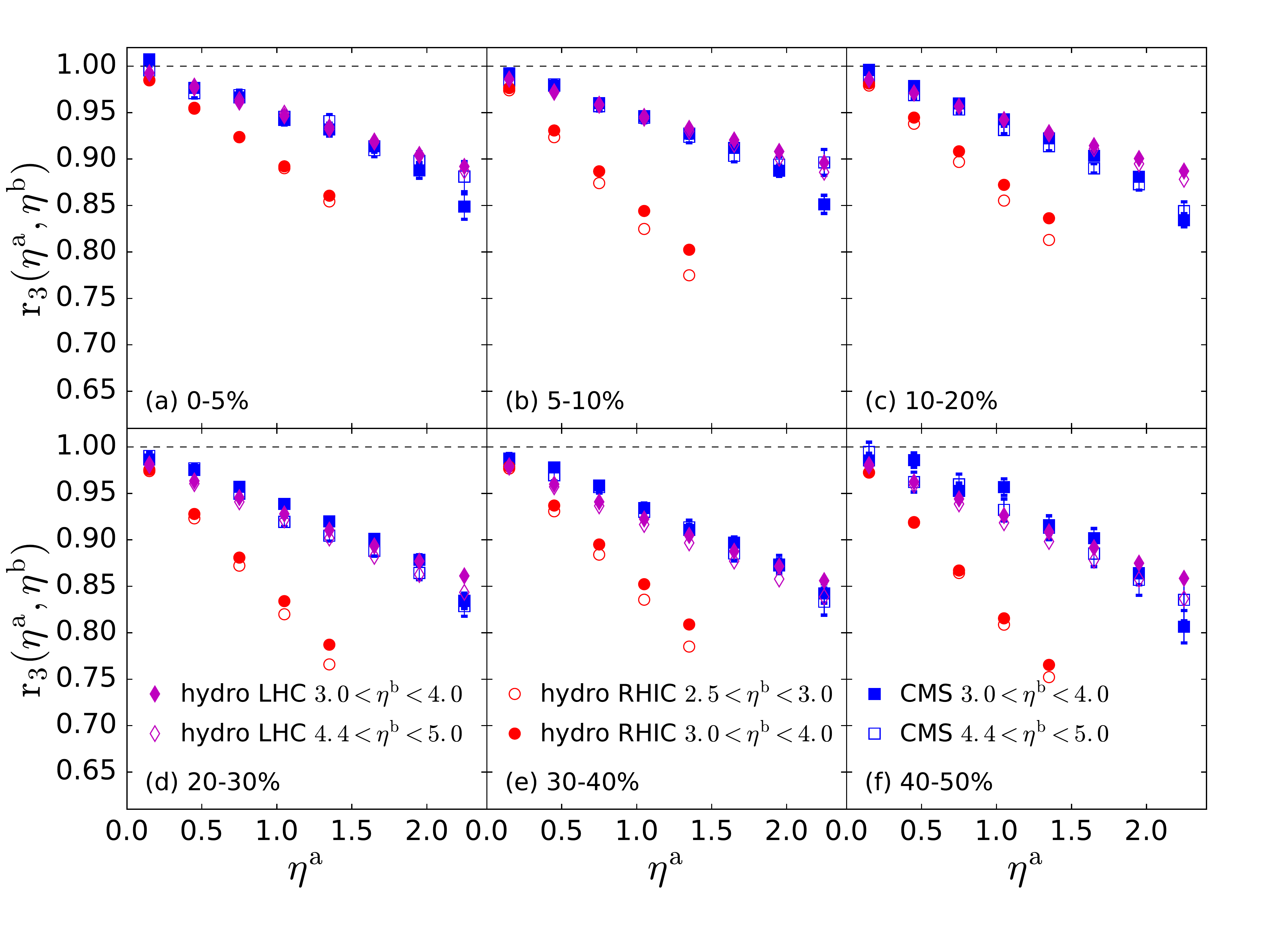}}
\caption{(Color online) Top panel: The factorization ratio $r_2$ as 
a function of space-time rapidity $\eta^a$ for two different 
reference rapidity bin $3.0<\eta^{b}< 4.0$ and $4.4<\eta^{b}< 5.0$ 
in Pb-Pb collisions at $\sqrt{s_{NN}}=2.76$ TeV (open and solid 
diamonds), and for $2.5<\eta^{b}< 3.0$ and $3.0<\eta^{b}<4.0$ in 
Au-Au collisions at $\sqrt{s_{NN}}=200$ GeV (open and solid circles) 
from event-by-event (3+1)D ideal hydrodynamics simulations compared 
with experimental data from CMS collaboration Ref.~\cite 
{Khachatryan:2015oea} for Pb-Pb collisions at $\sqrt{s_{NN}}=2.76$ 
TeV (empty and solid squares). The figures are from Ref.~\cite 
{Pang:2015zrq}.}
\label{fig_long_corr}
\end{figure}

AMPT uses HIJING to generate initial partons from hard and semi-hard 
scatterings and excited strings from soft interactions. The number 
of mini-jet partons per binary nucleon-nucleon collision in hard and 
semi-hard scatterings follow a Poisson distribution with the mean 
value given by the jet cross-section. The number of excited strings 
is equal to the number of participant nucleons in each event. 
Besides random fluctuations from mini-jet partons, the parton 
density fluctuates along longitudinal direction according to the 
length of strings. There are basically three types of strings:
\begin{enumerate}
\item Strings associated with each wounded nucleon (between a 
valence quark and a diquark),
\item Single strings between q-q pairs from quark annihilation and 
gluon fusion processes,
\item Strings between one hard parton from parton scatterings and 
valence quark or diquark in wounded nucleons.
\end{enumerate}
These strings finally fragment into the partons along the 
longitudinal direction and giving rise to fluctuating energy density 
distribution in $\eta_s$, see Fig.~\ref{fig_long_string} for the 
distribution of coloured strings in the longitudinal direction for a 
typical Pb-Pb collision at $\sqrt{s_{NN}}=2.76$ TeV. For more 
details about the longitudinal fluctuations and the visualization of 
parton density distribution in $\eta_s$ see Ref.~\cite{Pang:2015zrq}.

The idea of a gradual twist of the fireball (or torqued fireball) 
along the longitudinal direction is due to Piotr Bozek {\it et al.} 
\cite{Bozek:2010vz}. According to Bozek {\it et al.} the following 
ingredients are responsible for the appearance of the torque effect:
\begin{enumerate}
\item Statistical fluctuations of the transverse density of the 
sources (wounded nucleons), and
\item The asymmetric shape of the particle emission function, peaked 
in the forward (backward) rapidity for the forward (backward) moving 
wounded nucleons.
\end{enumerate}
We note that in this model, the initial energy density profile is 
parametrized in such a way that after the hydrodynamics evolution 
and the freeze-out the hadronic spectra produced at different 
rapidities match with the corresponding experimental data. Whereas 
in the case of AMPT initial condition we do not need to use such 
procedure in order to explain the corresponding experimental data, 
for example in Fig.~\ref{fig_long_corr} we show the comparison of 
experimental data of longitudinal correlation for Pb-Pb collisions 
from CMS collaboration \cite{Khachatryan:2015oea} and a (3+1)D 
hydrodynamics simulation result with AMPT initial condition. Note 
that with AMPT initial condition the experimental data is quite well 
described by the (3+1)D ideal hydrodynamics simulation. In the AMPT 
initial condition both longitudinal fluctuations and torque effects 
are present, the interplay of twist and fluctuation and the relative 
contribution of this two effects in heavy-ion collisions was studied 
within (3+1)D hydrodynamics model and AMPT in Ref.~\cite
{Pang:2014pxa}.

Many techniques have been proposed to study the longitudinal 
structure of final hadron production in heavy-ion collisions and the 
underlying mechanisms. For example, three-particle correlations were 
suggested to measure the twist effect \cite{Borghini:2002hm} in 
heavy-ion collisions at RHIC. One can also characterise the 
longitudinal fluctuations in terms of coefficients in the Chebyshev 
polynomials \cite{Bzdak:2012tp} and the Legendre polynomial 
expansion of two-particle correlations in pseudorapidity \cite 
{Monnai:2015sca, Bozek:2015tca}. The most intuitive method is to 
measure the forward-backward (FB) event plane angles or anisotropic 
flow differences \cite{Pang:2014pxa} with varying pseudorapidity 
gaps. These methods are used within the torqued fireball model \cite 
{Bozek:2010vz}, (3+1)D hydrodynamics model and the AMPT model \cite 
{Pang:2014pxa, Xiao:2012uw} to study the decorrelation of event 
plane angles or anisotropic flow along the pseudorapidity direction. 
Jia {\it et al.} \cite{Jia:2014ysa} also proposed an ``event-shape 
twist" technique to study the event plane decorrelation due to the 
twist in initial energy density distributions by selecting events 
with big FB event plane angle differences. Alternatively by 
selecting events with vanishingly small FB event plane angle 
differences, one can then eliminate the twist effect and the 
measured decorrelation of anisotropic flow with finite 
pseudorapidity gaps should be caused only by random fluctuations of 
event plane angles as was done in Ref.~\cite{Pang:2014pxa}. Before 
ending this section we note that the experimentally observed 
difference in the longitudinal correlation ($r_2$ and $r_3$) for 
different reference rapidity bin \cite{Khachatryan:2015oea} is not 
yet understood within theoretical model studies \cite{Pang:2015zrq}. 
We need further studies in order to understand those finer details.


\subsection{Flow in intense magnetic field}

The most strongest known magnetic field ($|\vec{B}|\sim10^{18} 
-10^{19}$ Gauss) in the universe is produced in laboratory 
experiments of Au-Au or Pb-Pb collisions such as at RHIC and at LHC. 
Previous theoretical studies show that the intensity of the produced 
magnetic field rises approximately linearly with the centre of mass 
energy ($\sqrt{s_{{\rm NN}}}$) of the colliding nucleons \cite 
{Bzdak:2011yy, Deng:2012pc}. The corresponding electric fields in 
such collisions also becomes very strong which is same order of 
magnitude as the magnetic field ($e\vec{B} \approx e\vec{E} \sim 10 
m_{\pi}^2$ for a typical Au-Au collision at top RHIC energy 
$\sqrt{s_{\rm NN}}=200$ GeV) \cite{Bloczynski:2012en}, where $m_\pi$ 
is the pion mass. Such intense electric and magnetic fields are 
strong enough to initiate the particle production from vacuum via 
Schwinger mechanism \cite{KEKIntense:2010}.

\begin{figure}[t!]
{\includegraphics[width=6cm]{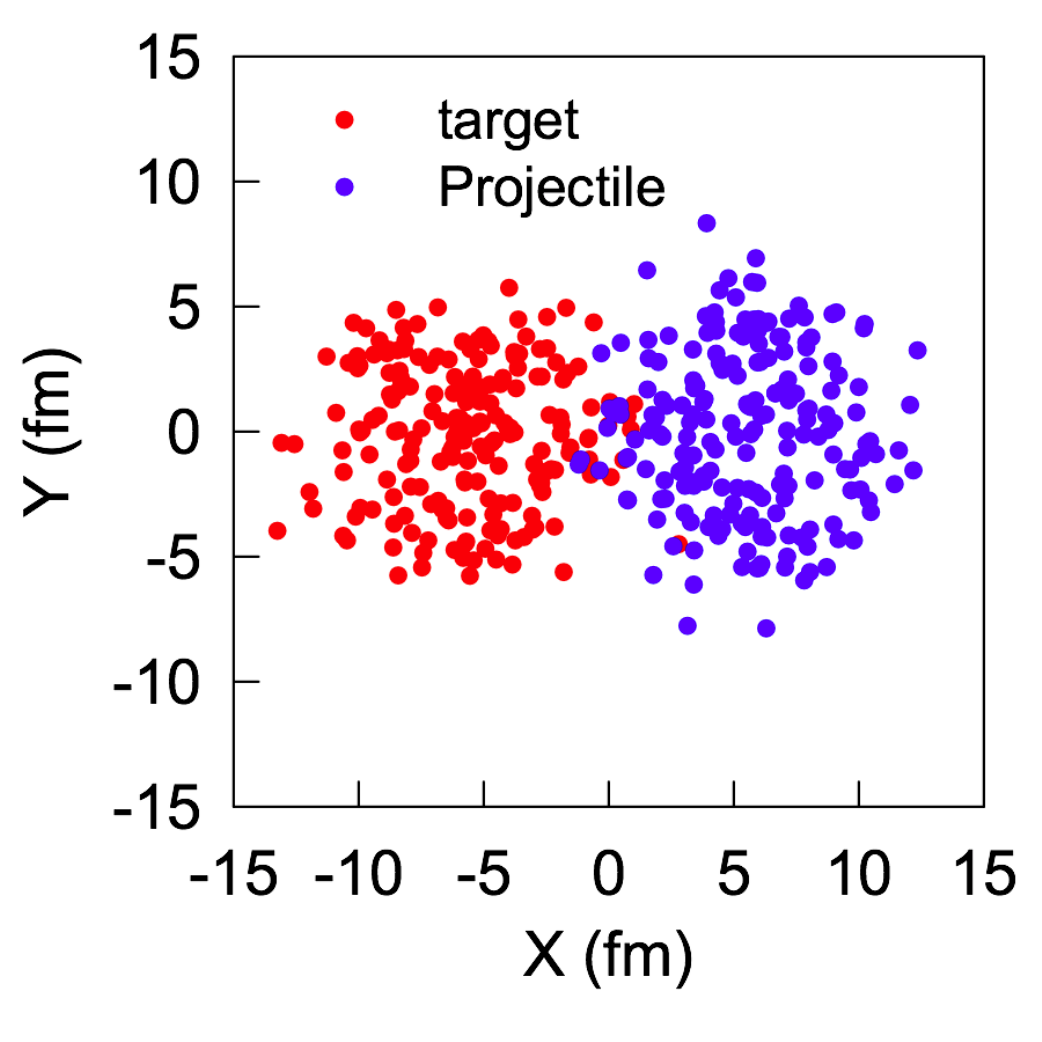}}
\caption{(Color online) Distribution of nucleons inside target and 
projectile nuclei in a typical Au-Au collision at $\sqrt{s_{\rm 
NN}}=200$ GeV for $b=12$ fm. Figure is from Ref.~\cite{Roy:2015coa}.}
\label{fig_EMField_dist}
\end{figure}

\begin{figure}[t!]
{\includegraphics[width=7cm]{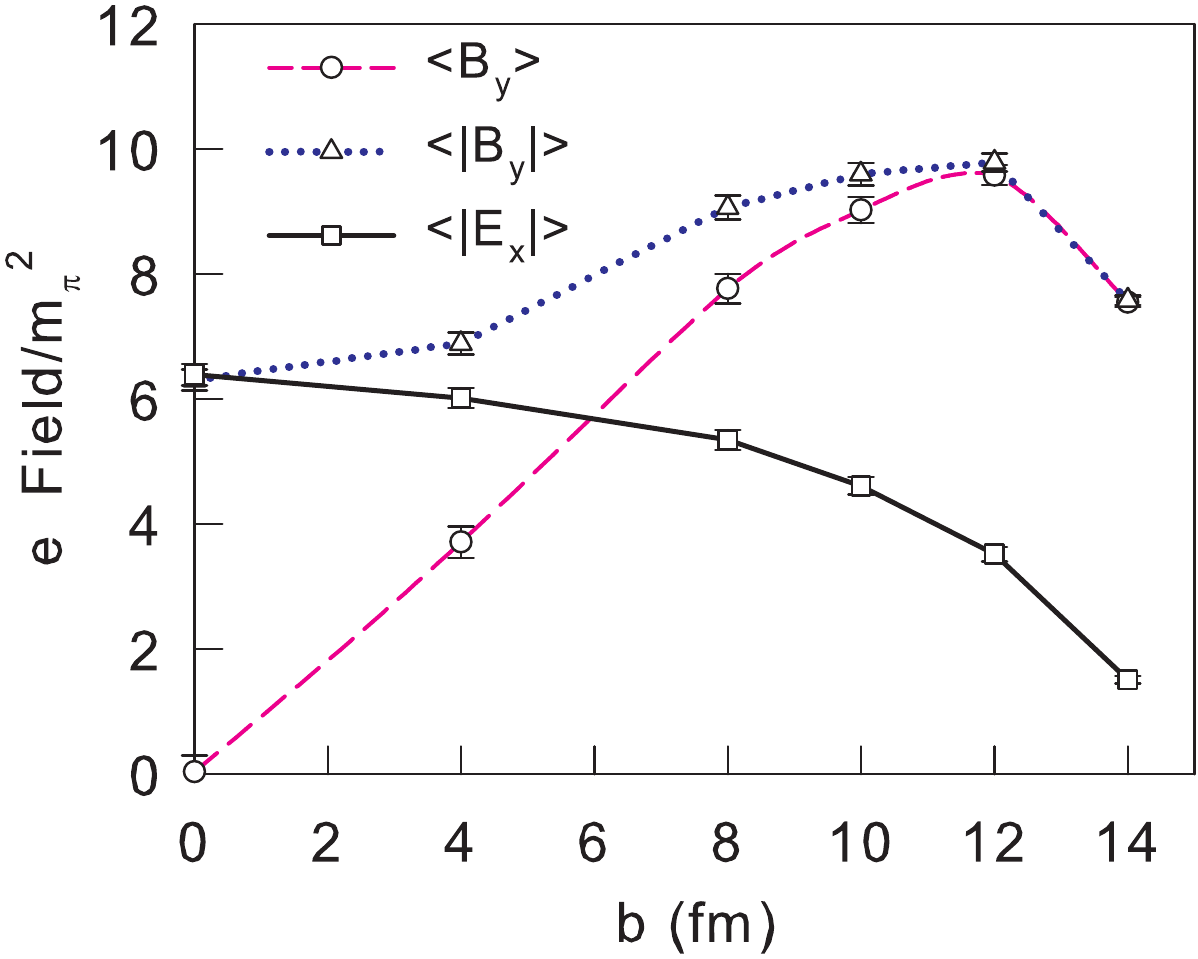}}
\caption{(Color online) Impact parameter dependence of event 
averaged magnetic and electric fields at the centre of the fireball 
for Au-Au collisions at $\sqrt{s_{\rm NN}}=200$ GeV. Figure is from 
Ref.~\cite{Roy:2015coa}.}
\label{fig_EMField_IP}
\end{figure}

The origin of such large electric and magnetic field is the 
relativistic velocities of the positive charge nucleus. Within a MC 
Glauber model the electric ($\vec{E}$) and magnetic ($\vec{B}$) 
field at position $\vec{r}$ and at time $t$ for a nucleus of charge 
$Ze$ moving with velocity $\vec{v}$ in straight line is given by 
\begin{align}
\vec{E} \left( \vec{r},t \right) &= \frac{e}{4\pi} \sum _{i=1}^{{N}_{proton}}{{Z}_{i}
\frac{\vec{{R}_{i}} - {R}_{i}\vec{{v}_{i}}}{{\left( {R}_{i}-\vec{{R}_{i}} 
\cdot\vec{{v}_{i}} \right)}^{3}}\left( 1-{v}_{i}^{2} \right)} , \label{eq:Efield_LW}\\
\vec {B} \left( \vec{r},t \right) &= \frac{e}{4\pi } \sum _{i=1}^{{N}_{proton}}{{Z}_{i}
\frac{\vec{{v}_{i}}\times\vec{{R}_{i}}}{{\left( {R}_{i}-\vec{{R}_{i}} 
\cdot\vec{{v}_{i}} \right)}^{3}}\left( 1-{v}_{i}^{2} \right)} .\label{eq:Bfield_LW}
\end{align}
Here $\vec{R_{i}}=\vec{x}-\vec{x_{i}}(t)$ is the distance from a 
proton at position $\vec{x_{i}}$ to $\vec{x}$ where the field is 
evaluated. In the above expression the summation index $i$ denotes 
the contribution of all protons inside the colliding nucleus, for 
example Fig.~\ref{fig_EMField_dist} shows the positions of nucleons 
inside the two Au nucleus for a typical peripheral collision 
calculated in MC-Glauber model. Due to the fluctuating proton 
position from event to event the electric and magnetic field becomes 
irregular both in direction and in magnitude in the transverse 
plane. Moreover, the magnetic field in the central collisions 
becomes non-zero for such initial random proton positions. This can 
be seen from Fig.~\ref {fig_EMField_IP} where the event averaged 
value of $B_y$ as a function of impact parameter $b$ is shown. The 
black dashed-dotted line corresponds to the average of the absolute 
magnitude of $B_y$ which is clearly non-zero even for $b=0$ fm 
collisions. Note that we have used the natural unit where 
$\hbar=c=k_b=\varepsilon_0=\mu_0=1$, with this choice the electric 
charge $e=\sqrt{\frac{4\pi}{137}}$ becomes a dimensionless number. 
In the limit $v \sim c $, the denominator in Eq.~(\ref 
{eq:Efield_LW}) and (\ref{eq:Bfield_LW}) becomes very small and we 
have large $\vec{E}$ and $\vec{B}$.

There are large number of theoretical predictions based on the 
expectation of the creation of large magnetic field in heavy-ion 
collisions such as chiral magnetic effect, chiral electric effect, 
chiral magnetic waves etc., the discussion of which is beyond the 
scope of this review. For more details, we refer the reader to the 
following references \cite{Kharzeev:2013ffa, Bzdak:2012ia, 
Kharzeev:2015kna, Tuchin:2014hza, Huang:2015oca}. Here we will 
concentrate on the possible effect of this large electro-magnetic 
field on the initial energy density and the subsequent hydrodynamics 
evolution of QGP produced in RHIC or LHC experiments. To the best of 
our knowledge, one of the first numerical study of the effect of 
magnetic fields on the hydrodynamics evolution in heavy-ion 
collisions are by Gursoy {\it et al.} \cite{Gursoy:2014aka} and by 
Hirono {\it et al.} \cite {Hirono:2014oda}. However, those studies 
were based on several assumptions and none of them have considered 
the full magneto-hydrodynamics solution for the QGP evolution. We 
note here that the electric and magnetic field might affect the 
initial energy density, subsequent hydrodynamics stage, and the 
freeze-out distribution functions provided that the field is strong 
enough and lives until freeze-out.

One of us has recently studied the importance of electromagnetic 
field energy density compared to the energy density of the QGP fluid 
for Au-Au collisions at $\sqrt{s_{NN}}=200$ GeV in Ref.~\cite 
{Roy:2015coa}. The ratio ($\sigma$) of the magnetic field energy 
density to the fluid energy density was found to be $\sim 1$ for 
peripheral collisions, but in central collisions $\sigma<<1$. It was 
also found that electric field also contributes similar energy 
density as magnetic field. Recent study by Tuchin \cite 
{Tuchin:2013apa} shows that the decay of the initial magnetic field 
can be substantially delayed in the case of finite electrical 
conductivity of QGP. Thus it becomes increasingly important to 
consider the electromagnetic field in the hydrodynamic evolution of 
heavy-ion collisions \cite{Tuchin:2013ie}. In Refs.~\cite 
{Roy:2015kma, Pu:2016ayh, Pu:2016bxy} analytic solution of 
relativistic hydrodynamics for simplified cases was obtained. 
Finding analytic solution for general initial conditions is very 
difficult and there are very few analytical solution that exists for 
relativistic magnetohydrodynamics. The only possible way is to use 
numerical methods to solve magnetohydrodynamic equations relevant 
for heavy-ion collisions. This is not an easy task to accomplish. 
Initial effort in this direction can be found in Ref.~\cite 
{Pang:2016yuh}. However, we note that the authors of Ref.~\cite 
{Pang:2016yuh} have solved usual hydrodynamics conservation 
equations (without magnetic field) by considering an external force 
originating due to the paramagnetic interaction of QGP with the 
magnetic field.
 
\begin{figure}[t!]
{\includegraphics[width=6cm]{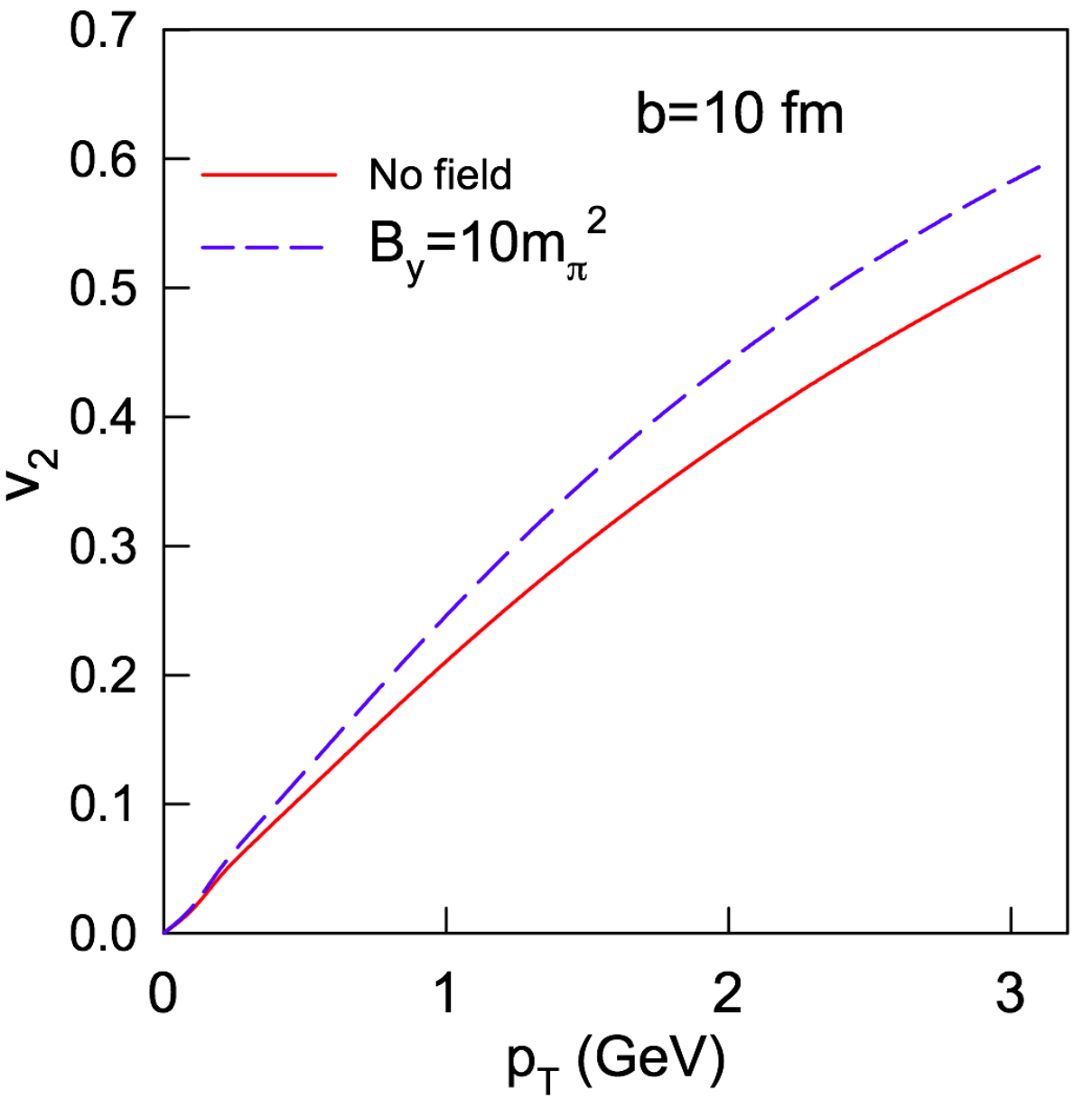}}
\caption{(Color online) Elliptic flow of $\pi^{-}$ as a function of 
$p_T$ for b=10 fm Au-Au collisions at $\sqrt{s_{\rm NN}}=200$ GeV. 
Red line corresponds to (2+1)D ideal hydrodynamics result without 
magnetic field, the blue dashed line correspond to $v_2$ with 
external magnetic field.}
\label{fig_v2pi_magfield}
\end{figure}

One of us has also recently solved the hydrodynamics equations where 
magnetic field is taken into account in the energy-momentum tensor 
of the fluid. A realistic space and time dependence of magnetic 
field is considered for an ideal fluid evolution in Au-Au collisions 
at $\sqrt{s_{NN}}=200$ GeV. It is found that in the presence of a 
finite electrical conductivity of QGP, the elliptic flow of $\pi^{-}$
increases noticeably, depending on the details of the magnitude of 
the magnetic field and the subsequent time evolution of the field. 
This is still a very new field of study and at present more detailed 
investigations are underway. In Fig.~\ref{fig_v2pi_magfield} we show 
the $v_2$ of $\pi^{-}$ obtained for impact parameter $b=10$ fm Au-Au 
collisions at $\sqrt{s_{NN}}=200$ GeV collisions. Initial value of 
magnetic field is taken to be $10m_{\pi}^{2}$ and the time variation 
of the magnetic field is obtained by parametrizing the results from 
Ref.~\cite{Tuchin:2013apa}. A realistic spatial profile for 
y-component of magnetic field was considered for the simulation. 
From Fig.~\ref {fig_v2pi_magfield} one can see that $v_2$ of 
$\pi^{-}$ is noticeably enhanced in the presence of magnetic field 
(blue dashed line) compared to the case of no magnetic field (red 
line).


\section{Outlook}

In this review article, we have discussed various aspects of 
relativistic dissipative hydrodynamics and its application to high 
energy heavy-ion collisions. While considerable success has been 
achieved in explaining many experimental observations, there are 
several issues that still needs further investigation. For example 
the experimentally measured longitudinal correlation of flow 
harmonics shows a splitting in the quantities corresponds to the 
correlation measure namely $r_2\left(\eta_a,\eta_b\right)$ and 
$r_3\left(\eta_a,\eta_b\right)$ for two different reference rapidity 
windows, which cannot be explained within a (3+1)D ideal 
hydrodynamics model with initial condition obtained from HIJING 
model. The reason behind this experimentally observed difference in 
$r_2\left(\eta_a,\eta_b\right)$ and $r_3\left(\eta_a,\eta_b\right)$ 
is still poorly understood. 

It was also argued in the present review that the effect of magnetic 
field might not be negligible on the hydrodynamics evolution of QGP 
produced in heavy-ion collisions. Particularly it may be important 
for the reason that the elliptic and higher-order flow harmonics 
might be affected under such strong magnetic field. However, at 
present there are some open issues in this regard: the electrical 
conductivity of the QGP might play an essential role in the temporal 
decay of the magnetic field. A poor knowledge of the temperature 
dependent electrical conductivity is one of the major source of 
uncertainties. In addition to that, the magnetic susceptibility of 
the QGP and the hadron resonance gas should be included for a 
realistic calculation.

Another unsolved problem is the experimentally measured $v_2$, $v_3$ 
in ultra-central collisions. Within error bars the magnitude of $v_2$
and $v_3$ are observed to be same for ultra-central collisions 
(0-0.2\%). Although viscous hydrodynamics model with MC-KLN initial 
condition considering nucleon-nucleon correlations produces quite 
close result to the experimental measurement, it is still not fully 
explained within the given error. Another puzzling aspect of high 
energy collisions is the observation of flow like behaviour in small 
systems. Initial study shows that the experimentally observed flow 
harmonics and other bulk observables for high multiplicity p-p and 
p-Pb collisions can be well described within viscous hydrodynamics 
model simulation. A detailed theoretical explanation of how such 
small system behave collectively is still not well understood. There 
are also possibility for non-hydrodynamical origin of this observed 
flow in such small system. This is a topic of current research and 
we hope we will have more clear theoretical understanding within 
next few years when further studies including alternative 
possibilities will be available. 

Finally we note that the field of high energy collisions is a very 
active area of research, we have not covered all aspects of the 
recent developments in the field related to 
hydrodynamics/collectivity of the QGP. For example the 
event-by-event distribution of flow harmonics \cite{Aad:2013xma} and 
their correlations \cite{Bilandzic:2013kga, ALICE:2016kpq} emerges 
as a promising observable to better constrain the initial conditions 
and the shear viscosity of the QGP. The event-by-event study of 
photon and dilepton production within viscous hydrodynamics provide 
us another window to look at the early stages of the heavy-ion 
collisions \cite{Chatterjee:2014nta, Shen:2014lpa}. We hope that 
through all these ongoing experimental and 
theoretical/phenomenological studies we will have much refined 
understanding about the collective behaviour and the transport 
properties of the QGP in the near future.


\section*{Conflict of Interest}

The authors declare that there is no conflict of interest regarding 
the publication of this article.


\begin{acknowledgments}

A.J. was supported by the Frankfurt Institute for Advanced Studies 
(FIAS), Germany. V.R. is partially supported by the Alexander von 
Humboldt foundation and J. W. Goethe university, Frankfurt, Germany.

\end{acknowledgments}



\begin{thebibliography}{99}

\bibitem{Lee} 
  T.~D.~Lee and G.~C.~Wick,
  ``Vacuum Stability and Vacuum Excitation in a Spin 0 Field Theory,''
  Phys.\ Rev.\ D {\bf 9}, 2291 (1974).

\bibitem{Collins:1974ky} 
  J.~C.~Collins and M.~J.~Perry,
  ``Superdense Matter: Neutrons Or Asymptotically Free Quarks?,''
  Phys.\ Rev.\ Lett.\  {\bf 34}, 1353 (1975).

\bibitem{Itoh:1970uw} 
  N.~Itoh,
  ``Hydrostatic Equilibrium of Hypothetical Quark Stars,''
  Prog.\ Theor.\ Phys.\  {\bf 44}, 291 (1970).

\bibitem{Baumgardt:1975qv} 
  H.~G.~Baumgardt, J.~U.~Schott, Y.~Sakamoto, E.~Schopper, H.~Stoecker, J.~Hofmann, W.~Scheid and W.~Greiner,
  ``Shock Waves and MACH Cones in Fast Nucleus-Nucleus Collisions,''
  Z.\ Phys.\ A {\bf 273}, 359 (1975).

\bibitem{CERN} 
  See the CERN press release and included references given in 
  http://newstate-matter.web.cern.ch/newstate-matter/Science.html

\bibitem{Tannenbaum:2006ch} 
  M.~J.~Tannenbaum,
  ``Recent results in relativistic heavy ion collisions: From `a new state of matter' to `the perfect fluid',''
  Rept.\ Prog.\ Phys.\  {\bf 69}, 2005 (2006),
  [nucl-ex/0603003].

\bibitem{Kolb}
  P.~Kolb and U.~Heinz, {\it Quark Gluon Plasma 3}, eds. R.~C.~Hwa and X.~N.~Wang 
  (World Scientific, Singapore, 2003).

\bibitem{Gyulassy}
  M.~Gyulassy, I.~Vitev, X.~Wang, and B.~W.~Zhang, {\it Quark Gluon Plasma 3}, eds. R.~C.~Hwa and X.~N.~Wang 
  (World Scientific, Singapore, 2003).

\bibitem{Tomasik}
  B.~Tom\"asik and U.~Wiedemann, {\it Quark Gluon Plasma 3}, eds. R.~C.~Hwa and X.~N.~Wang 
  (World Scientific, Singapore, 2003).

\bibitem{Muller}
 B.~M\"uller, {\it Lecture Notes in Physics}, Vol. 225 (Springer, New York, 1985).

\bibitem{Arsene:2004fa} 
  I.~Arsene {\it et al.}  [BRAHMS Collaboration],
  ``Quark gluon plasma and color glass condensate at RHIC? The Perspective from the BRAHMS experiment,''
  Nucl.\ Phys.\ A {\bf 757}, 1 (2005),
  [nucl-ex/0410020].

\bibitem{Adcox:2004mh} 
  K.~Adcox {\it et al.}  [PHENIX Collaboration],
  ``Formation of dense partonic matter in relativistic nucleus-nucleus collisions at RHIC: Experimental evaluation by the PHENIX collaboration,''
  Nucl.\ Phys.\ A {\bf 757}, 184 (2005),
  [nucl-ex/0410003].

\bibitem{Back:2004je} 
  B.~B.~Back, M.~D.~Baker, M.~Ballintijn, D.~S.~Barton, B.~Becker, R.~R.~Betts, A.~A.~Bickley and R.~Bindel {\it et al.},
  ``The PHOBOS perspective on discoveries at RHIC,''
  Nucl.\ Phys.\ A {\bf 757}, 28 (2005),
  [nucl-ex/0410022].

\bibitem{Adams:2005dq} 
  J.~Adams {\it et al.}  [STAR Collaboration],
  ``Experimental and theoretical challenges in the search for the quark gluon plasma: The STAR Collaboration's critical assessment of the evidence from RHIC collisions,''
  Nucl.\ Phys.\ A {\bf 757}, 102 (2005),
  [nucl-ex/0501009].

\bibitem{Aamodt:2010pa} 
  KAamodt {\it et al.}  [ALICE Collaboration],
  ``Elliptic flow of charged particles in Pb-Pb collisions at 2.76 TeV,''
  Phys.\ Rev.\ Lett.\  {\bf 105}, 252302 (2010),
  [arXiv:1011.3914 [nucl-ex]].

\bibitem{Aamodt:2010pb} 
  KAamodt {\it et al.}  [ALICE Collaboration],
  ``Charged-particle multiplicity density at mid-rapidity in central Pb-Pb collisions at $\sqrt{s_{NN}} = 2.76$ TeV,''
  Phys.\ Rev.\ Lett.\  {\bf 105}, 252301 (2010),
  [arXiv:1011.3916 [nucl-ex]].

\bibitem{Aamodt:2010cz} 
  K.~Aamodt {\it et al.}  [ALICE Collaboration],
  ``Centrality dependence of the charged-particle multiplicity density at mid-rapidity in Pb-Pb collisions at $\sqrt{s_{NN}}=2.76$ TeV,''
  Phys.\ Rev.\ Lett.\  {\bf 106}, 032301 (2011),
  [arXiv:1012.1657 [nucl-ex]].

\bibitem{ALICE:2011ab} 
  K.~Aamodt {\it et al.}  [ALICE Collaboration],
  ``Higher harmonic anisotropic flow measurements of charged particles in Pb-Pb collisions at $\sqrt{s_{NN}}$=2.76 TeV,''
  Phys.\ Rev.\ Lett.\  {\bf 107}, 032301 (2011),
  [arXiv:1105.3865 [nucl-ex]].

\bibitem{gsi.de}
  http://www.gsi.de/fair/experiments/CBM/1intro.html

\bibitem{duke.edu} 
  http://qgp.phy.duke.edu

\bibitem{Stoecker:1986ci} 
  H.~Stoecker and W.~Greiner,
  ``High-Energy Heavy Ion Collisions: Probing the Equation of State of Highly Excited Hadronic Matter,''
  Phys.\ Rept.\  {\bf 137}, 277 (1986).

\bibitem{Rischke:1995ir} 
  D.~H.~Rischke, S.~Bernard and J.~A.~Maruhn,
  ``Relativistic hydrodynamics for heavy ion collisions. 1. General aspects and expansion into vacuum,''
  Nucl.\ Phys.\ A {\bf 595}, 346 (1995),
  [nucl-th/9504018].

\bibitem{Rischke:1995mt} 
  D.~H.~Rischke, Y.~Pursun and J.~A.~Maruhn,
  ``Relativistic hydrodynamics for heavy ion collisions. 2. Compression of nuclear matter and the phase transition to the quark - gluon plasma,''
  Nucl.\ Phys.\ A {\bf 595}, 383 (1995)
  [Erratum-ibid.\ A {\bf 596}, 717 (1996)],
  [nucl-th/9504021].

\bibitem{Shuryak:2003xe} 
  E.~Shuryak,
  ``Why does the quark gluon plasma at RHIC behave as a nearly ideal fluid?,''
  Prog.\ Part.\ Nucl.\ Phys.\  {\bf 53}, 273 (2004),
  [hep-ph/0312227].

\bibitem{Romatschke:2007mq} 
  P.~Romatschke and U.~Romatschke,
  ``Viscosity Information from Relativistic Nuclear Collisions: How Perfect is the Fluid Observed at RHIC?,''
  Phys.\ Rev.\ Lett.\  99 (2007) 172301,
  [arXiv:0706.1522 [nucl-th]].

\bibitem{Song:2010mg} 
  H.~Song, S.~A.~Bass, U.~Heinz, T.~Hirano and C.~Shen,
  ``200 A GeV Au+Au collisions serve a nearly perfect quark-gluon liquid,''
  Phys.\ Rev.\ Lett.\  106 (2011) 192301,
 [arXiv:1011.2783 [nucl-th]].

\bibitem{Luzum:2010ag} 
  M.~Luzum,
  ``Elliptic flow at energies available at the CERN Large Hadron Collider: Comparing heavy-ion data to viscous hydrodynamic predictions,''
  Phys.\ Rev.\ C 83 (2011) 044911,
  [arXiv:1011.5173 [nucl-th]].

\bibitem{Qiu:2011hf} 
  Z.~Qiu, C.~Shen and U.~W.~Heinz,
  ``Hydrodynamic elliptic and triangular flow in Pb-Pb collisions at sqrt(s)=2.76ATeV,''
  Phys.\ Lett.\ B 707 (2012) 151,
  [arXiv:1110.3033 [nucl-th]].

\bibitem{Policastro:2001yc} 
  G.~Policastro, D.~T.~Son and A.~O.~Starinets,
  ``The Shear viscosity of strongly coupled N=4 supersymmetric Yang-Mills plasma,''
  Phys.\ Rev.\ Lett.\  {\bf 87}, 081601 (2001),
  [hep-th/0104066].

\bibitem{Kovtun:2004de} 
  P.~Kovtun, D.~T.~Son and A.~O.~Starinets,
  ``Viscosity in strongly interacting quantum field theories from black hole physics,''
  Phys.\ Rev.\ Lett.\  {\bf 94}, 111601 (2005),
  [hep-th/0405231].

\bibitem{Schaefer:2014awa} 
  T.~Schaefer,
  ``Fluid Dynamics and Viscosity in Strongly Correlated Fluids,''
  Ann.\ Rev.\ Nucl.\ Part.\ Sci.\  {\bf 64}, 125 (2014),
  [arXiv:1403.0653 [hep-ph]].

\bibitem{Chaudhuri:2013yna} 
  A.~K.~Chaudhuri,
  ``Viscous Hydrodynamic Model for Relativistic Heavy Ion Collisions,''
  Adv.\ High Energy Phys.\  {\bf 2013}, 693180 (2013).

\bibitem {Ibanez}
  J.M. Ib\'{a}\~{n}ez, in 
  {\it Current Trends in Relativistic Astrophysics: Theoretical, Numerical, Observational},
  Vol. 617, Lecture Notes in Physics, (Springer, Berlin, 2003).
  L. Fern\'{a}ndez-Jambrina and L.M. Gonz\'{a}lez-Romero (eds.).

\bibitem{Danielewicz:1984ww} 
  P.~Danielewicz and M.~Gyulassy,
  ``Dissipative Phenomena in Quark Gluon Plasmas,''
  Phys.\ Rev.\ D {\bf 31}, 53 (1985).

\bibitem{Eckart:1940zz} 
  C.~Eckart,
  ``The Thermodynamics of Irreversible Processes. 1. The Simple Fluid,''
  Phys.\ Rev.\  {\bf 58}, 267 (1940).

\bibitem{Landau} 
  L.D. Landau and E.M. Lifshitz, {\it Fluid Mechanics}
  (Butterworth-Heinemann, Oxford, 1987).

\bibitem{Hiscock:1983zz} 
  W.~A.~Hiscock and L.~Lindblom,
  ``Stability and causality in dissipative relativistic fluids,''
  Annals Phys.\  {\bf 151}, 466 (1983).

\bibitem{Hiscock:1985zz} 
  W.~A.~Hiscock and L.~Lindblom,
  ``Generic instabilities in first-order dissipative relativistic fluid theories,''
  Phys.\ Rev.\ D {\bf 31}, 725 (1985).

\bibitem{Hiscock:1987zz} 
  W.~A.~Hiscock and L.~Lindblom,
  ``Linear plane waves in dissipative relativistic fluids,''
  Phys.\ Rev.\ D {\bf 35}, 3723 (1987).

\bibitem{Israel:1979wp} 
  W.~Israel and J.~M.~Stewart,
  ``Transient relativistic thermodynamics and kinetic theory,''
  Annals Phys.\  {\bf 118}, 341 (1979).

\bibitem{Huovinen:2008te} 
  P.~Huovinen and D.~Molnar,
  ``The Applicability of causal dissipative hydrodynamics to relativistic heavy ion collisions,''
  Phys.\ Rev.\ C {\bf 79}, 014906 (2009),
  [arXiv:0808.0953 [nucl-th]].

\bibitem{Fermi}
  E.~Fermi,
  `{\it Thermodynamics}
  (Dover Publications, Inc. New York, 1956).

\bibitem{Reif}
  F.~Reif,
  {\it Fundamentals of Statistical and Thermal Physics} 
  (McGraw-Hill, 1965).

\bibitem{Reichl}
  L.~Reichl,
  {\it A Modern Course in Statistical Physics} 
  (Wiley-VCH, 2004).

\bibitem{Weinberg}
  S.~Weinberg,
  {\it Gravitation and Cosmology} 
  (Wiley, 1972).

\bibitem{Israel:1976tn} 
  W.~Israel,
  ``Nonstationary irreversible thermodynamics: A Causal relativistic theory,''
  Annals Phys.\  {\bf 100}, 310 (1976).

\bibitem{Denicol:2008ha} 
  G.~S.~Denicol, T.~Kodama, T.~Koide and P.~.Mota,
  ``Stability and Causality in relativistic dissipative hydrodynamics,''
  J.\ Phys.\ G {\bf 35}, 115102 (2008),
  [arXiv:0807.3120 [hep-ph]].

\bibitem{Pu:2009fj} 
  S.~Pu, T.~Koide and D.~H.~Rischke,
  ``Does stability of relativistic dissipative fluid dynamics imply causality?,''
  Phys.\ Rev.\ D {\bf 81}, 114039 (2010),
  [arXiv:0907.3906 [hep-ph]].

\bibitem{Grad}
  H.~Grad,
  ``On the kinetic theory of rarefied gases,'' 
  Comm. Pure Appl. Math. {\bf 2}, 331 (1949).

\bibitem{Muller:1967zza} 
  I.~Muller,
  ``Zum Paradoxon der Warmeleitungstheorie,''
  Z.\ Phys.\  {\bf 198}, 329 (1967).

\bibitem{Muller:1999in} 
  I.~Muller,
  ``Speeds of propagation in classical and relativistic extended thermodynamics,''
  Living Rev.\ Rel.\  {\bf 2}, 1 (1999).

\bibitem{Jou}
  D.~Jou, J.~Casas-V\'azquez, and G.~Lebon,
  {\it Extended Irreversible Thermodynamics} 
  (second edition, Springer-Verlag, Berlin, 1996).

\bibitem{Carter} 
  B.~Carter,
  ``Convective variational approach to relativistic thermodynamics of dissipative fluids,''
  Proc. R. Soc. London, Ser A, {\bf 433}, 45 (1991)

\bibitem{Grmela:1997zz} 
  M.~Grmela and H.~C.~Ottinger,
  ``Dynamics and thermodynamics of complex fluids 1. Development of a general formalism,''
  Phys.\ Rev.\ E {\bf 56}, 6620 (1997).

\bibitem{deGroot}
  S.R. de Groot, W.A. van Leeuwen, and Ch.G. van Weert, 
  {\it Relativistic Kinetic Theory --- Principles and Applications} 
  (North-Holland, Amsterdam, 1980).

\bibitem{Anderson_Witting}
  J.~L.~Anderson and H.~R.~Witting
  ``A relativistic relaxation-time for the Boltzmann equation,''
  Physica \textbf{74}, 466 (1974).

\bibitem{Jaiswal:2013fc} 
  A.~Jaiswal, R.~S.~Bhalerao and S.~Pal,
  ``Complete relativistic second-order dissipative hydrodynamics from the entropy principle,''
  Phys.\  Rev.\ C {\bf 87}, 021901(R) (2013),
  [arXiv:1302.0666 [nucl-th]].

\bibitem{Denicol:2012cn} 
  G.~S.~Denicol, H.~Niemi, E.~Molnar and D.~H.~Rischke,
  ``Derivation of transient relativistic fluid dynamics from the Boltzmann equation,''
  Phys.\ Rev.\ D {\bf 85}, 114047 (2012),
  [arXiv:1202.4551 [nucl-th]].

\bibitem{Denicol:2010xn} 
  G.~S.~Denicol, T.~Koide and D.~H.~Rischke,
  ``Dissipative relativistic fluid dynamics: a new way to derive the equations of motion from kinetic theory,''
  Phys.\ Rev.\ Lett.\  {\bf 105}, 162501 (2010),
  [arXiv:1004.5013 [nucl-th]].

\bibitem{Jaiswal:2012qm} 
  A.~Jaiswal, R.~S.~Bhalerao and S.~Pal,
  ``New relativistic dissipative fluid dynamics from kinetic theory,''
  Phys.\  Lett.\ B {\bf 720}, 347 (2013),
  [arXiv:1204.3779 [nucl-th]].

\bibitem{Jaiswal:2012dd} 
  A.~Jaiswal, R.~S.~Bhalerao and S.~Pal,
  ``Boltzmann equation with a nonlocal collision term and the resultant dissipative fluid dynamics,''
  J.\ Phys.\ Conf.\ Ser.\  {\bf 422}, 012003 (2013),
  [arXiv:1210.8427 [nucl-th]].

\bibitem{Bhalerao:2013aha} 
  R.~S.~Bhalerao, A.~Jaiswal, S.~Pal and V.~Sreekanth,
  ``Particle production in relativistic heavy-ion collisions: A consistent hydrodynamic approach,''
  Phys.\ Rev.\ C {\bf 88}, 044911 (2013),
  [arXiv:1305.4146 [nucl-th]].
  
\bibitem{Betz:2008me} 
  B.~Betz, D.~Henkel and D.~H.~Rischke,
  ``Complete second-order dissipative fluid dynamics,''
  J.\ Phys.\ G {\bf 36}, 064029 (2009);
  ``From kinetic theory to dissipative fluid dynamics,''
  Prog.\ Part.\ Nucl.\ Phys.\  {\bf 62} 556 (2009), 
  [arXiv:0812.1440 [nucl-th]].

\bibitem{Muronga:2003ta} 
  A.~Muronga,
  ``Causal theories of dissipative relativistic fluid dynamics for nuclear collisions,''
  Phys.\ Rev.\ C {\bf 69}, 034903 (2004),
  [nucl-th/0309055].

\bibitem{El:2008yy} 
  A.~El, A.~Muronga, Z.~Xu and C.~Greiner,
  ``Shear viscosity and out of equilibrium dissipative hydrodynamics,''
  Phys.\ Rev.\ C {\bf 79}, 044914 (2009),
  [arXiv:0812.2762 [hep-ph]].

\bibitem{Jaiswal:2013npa} 
  A.~Jaiswal,
  ``Relativistic dissipative hydrodynamics from kinetic theory with relaxation time approximation,''
  Phys.\  Rev.\ C {\bf 87}, 051901(R) (2013),
  [arXiv:1302.6311 [nucl-th]].
  
\bibitem{Jaiswal:2014isa} 
  A.~Jaiswal, R.~Ryblewski and M.~Strickland,
  ``Transport coefficients for bulk viscous evolution in the relaxation time approximation,''
  Phys.\ Rev.\ C {\bf 90}, 044908 (2014),
  [arXiv:1407.7231 [hep-ph]].
  
\bibitem{Florkowski:2015lra} 
  W.~Florkowski, A.~Jaiswal, E.~Maksymiuk, R.~Ryblewski and M.~Strickland,
  ``Relativistic quantum transport coefficients for second-order viscous hydrodynamics,''
  Phys.\ Rev.\ C {\bf 91}, 054907 (2015),
  [arXiv:1503.03226 [nucl-th]].

\bibitem{Bhalerao:2013pza} 
  R.~S.~Bhalerao, A.~Jaiswal, S.~Pal and V.~Sreekanth,
  ``Relativistic viscous hydrodynamics for heavy-ion collisions: A comparison between Chapman-Enskog and Grad's methods,''
  Phys.\ Rev.\ C {\bf 89}, 054903 (2014),
  arXiv:1312.1864 [nucl-th].

\bibitem{Jaiswal:2015mxa} 
  A.~Jaiswal, B.~Friman and K.~Redlich,
  ``Relativistic second-order dissipative hydrodynamics at finite chemical potential,''
  Phys.\ Lett.\ B {\bf 751}, 548 (2015),
  [arXiv:1507.02849 [nucl-th]].

\bibitem{Jaiswal:2016pmi} 
  A.~Jaiswal, B.~Friman and K.~Redlich,
  ``Relativistic second-order dissipative fluid dynamics at finite chemical potential,''
  EPJ Web Conf.\  {\bf 120}, 03008 (2016).

\bibitem{Jaiswal:2013vta} 
  A.~Jaiswal,
  ``Relativistic third-order dissipative fluid dynamics from kinetic theory,''
  Phys.\  Rev.\ C {\bf 88}, 021903(R) (2013),
  [arXiv:1305.3480 [nucl-th]].

\bibitem{Jaiswal:2014raa} 
  A.~Jaiswal,
  ``Relaxation-time approximation and relativistic third-order viscous hydrodynamics from kinetic theory,''
  Nucl.\ Phys.\ A {\bf 931}, 1205 (2014),
  [arXiv:1407.0837 [nucl-th]].
  
\bibitem{Chattopadhyay:2014lya} 
  C.~Chattopadhyay, A.~Jaiswal, S.~Pal and R.~Ryblewski,
  ``Relativistic third-order viscous corrections to the entropy four-current from kinetic theory,''
  Phys.\ Rev.\ C {\bf 91}, 024917 (2015),
  [arXiv:1411.2363 [nucl-th]].
 
\bibitem{El:2009vj} 
  A.~El, Z.~Xu and C.~Greiner,
  ``Extension of relativistic dissipative hydrodynamics to third order,''
  Phys.\ Rev.\ C {\bf 81}, 041901 (2010),
  [arXiv:0907.4500 [hep-ph]].
  
\bibitem{Bazow:2013ifa} 
  D.~Bazow, U.~W.~Heinz and M.~Strickland,
  ``Second-order (2+1)-dimensional anisotropic hydrodynamics,''
  Phys.\ Rev.\ C {\bf 90}, 054910 (2014)
  [arXiv:1311.6720 [nucl-th]].

\bibitem{Tinti:2013vba} 
  L.~Tinti and W.~Florkowski,
  ``Projection method and new formulation of leading-order anisotropic hydrodynamics,''
  Phys.\ Rev.\ C {\bf 89}, 034907 (2014),
  [arXiv:1312.6614 [nucl-th]].

\bibitem{Florkowski:2013lza} 
  W.~Florkowski, R.~Ryblewski and M.~Strickland,
  ``Anisotropic Hydrodynamics for Rapidly Expanding Systems,''
  Nucl.\ Phys.\ A {\bf 916}, 249 (2013),
  [arXiv:1304.0665 [nucl-th]].

\bibitem{Florkowski:2010cf} 
  W.~Florkowski and R.~Ryblewski,
  ``Highly-anisotropic and strongly-dissipative hydrodynamics for early stages of relativistic heavy-ion collisions,''
  Phys.\ Rev.\ C {\bf 83}, 034907 (2011),
  [arXiv:1007.0130 [nucl-th]].
 
\bibitem{Martinez:2010sc} 
  M.~Martinez and M.~Strickland,
  ``Dissipative Dynamics of Highly Anisotropic Systems,''
  Nucl.\ Phys.\ A {\bf 848}, 183 (2010),
  [arXiv:1007.0889 [nucl-th]].

\bibitem{Romatschke:2003ms} 
  P.~Romatschke and M.~Strickland,
  ``Collective modes of an anisotropic quark gluon plasma,''
  Phys.\ Rev.\ D {\bf 68}, 036004 (2003),
  [hep-ph/0304092].

\bibitem{Kikuchi:2015swa}
  Y.~Kikuchi, K.~Tsumura and T.~Kunihiro,
  ``Derivation of second-order relativistic hydrodynamics for reactive multicomponent systems,''
  Phys.\ Rev.\ C {\bf 92}, 064909 (2015),
  [arXiv:1507.04894 [hep-ph]].

\bibitem{Tsumura:2015fxa} 
  K.~Tsumura, Y.~Kikuchi and T.~Kunihiro,
  ``Relativistic Causal Hydrodynamics Derived from Boltzmann Equation: a novel reduction theoretical approach,''
  Phys.\ Rev.\ D {\bf 92}, 085048 (2015),
  [arXiv:1506.00846 [hep-ph]].
  
\bibitem{Tsumura:2012kp} 
  K.~Tsumura and T.~Kunihiro,
  ``Derivation of relativistic hydrodynamic equations consistent with relativistic Boltzmann equation by renormalization-group method,''
  Eur.\ Phys.\ J.\ A {\bf 48}, 162 (2012),
  [arXiv:1206.1929 [nucl-th]].
  
\bibitem{Tsumura:2012gq} 
  K.~Tsumura and T.~Kunihiro,
  ``New forms of non-relativistic and relativistic hydrodynamic equations as derived by the renormalization-group method 
  - possible functional ansatz in the moment method consistent with Chapman-Enskog theory -,''
  Prog.\ Theor.\ Phys.\ Suppl.\  {\bf 195}, 19 (2012),
  [arXiv:1205.5843 [nucl-th]].

\bibitem{York:2008rr} 
  M.~A.~York and G.~D.~Moore,
  ``Second order hydrodynamic coefficients from kinetic theory,''
  Phys.\ Rev.\ D {\bf 79}, 054011 (2009),
  [arXiv:0811.0729 [hep-ph]].

\bibitem{Baier:2007ix} 
  R.~Baier, P.~Romatschke, D.~T.~Son, A.~O.~Starinets and M.~A.~Stephanov,
  ``Relativistic viscous hydrodynamics, conformal invariance, and holography,''
  JHEP {\bf 0804}, 100 (2008),
  [arXiv:0712.2451 [hep-th]].

\bibitem{Bhattacharyya:2008jc} 
  S.~Bhattacharyya, V.~E.~Hubeny, S.~Minwalla and M.~Rangamani,
  ``Nonlinear Fluid Dynamics from Gravity,''
  JHEP {\bf 0802}, 045 (2008),
  [arXiv:0712.2456 [hep-th]].

\bibitem{Chaudhuri:2012yt} 
  A.~K.~Chaudhuri,
  ``A short course on Relativistic Heavy Ion Collisions,''
  arXiv:1207.7028 [nucl-th].

\bibitem{Romatschke:2009im} 
  P.~Romatschke,
  ``New Developments in Relativistic Viscous Hydrodynamics,''
  Int.\ J.\ Mod.\ Phys.\ E {\bf 19}, 1 (2010),
  arXiv:0902.3663 [hep-ph].

\bibitem{Heinz:2009xj} 
  U.~W.~Heinz,
  ``Early collective expansion: Relativistic hydrodynamics and the transport properties of QCD matter,''
  Landolt-Bornstein {\bf 23}, 240 (2010),
  [arXiv:0901.4355 [nucl-th]].

\bibitem{Ollitrault:2008zz} 
  J.~Y.~Ollitrault,
  ``Relativistic hydrodynamics for heavy-ion collisions,''
  Eur.\ J.\ Phys.\  {\bf 29}, 275 (2008),
  [arXiv:0708.2433 [nucl-th]].

\bibitem{Teaney:2009qa} 
  D.~A.~Teaney,
  ``Viscous Hydrodynamics and the Quark Gluon Plasma,''
  arXiv:0905.2433 [nucl-th].

\bibitem{deSouza:2015ena} 
  R.~Derradi de Souza, T.~Koide and T.~Kodama,
  ``Hydrodynamic Approaches in Relativistic Heavy Ion Reactions,''
  Prog.\ Part.\ Nucl.\ Phys.\  {\bf 86}, 35 (2016),
  [arXiv:1506.03863 [nucl-th]].

\bibitem{Bialas:1976ed} 
  A.~Bialas, M.~Bleszynski and W.~Czyz,
  ``Multiplicity Distributions in Nucleus-Nucleus Collisions at High-Energies,''
  Nucl.\ Phys.\ B {\bf 111}, 461 (1976).

\bibitem{Miller:2007ri} 
  M.~L.~Miller, K.~Reygers, S.~J.~Sanders and P.~Steinberg,
  ``Glauber modeling in high energy nuclear collisions,''
  Ann.\ Rev.\ Nucl.\ Part.\ Sci.\  {\bf 57}, 205 (2007),
  [nucl-ex/0701025].
  
\bibitem{Roy:2010zd} 
  V.~Roy and A.~K.~Chaudhuri,
  Phys.\ Rev.\ C {\bf 81}, 067901 (2010)
  doi:10.1103/PhysRevC.81.067901
  [arXiv:1003.5791 [nucl-th]].

\bibitem{Holopainen:2012id} 
  H.~Holopainen and P.~Huovinen,
  ``Dynamical Freeze-out in Event-by-Event Hydrodynamics,''
  J.\ Phys.\ Conf.\ Ser.\  {\bf 389}, 012018 (2012),
  [arXiv:1207.7331 [hep-ph]].

\bibitem{McLerran:1993ni} 
  L.~D.~McLerran and R.~Venugopalan,
  ``Computing quark and gluon distribution functions for very large nuclei,''
  Phys.\ Rev.\ D {\bf 49}, 2233 (1994),
  [hep-ph/9309289].

\bibitem{McLerran:1993ka} 
  L.~D.~McLerran and R.~Venugopalan,
  ``Gluon distribution functions for very large nuclei at small transverse momentum,''
  Phys.\ Rev.\ D {\bf 49}, 3352 (1994),
  [hep-ph/9311205].

\bibitem{Romatschke:2005pm} 
  P.~Romatschke and R.~Venugopalan,
  ``Collective non-Abelian instabilities in a melting color glass condensate,''
  Phys.\ Rev.\ Lett.\  {\bf 96}, 062302 (2006),
  [hep-ph/0510121].
  
\bibitem{Fukushima:2006ax} 
  K.~Fukushima, F.~Gelis and L.~McLerran,
  ``Initial Singularity of the Little Bang,''
  Nucl.\ Phys.\ A {\bf 786}, 107 (2007),
  [hep-ph/0610416].

\bibitem{Attems:2012js} 
  M.~Attems, A.~Rebhan and M.~Strickland,
  ``Instabilities of an anisotropically expanding non-Abelian plasma: 3D+3V discretized hard-loop simulations,''
  Phys.\ Rev.\ D {\bf 87}, 025010 (2013),
  [arXiv:1207.5795 [hep-ph]].

\bibitem{Dumitru:2007qr} 
  A.~Dumitru, E.~Molnar and Y.~Nara,
  ``Entropy production in high-energy heavy-ion collisions and the correlation of shear viscosity and thermalization time,''
  Phys.\ Rev.\ C {\bf 76}, 024910 (2007),
  [arXiv:0706.2203 [nucl-th]].

\bibitem{Kharzeev:2000ph} 
  D.~Kharzeev and M.~Nardi,
  ``Hadron production in nuclear collisions at RHIC and high density QCD,''
  Phys.\ Lett.\ B {\bf 507}, 121 (2001),
  [nucl-th/0012025].

\bibitem{Kharzeev:2001gp} 
  D.~Kharzeev and E.~Levin,
  ``Manifestations of high density QCD in the first RHIC data,''
  Phys.\ Lett.\ B {\bf 523}, 79 (2001),
  [nucl-th/0108006].
  
\bibitem{Kharzeev:2001yq} 
  D.~Kharzeev, E.~Levin and M.~Nardi,
  ``The Onset of classical QCD dynamics in relativistic heavy ion collisions,''
  Phys.\ Rev.\ C {\bf 71}, 054903 (2005),
  [hep-ph/0111315].
  
\bibitem{Drescher:2006ca} 
  H.-J.~Drescher and Y.~Nara,
  ``Effects of fluctuations on the initial eccentricity from the Color Glass Condensate in heavy ion collisions,''
  Phys.\ Rev.\ C {\bf 75}, 034905 (2007),
  [nucl-th/0611017].

\bibitem{Drescher:2007ax} 
  H.~J.~Drescher and Y.~Nara,
  ``Eccentricity fluctuations from the color glass condensate at RHIC and LHC,''
  Phys.\ Rev.\ C {\bf 76}, 041903 (2007),
  [arXiv:0707.0249 [nucl-th]].

\bibitem{Voloshin:2003ud} 
  S.~A.~Voloshin,
  ``Transverse radial expansion in nuclear collisions and two particle correlations,''
  Phys.\ Lett.\ B {\bf 632}, 490 (2006),
  [nucl-th/0312065].

\bibitem{Shuryak:2007fu} 
  E.~V.~Shuryak,
  ``On the origin of the 'Ridge' phenomenon induced by jets in heavy ion collisions,''
  Phys.\ Rev.\ C {\bf 76}, 047901 (2007),
  [arXiv:0706.3531 [nucl-th]].

\bibitem{McLerran:1994vd} 
  L.~D.~McLerran and R.~Venugopalan,
  ``Green's functions in the color field of a large nucleus,''
  Phys.\ Rev.\ D {\bf 50}, 2225 (1994),
  [hep-ph/9402335].

\bibitem{JalilianMarian:1996xn} 
  J.~Jalilian-Marian, A.~Kovner, L.~D.~McLerran and H.~Weigert,
  ``The Intrinsic glue distribution at very small x,''
  Phys.\ Rev.\ D {\bf 55}, 5414 (1997),
  [hep-ph/9606337].

\bibitem{Kovchegov:1996ty} 
  Y.~V.~Kovchegov,
  ``NonAbelian Weizsacker-Williams field and a two-dimensional effective color charge density for a very large nucleus,''
  Phys.\ Rev.\ D {\bf 54}, 5463 (1996),
  [hep-ph/9605446].

\bibitem{Kovner:1995ja} 
  A.~Kovner, L.~D.~McLerran and H.~Weigert,
  ``Gluon production from nonAbelian Weizsacker-Williams fields in nucleus-nucleus collisions,''
  Phys.\ Rev.\ D {\bf 52}, 6231 (1995),
  [hep-ph/9502289].
  
\bibitem{Kovner:1995ts} 
  A.~Kovner, L.~D.~McLerran and H.~Weigert,
  ``Gluon production at high transverse momentum in the McLerran-Venugopalan model of nuclear structure functions,''
  Phys.\ Rev.\ D {\bf 52}, 3809 (1995),
  [hep-ph/9505320].

\bibitem{Krasnitz:1999wc} 
  A.~Krasnitz and R.~Venugopalan,
  ``The Initial energy density of gluons produced in very high-energy nuclear collisions,''
  Phys.\ Rev.\ Lett.\  {\bf 84}, 4309 (2000),
  [hep-ph/9909203].

\bibitem{Krasnitz:2000gz} 
  A.~Krasnitz and R.~Venugopalan,
  ``The Initial gluon multiplicity in heavy ion collisions,''
  Phys.\ Rev.\ Lett.\  {\bf 86}, 1717 (2001),
  [hep-ph/0007108].

\bibitem{Lappi:2003bi} 
  T.~Lappi,
  ``Production of gluons in the classical field model for heavy ion collisions,''
  Phys.\ Rev.\ C {\bf 67}, 054903 (2003),
  [hep-ph/0303076].

\bibitem{Gale:2012rq} 
  C.~Gale, S.~Jeon, B.~Schenke, P.~Tribedy and R.~Venugopalan,
  ``Event-by-event anisotropic flow in heavy-ion collisions from combined Yang-Mills and viscous fluid dynamics,''
  Phys.\ Rev.\ Lett.\  {\bf 110}, 012302 (2013),
  [arXiv:1209.6330 [nucl-th]].

\bibitem{Steinheimer:2007iy} 
  J.~Steinheimer, M.~Bleicher, H.~Petersen, S.~Schramm, H.~Stocker and D.~Zschiesche,
  ``(3+1)-dimensional hydrodynamic expansion with a critical point from realistic initial conditions,''
  Phys.\ Rev.\ C {\bf 77}, 034901 (2008),
  [arXiv:0710.0332 [nucl-th]].
  
\bibitem{Pang:2012he} 
  L.~Pang, Q.~Wang and X.~N.~Wang,
  ``Effects of initial flow velocity fluctuation in event-by-event (3+1)D hydrodynamics,''
  Phys.\ Rev.\ C {\bf 86}, 024911 (2012),
  [arXiv:1205.5019 [nucl-th]].

\bibitem{Bhalerao:2015iya} 
  R.~S.~Bhalerao, A.~Jaiswal and S.~Pal,
  ``Collective flow in event-by-event partonic transport plus hydrodynamics hybrid approach,''
  Phys.\ Rev.\ C {\bf 92}, 014903 (2015),
  [arXiv:1503.03862 [nucl-th]].

\bibitem{Zhang:1999bd} 
  B.~Zhang, C.~M.~Ko, B.~A.~Li and Z.~w.~Lin,
  ``A multiphase transport model for nuclear collisions at RHIC,''
  Phys.\ Rev.\ C {\bf 61}, 067901 (2000),
  [nucl-th/9907017].
  
\bibitem{Lin:2004en} 
  Z.~W.~Lin, C.~M.~Ko, B.~A.~Li, B.~Zhang and S.~Pal,
  ``A Multi-phase transport model for relativistic heavy ion collisions,''
  Phys.\ Rev.\ C {\bf 72}, 064901 (2005),
  [nucl-th/0411110].
  
\bibitem{Wang:1991hta} 
  X.~N.~Wang and M.~Gyulassy,
  ``HIJING: A Monte Carlo model for multiple jet production in p p, p A and A A collisions,''
  Phys.\ Rev.\ D {\bf 44}, 3501 (1991).
  
\bibitem{Gyulassy:1994ew} 
  M.~Gyulassy and X.~N.~Wang,
  ``HIJING 1.0: A Monte Carlo program for parton and particle production in high-energy hadronic and nuclear collisions,''
  Comput.\ Phys.\ Commun.\  {\bf 83}, 307 (1994),
  [nucl-th/9502021].

\bibitem{vanderSchee:2013pia} 
  W.~van der Schee, P.~Romatschke and S.~Pratt,
  ``Fully Dynamical Simulation of Central Nuclear Collisions,''
  Phys.\ Rev.\ Lett.\  {\bf 111}, 222302 (2013),
  [arXiv:1307.2539].

\bibitem{deHaro:2000vlm} 
  S.~de Haro, S.~N.~Solodukhin and K.~Skenderis,
  ``Holographic reconstruction of space-time and renormalization in the AdS / CFT correspondence,''
  Commun.\ Math.\ Phys.\  {\bf 217}, 595 (2001),
  [hep-th/0002230].

\bibitem{Romatschke:2013re} 
  P.~Romatschke and J.~D.~Hogg,
  ``Pre-Equilibrium Radial Flow from Central Shock-Wave Collisions in AdS5,''
  JHEP {\bf 1304}, 048 (2013),
  [arXiv:1301.2635 [hep-th]].

\bibitem{Vredevoogd:2008id} 
  J.~Vredevoogd and S.~Pratt,
  ``Universal Flow in the First Stage of Relativistic Heavy Ion Collisions,''
  Phys.\ Rev.\ C {\bf 79}, 044915 (2009),
  [arXiv:0810.4325 [nucl-th]].

\bibitem{Casalderrey-Solana:2013aba} 
  J.~Casalderrey-Solana, M.~P.~Heller, D.~Mateos and W.~van der Schee,
  ``From full stopping to transparency in a holographic model of heavy ion collisions,''
  Phys.\ Rev.\ Lett.\  {\bf 111}, 181601 (2013),
  [arXiv:1305.4919 [hep-th]].

\bibitem{Grumiller:2008va} 
  D.~Grumiller and P.~Romatschke,
  ``On the collision of two shock waves in AdS(5),''
  JHEP {\bf 0808}, 027 (2008),
  [arXiv:0803.3226 [hep-th]].

\bibitem{Taliotis:2010pi} 
  A.~Taliotis,
  ``Heavy Ion Collisions with Transverse Dynamics from Evolving AdS Geometries,''
  JHEP {\bf 1009}, 102 (2010),
  [arXiv:1004.3500 [hep-th]].

\bibitem{Chesler:2010bi} 
  P.~M.~Chesler and L.~G.~Yaffe,
  ``Holography and colliding gravitational shock waves in asymptotically $AdS_5$ spacetime,''
  Phys.\ Rev.\ Lett.\  {\bf 106}, 021601 (2011),
  [arXiv:1011.3562 [hep-th]].

\bibitem{vanderSchee:2012qj} 
  W.~van der Schee,
  ``Holographic thermalization with radial flow,''
  Phys.\ Rev.\ D {\bf 87}, 061901 (2013),
  [arXiv:1211.2218 [hep-th]].

\bibitem{Chesler:2013lia} 
  P.~M.~Chesler and L.~G.~Yaffe,
  ``Numerical solution of gravitational dynamics in asymptotically anti-de Sitter spacetimes,''
  JHEP {\bf 1407}, 086 (2014),
  [arXiv:1309.1439 [hep-th]].

\bibitem{Attems:2016tby} 
  M.~Attems, J.~Casalderrey-Solana, D.~Mateos, D.~Santos-Oliván, C.~F.~Sopuerta, M.~Triana and M.~Zilhão,
  ``Collisions in Non-conformal Theories: Hydrodynamization without Equilibration,''
  arXiv:1604.06439 [hep-th].

\bibitem{Huovinen:2009yb} 
  P.~Huovinen and P.~Petreczky,
  ``QCD Equation of State and Hadron Resonance Gas,''
  Nucl.\ Phys.\ A {\bf 837}, 26 (2010),
  [arXiv:0912.2541 [hep-ph]].

\bibitem{Boyd:1996bx} 
  G.~Boyd, J.~Engels, F.~Karsch, E.~Laermann, C.~Legeland, M.~Lutgemeier and B.~Petersson,
  ``Thermodynamics of SU(3) lattice gauge theory,''
  Nucl.\ Phys.\ B {\bf 469}, 419 (1996),
  [hep-lat/9602007].
  
\bibitem{Bazavov:2009zn} 
  A.~Bazavov {\it et al.},
  ``Equation of state and QCD transition at finite temperature,''
  Phys.\ Rev.\ D {\bf 80}, 014504 (2009),
  [arXiv:0903.4379 [hep-lat]].

\bibitem{Cooper:1974mv} 
  F.~Cooper and G.~Frye,
  ``Comment on the Single Particle Distribution in the Hydrodynamic and Statistical Thermodynamic Models of Multiparticle Production,''
  Phys.\ Rev.\ D {\bf 10}, 186 (1974).

\bibitem{Huovinen:2012is} 
  P.~Huovinen and H.~Petersen,
  Eur.\ Phys.\ J.\ A {\bf 48}, 171 (2012),
  [arXiv:1206.3371 [nucl-th]].

\bibitem{Sollfrank:1990qz} 
  J.~Sollfrank, P.~Koch and U.~W.~Heinz,
  ``The Influence of resonance decays on the P(t) spectra from heavy ion collisions,''
  Phys.\ Lett.\ B {\bf 252}, 256 (1990).

\bibitem{Victor:thesis}
 ``Dissipative fluid dynamics for ultra-relativistic nuclear collisions",Phd thesis by Victor Roy,
www.hbni.ac.in/phdthesis/phys/PHYS04200704003.pdf

\bibitem{Aichelin:1986wa} 
  J.~Aichelin and H.~Stoecker,
  ``Quantum molecular dynamics. A Novel approach to N body correlations in heavy ion collisions,''
  Phys.\ Lett.\ B {\bf 176}, 14 (1986).

\bibitem{Sorge:1989dy} 
  H.~Sorge, H.~Stoecker and W.~Greiner,
  ``Poincare Invariant Hamiltonian Dynamics: Modeling Multi - Hadronic Interactions in a Phase Space Approach,''
  Annals Phys.\  {\bf 192}, 266 (1989).

\bibitem{Ehehalt:1996uq} 
  W.~Ehehalt and W.~Cassing,
  ``Relativistic transport approach for nucleus nucleus collisions from SIS to SPS energies,''
  Nucl.\ Phys.\ A {\bf 602}, 449 (1996).

\bibitem{Bass:1998ca} 
  S.~A.~Bass {\it et al.},
  ``Microscopic models for ultrarelativistic heavy ion collisions,''
  Prog.\ Part.\ Nucl.\ Phys.\  {\bf 41}, 255 (1998),
  [nucl-th/9803035].

\bibitem{Bleicher:1999xi} 
  M.~Bleicher {\it et al.},
  ``Relativistic hadron hadron collisions in the ultrarelativistic quantum molecular dynamics model,''
  J.\ Phys.\ G {\bf 25}, 1859 (1999),
  [hep-ph/9909407].

\bibitem{Bass:2000ib} 
  S.~A.~Bass and A.~Dumitru,
  ``Dynamics of hot bulk QCD matter: From the quark gluon plasma to hadronic freezeout,''
  Phys.\ Rev.\ C {\bf 61}, 064909 (2000),
  [nucl-th/0001033].

\bibitem{Teaney:2001gc} 
  D.~Teaney, J.~Lauret and E.~V.~Shuryak,
  ``Hydro+cascade, flow, the equation of state, predictions and data,''
  Nucl.\ Phys.\ A {\bf 698}, 479 (2002),
  [nucl-th/0104041].

\bibitem{Hirano:2005xf} 
  T.~Hirano, U.~W.~Heinz, D.~Kharzeev, R.~Lacey and Y.~Nara,
  ``Hadronic dissipative effects on elliptic flow in ultrarelativistic heavy-ion collisions,''
  Phys.\ Lett.\ B {\bf 636}, 299 (2006),
  [nucl-th/0511046].

\bibitem{Nonaka:2006yn} 
  C.~Nonaka and S.~A.~Bass,
  ``Space-time evolution of bulk QCD matter,''
  Phys.\ Rev.\ C {\bf 75}, 014902 (2007),
  [nucl-th/0607018].

\bibitem{Song:2010aq} 
  H.~Song, S.~A.~Bass and U.~Heinz,
  ``Viscous QCD matter in a hybrid hydrodynamic+Boltzmann approach,''
  Phys.\ Rev.\ C {\bf 83}, 024912 (2011),
  [arXiv:1012.0555 [nucl-th]].

\bibitem{Heinz:2013th} 
  U.~W. Heinz and R.~Snellings,
  ``Collective flow and viscosity in relativistic heavy-ion collisions,''
  Ann.\ Rev.\ Nucl.\ Part.\ Sci.\  {\bf 63}, 123 (2013),
  arXiv:1301.2826 [nucl-th].

\bibitem{Teaney:2003kp} 
  D.~Teaney,
  ``The Effects of viscosity on spectra, elliptic flow, and HBT radii,''
  Phys.\ Rev.\ C {\bf 68}, 034913 (2003),
  [nucl-th/0301099].

\bibitem{Chaudhuri:2010hr} 
  A.~K.~Chaudhuri and V.~Roy,
  ``Charged particle's $p_T$ spectra and elliptic flow in $\sqrt{s_{NN}}$=200 GeV Au+Au collisions: QGP versus hadronic resonance gas,''
  arXiv:1009.5223 [nucl-th].

\bibitem{Lacey:2006bc} 
  R.~A.~Lacey {\it et al.},
  ``Has the QCD Critical Point been Signaled by Observations at RHIC?,''
  Phys.\ Rev.\ Lett.\  {\bf 98}, 092301 (2007),
  [nucl-ex/0609025].
  
\bibitem{Drescher:2007cd} 
  H.~J.~Drescher, A.~Dumitru, C.~Gombeaud and J.~Y.~Ollitrault,
  ``The Centrality dependence of elliptic flow, the hydrodynamic limit, and the viscosity of hot QCD,''
  Phys.\ Rev.\ C {\bf 76}, 024905 (2007),
  [arXiv:0704.3553 [nucl-th]].

\bibitem{Gavin:2006xd} 
  S.~Gavin and M.~Abdel-Aziz,
  ``Measuring Shear Viscosity Using Transverse Momentum Correlations in Relativistic Nuclear Collisions,''
  Phys.\ Rev.\ Lett.\  {\bf 97}, 162302 (2006),
  [nucl-th/0606061].

\bibitem{vanHees:2008gj} 
  H.~van Hees, M.~Mannarelli, V.~Greco and R.~Rapp,
  ``T-matrix approach to heavy quark diffusion in the QGP,''
  Eur.\ Phys.\ J.\ C {\bf 61}, 799 (2009),
  [arXiv:0808.3710 [hep-ph]].

\bibitem{Chen:2009sm} 
  J.~W.~Chen, H.~Dong, K.~Ohnishi and Q.~Wang,
  ``Shear Viscosity of a Gluon Plasma in Perturbative QCD,''
  Phys.\ Lett.\ B {\bf 685}, 277 (2010),
  [arXiv:0907.2486 [nucl-th]].

\bibitem{Xu:2007ns} 
  Z.~Xu and C.~Greiner,
  ``Shear viscosity in a gluon gas,''
  Phys.\ Rev.\ Lett.\  {\bf 100}, 172301 (2008),
  [arXiv:0710.5719 [nucl-th]].
  
\bibitem{Meyer:2007ic} 
  H.~B.~Meyer,
  ``A Calculation of the shear viscosity in SU(3) gluodynamics,''
  Phys.\ Rev.\ D {\bf 76}, 101701 (2007),
  [arXiv:0704.1801 [hep-lat]].

\bibitem{Demir:2008tr} 
  N.~Demir and S.~A.~Bass,
  ``Shear-Viscosity to Entropy-Density Ratio of a Relativistic Hadron Gas,''
  Phys.\ Rev.\ Lett.\  {\bf 102}, 172302 (2009),
  [arXiv:0812.2422 [nucl-th]].

\bibitem{Xu:2011fe} 
  J.~Xu and C.~M.~Ko,
  ``Triangular flow in heavy ion collisions in a multiphase transport model,''
  Phys.\ Rev.\ C {\bf 84}, 014903 (2011),
  [arXiv:1103.5187 [nucl-th]].

\bibitem{Bozek:2011wa} 
  P.~Bozek,
  ``Components of the elliptic flow in Pb-Pb collisions at s**(1/2) = 2.76-TeV,''
  Phys.\ Lett.\ B {\bf 699}, 283 (2011),
  [arXiv:1101.1791 [nucl-th]].

\bibitem{Schenke:2011tv} 
  B.~Schenke, S.~Jeon and C.~Gale,
  ``Anisotropic flow in $\sqrt{s}=2.76$ TeV Pb+Pb collisions at the LHC,''
  Phys.\ Lett.\ B {\bf 702}, 59 (2011),
  [arXiv:1102.0575 [hep-ph]].

\bibitem{Luzum:2008cw} 
  M.~Luzum and P.~Romatschke,
  ``Conformal Relativistic Viscous Hydrodynamics: Applications to RHIC results at s(NN)**(1/2) = 200-GeV,''
  Phys.\ Rev.\ C {\bf 78}, 034915 (2008)
  Erratum: [Phys.\ Rev.\ C {\bf 79}, 039903 (2009)],
  [arXiv:0804.4015 [nucl-th]].

\bibitem{Kisiel:2005hn} 
  A.~Kisiel, T.~Taluc, W.~Broniowski and W.~Florkowski,
  ``THERMINATOR: THERMal heavy-IoN generATOR,''
  Comput.\ Phys.\ Commun.\  {\bf 174}, 669 (2006),
  [nucl-th/0504047].

\bibitem{Roy:2011xt} 
  V.~Roy and A.~K.~Chaudhuri,
  Phys.\ Lett.\ B {\bf 703}, 313 (2011)
  doi:10.1016/j.physletb.2011.08.006
  [arXiv:1103.2870 [nucl-th]].

\bibitem{Meyer:2007dy} 
  H.~B.~Meyer,
  ``A Calculation of the bulk viscosity in SU(3) gluodynamics,''
  Phys.\ Rev.\ Lett.\  {\bf 100}, 162001 (2008),
  [arXiv:0710.3717 [hep-lat]].
  
\bibitem{Paech:2006st} 
  K.~Paech and S.~Pratt,
  ``Origins of bulk viscosity in relativistic heavy ion collisions,''
  Phys.\ Rev.\ C {\bf 74}, 014901 (2006),
  [nucl-th/0604008].

\bibitem{Ryu:2015vwa} 
  S.~Ryu, J.-F.~Paquet, C.~Shen, G.~S.~Denicol, B.~Schenke, S.~Jeon and C.~Gale,
  ``Importance of the Bulk Viscosity of QCD in Ultrarelativistic Heavy-Ion Collisions,''
  Phys.\ Rev.\ Lett.\  {\bf 115}, 132301 (2015),
  [arXiv:1502.01675 [nucl-th]].

\bibitem{Noronha-Hostler:2013gga} 
  J.~Noronha-Hostler, G.~S.~Denicol, J.~Noronha, R.~P.~G.~Andrade and F.~Grassi,
  ``Bulk Viscosity Effects in Event-by-Event Relativistic Hydrodynamics,''
  Phys.\ Rev.\ C {\bf 88}, 044916 (2013),
  [arXiv:1305.1981 [nucl-th]].

\bibitem{Song:2009rh} 
  H.~Song and U.~W.~Heinz,
  ``Interplay of shear and bulk viscosity in generating flow in heavy-ion collisions,''
  Phys.\ Rev.\ C {\bf 81}, 024905 (2010),
  [arXiv:0909.1549 [nucl-th]].

\bibitem{Roy:2011pk} 
  V.~Roy and A.~K.~Chaudhuri,
  ``2+1 dimensional hydrodynamics including bulk viscosity: A Systematics study,''
  Phys.\ Rev.\ C {\bf 85}, 024909 (2012)
  [Erratum-ibid.\ C {\bf 85}, 049902 (2012)],
  [arXiv:1109.1630 [nucl-th]].

\bibitem{Noronha-Hostler:2014dqa} 
  J.~Noronha-Hostler, J.~Noronha and F.~Grassi,
  ``Bulk viscosity-driven suppression of shear viscosity effects on the 
  flow harmonics at energies available at the BNL Relativistic Heavy Ion Collider,''
  Phys.\ Rev.\ C {\bf 90}, no. 3, 034907 (2014),
  [arXiv:1406.3333 [nucl-th]].

 
\bibitem{Denicol:2009am} 
  G.~S.~Denicol, T.~Kodama, T.~Koide and P.~.Mota,
  ``Effect of bulk viscosity on Elliptic Flow near QCD phase transition,''
  Phys.\ Rev.\ C {\bf 80}, 064901 (2009),
  [arXiv:0903.3595 [hep-ph]].

\bibitem{Loizides:2016tew} 
  C.~Loizides,
  ``Experimental overview on small collision systems at the LHC,''
  arXiv:1602.09138 [nucl-ex].


\bibitem{Ortiz:2013yxa} 
  A.~Ortiz Velasquez, P.~Christiansen, E.~Cuautle Flores, I.~Maldonado Cervantes and G.~Pai?,
  ``Color Reconnection and Flowlike Patterns in $pp$ Collisions,''
  Phys.\ Rev.\ Lett.\  {\bf 111}, 042001 (2013),
  [arXiv:1303.6326 [hep-ph]].

\bibitem{Gyulassy:2014cfa} 
  M.~Gyulassy, P.~Levai, I.~Vitev and T.~S.~Biro,
  ``Non-Abelian Bremsstrahlung and Azimuthal Asymmetries in High Energy $p+A$ Reactions,''
  Phys.\ Rev.\ D {\bf 90}, 054025 (2014),
  [arXiv:1405.7825 [hep-ph]].

\bibitem{Chatrchyan:2013nka} 
  S.~Chatrchyan {\it et al.} [CMS Collaboration],
  ``Multiplicity and transverse momentum dependence of two- and four-particle correlations in pPb and PbPb collisions,''
  Phys.\ Lett.\ B {\bf 724}, 213 (2013),
  [arXiv:1305.0609 [nucl-ex]].

\bibitem{Chatrchyan:2013eya} 
  S.~Chatrchyan {\it et al.} [CMS Collaboration],
  ``Study of the production of charged pions, kaons, and protons in pPb collisions at $\sqrt{s_{NN}} =\  $ 5.02 $\,\text {TeV}$,''
  Eur.\ Phys.\ J.\ C {\bf 74}, 2847 (2014),
  [arXiv:1307.3442 [hep-ex]].

\bibitem{ABELEV:2013wsa} 
  B.~B.~Abelev {\it et al.} [ALICE Collaboration],
  ``Long-range angular correlations of $\rm \pi$, K and p in p-Pb collisions at $\sqrt{s_{\rm NN}}$ = 5.02 TeV,''
  Phys.\ Lett.\ B {\bf 726}, 164 (2013),
  [arXiv:1307.3237 [nucl-ex]].

\bibitem{Bozek:2014wpa} 
  P.~Bożek and W.~Broniowski,
  ``Collective flow in small systems,''
  Nucl.\ Phys.\ A {\bf 931}, 883 (2014),
  [arXiv:1407.6478 [nucl-th]].

\bibitem{Kalaydzhyan:2015xba} 
  T.~Kalaydzhyan and E.~Shuryak,
  ``Collective flow in high-multiplicity proton-proton collisions,''
  Phys.\ Rev.\ C {\bf 91}, 054913 (2015),
  [arXiv:1503.05213 [hep-ph]].

\bibitem{Ghosh:2014eqa} 
  P.~Ghosh, S.~Muhuri, J.~K.~Nayak and R.~Varma,
  ``Indication of transverse radial flow in high-multiplicity proton-proton collisions at the Large Hadron Collider,''
  J.\ Phys.\ G {\bf 41}, 035106 (2014),
  [arXiv:1402.6813 [hep-ph]].

\bibitem{Chatrchyan:2012qb}
  S.~Chatrchyan {\it et al.}  [CMS Collaboration],
  ``Study of the inclusive production of charged pions, kaons, and protons in $pp$ collisions at $\sqrt{s}=0.9$, 2.76, and 7 TeV,''
  Eur.\ Phys.\ J.\ C {\bf 72}, 2164 (2012),
  [arXiv:1207.4724 [hep-ex]].

\bibitem{Bozek:2010pb} 
  P.~Bozek,
  ``Elliptic flow in proton-proton collisions at $sqrt(S) = 7$ TeV,''
  Eur.\ Phys.\ J.\ C {\bf 71}, 1530 (2011),
  [arXiv:1010.0405 [hep-ph]].

\bibitem{Prasad:2009bx} 
  S.~K.~Prasad, V.~Roy, S.~Chattopadhyay and A.~K.~Chaudhuri,
  ``Elliptic flow ($v_2$) in pp collisions at energies available at the CERN Large Hadron Collider: A hydrodynamical approach,''
  Phys.\ Rev.\ C {\bf 82}, 024909 (2010),
  [arXiv:0910.4844 [nucl-th]].
  
\bibitem{Bzdak:2013zma} 
  A.~Bzdak, B.~Schenke, P.~Tribedy and R.~Venugopalan,
  ``Initial state geometry and the role of hydrodynamics in proton-proton, proton-nucleus and deuteron-nucleus collisions,''
  Phys.\ Rev.\ C {\bf 87}, 064906 (2013),
  [arXiv:1304.3403 [nucl-th]].
  
\bibitem{Shuryak:2013ke} 
  E.~Shuryak and I.~Zahed,
  ``High-multiplicity pp and pA collisions: Hydrodynamics at its edge,''
  Phys.\ Rev.\ C {\bf 88}, 044915 (2013),
  [arXiv:1301.4470 [hep-ph]].

\bibitem{CMS:2013bza} 
  S.~Chatrchyan {\it et al.} [CMS Collaboration],
  ``Studies of azimuthal dihadron correlations in ultra-central PbPb collisions at $\sqrt{s_{NN}} =$ 2.76 TeV,''
  JHEP {\bf 1402}, 088 (2014),
  [arXiv:1312.1845 [nucl-ex]].
  
\bibitem{Shen:2015qta} 
  C.~Shen, Z.~Qiu and U.~Heinz,
  ``Shape and flow fluctuations in ultracentral Pb + Pb collisions at the energies available at the CERN Large Hadron Collider,''
  Phys.\ Rev.\ C {\bf 92}, 014901 (2015),
  [arXiv:1502.04636 [nucl-th]].
  
\bibitem{Denicol:talk_ECT}
  Fluid dynamical description of heavy ion collisions, Trento, Italy. \\
  $http://www.ectstar.eu/sites/www.ectstar.eu/files/talks/Denicol_Trento_2014.pdf$
  
\bibitem{Alver:2010gr} 
  B.~Alver and G.~Roland,
  ``Collision geometry fluctuations and triangular flow in heavy-ion collisions,''
  Phys.\ Rev.\ C {\bf 81}, 054905 (2010)
  Erratum: [Phys.\ Rev.\ C {\bf 82}, 039903 (2010)],
  [arXiv:1003.0194 [nucl-th]].

\bibitem{Broniowski:2007ft} 
  W.~Broniowski, P.~Bozek and M.~Rybczynski,
  ``Fluctuating initial conditions in heavy-ion collisions from the Glauber approach,''
  Phys.\ Rev.\ C {\bf 76}, 054905 (2007),
  [arXiv:0706.4266 [nucl-th]].

\bibitem{Hirano:2009ah} 
  T.~Hirano and Y.~Nara,
  ``Eccentricity fluctuation effects on elliptic flow in relativistic heavy ion collisions,''
  Phys.\ Rev.\ C {\bf 79}, 064904 (2009),
  [arXiv:0904.4080 [nucl-th]].

\bibitem{Roy:2012pn} 
  V.~Roy, B.~Mohanty and A.~K.~Chaudhuri,
  ``Elliptic and Hexadecapole flow of charged hadron in viscous hydrodynamics with Glauber and Color Glass Condensate initial conditions for Pb-Pb collision at $\sqrt{s_{NN}}$=2.76 TeV,''
  J.\ Phys.\ G {\bf 40}, 065103 (2013),
  [arXiv:1210.1700 [nucl-th]].

\bibitem{Petersen:2008dd} 
  H.~Petersen, J.~Steinheimer, G.~Burau, M.~Bleicher and H.~Stocker,
  ``A Fully Integrated Transport Approach to Heavy Ion Reactions with an Intermediate Hydrodynamic Stage,''
  Phys.\ Rev.\ C {\bf 78}, 044901 (2008),
  [arXiv:0806.1695 [nucl-th]].

\bibitem{Werner:2010aa} 
  K.~Werner, I.~Karpenko, T.~Pierog, M.~Bleicher and K.~Mikhailov,
  ``Event-by-Event Simulation of the Three-Dimensional Hydrodynamic Evolution from Flux Tube Initial Conditions in Ultrarelativistic Heavy Ion Collisions,''
  Phys.\ Rev.\ C {\bf 82}, 044904 (2010),
  [arXiv:1004.0805 [nucl-th]].

\bibitem{Schenke:2012wb} 
  B.~Schenke, P.~Tribedy and R.~Venugopalan,
  ``Fluctuating Glasma initial conditions and flow in heavy ion collisions,''
  Phys.\ Rev.\ Lett.\  {\bf 108}, 252301 (2012),
  [arXiv:1202.6646 [nucl-th]].

\bibitem{Ma:2010dv} 
  G.~L.~Ma and X.~N.~Wang,
  ``Jets, Mach cone, hot spots, ridges, harmonic flow, dihadron and $\gamma$-hadron correlation in high-energy heavy-ion collisions,''
  Phys.\ Rev.\ Lett.\  {\bf 106}, 162301 (2011),
  [arXiv:1011.5249 [nucl-th]].

\bibitem{Aamodt:2011by} 
  K.~Aamodt {\it et al.} [ALICE Collaboration],
  ``Harmonic decomposition of two-particle angular correlations in Pb-Pb collisions at $\sqrt{s_{NN}}=$ 2.76 TeV,''
  Phys.\ Lett.\ B {\bf 708}, 249 (2012),
  [arXiv:1109.2501 [nucl-ex]].

\bibitem{Pang:2015zrq} 
  L.~G.~Pang, H.~Petersen, G.~Y.~Qin, V.~Roy and X.~N.~Wang,
  ``Decorrelation of anisotropic flow along the longitudinal direction,''
  Eur.\ Phys.\ J.\ A {\bf 52}, 97 (2016),
  [arXiv:1511.04131 [nucl-th]].

\bibitem{Pang:2014pxa} 
  L.~G.~Pang, G.~Y.~Qin, V.~Roy, X.~N.~Wang and G.~L.~Ma,
  ``Longitudinal decorrelation of anisotropic flows in heavy-ion collisions at the CERN Large Hadron Collider,''
  Phys.\ Rev.\ C {\bf 91}, 044904 (2015),
  [arXiv:1410.8690 [nucl-th]].

\bibitem{Broniowski:2015oif} 
  W.~Broniowski and P.~Bozek,
  ``A simple model for rapidity fluctuations in the initial state of ultra-relativistic heavy-ion collisions,''
  arXiv:1512.01945 [nucl-th].

\bibitem{Bozek:2010vz} 
  P.~Bozek, W.~Broniowski and J.~Moreira,
  ``Torqued fireballs in relativistic heavy-ion collisions,''
  Phys.\ Rev.\ C {\bf 83}, 034911 (2011),
  [arXiv:1011.3354 [nucl-th]].

\bibitem{Jia:2014ysa} 
  J.~Jia and P.~Huo,
  ``Forward-backward eccentricity and participant-plane angle fluctuations and their influences on longitudinal dynamics of collective flow,''
  Phys.\ Rev.\ C {\bf 90}, 034915 (2014),
  [arXiv:1403.6077 [nucl-th]].

\bibitem{Khachatryan:2015oea} 
  V.~Khachatryan {\it et al.} [CMS Collaboration],
  ``Evidence for transverse momentum and pseudorapidity dependent event plane fluctuations in PbPb and pPb collisions,''
  Phys.\ Rev.\ C {\bf 92}, 034911 (2015),
  [arXiv:1503.01692 [nucl-ex]].

\bibitem{Borghini:2002hm} 
  N.~Borghini, P.~M.~Dinh and J.~Y.~Ollitrault,
  ``Analysis of directed flow from three particle correlations,''
  Nucl.\ Phys.\ A {\bf 715}, 629 (2003),
  [nucl-th/0208014].

\bibitem{Bzdak:2012tp} 
  A.~Bzdak and D.~Teaney,
  ``Longitudinal fluctuations of the fireball density in heavy-ion collisions,''
  Phys.\ Rev.\ C {\bf 87}, 024906 (2013)
  [arXiv:1210.1965 [nucl-th]].

\bibitem{Monnai:2015sca} 
  A.~Monnai and B.~Schenke,
  ``Pseudorapidity correlations in heavy ion collisions from viscous fluid dynamics,''
  Phys.\ Lett.\ B {\bf 752}, 317 (2016),
  [arXiv:1509.04103 [nucl-th]].
  
\bibitem{Bozek:2015tca} 
  P.~Bozek, W.~Broniowski and A.~Olszewski,
  ``Two-particle correlations in pseudorapidity in a hydrodynamic model,''
  Phys.\ Rev.\ C {\bf 92}, 054913 (2015),
  [arXiv:1509.04124 [nucl-th]].
  
\bibitem{Xiao:2012uw} 
  K.~Xiao, F.~Liu and F.~Wang,
  ``Event-plane decorrelation over pseudorapidity and its effect on azimuthal anisotropy measurements in relativistic heavy-ion collisions,''
  Phys.\ Rev.\ C {\bf 87}, 011901 (2013),
  [arXiv:1208.1195 [nucl-th]].

\bibitem{Bzdak:2011yy} 
  A.~Bzdak and V.~Skokov,
  ``Event-by-event fluctuations of magnetic and electric fields in heavy ion collisions,''
  Phys.\ Lett.\ B {\bf 710}, 171 (2012),
  [arXiv:1111.1949 [hep-ph]].

\bibitem{Deng:2012pc} 
  W.~T.~Deng and X.~G.~Huang,
  ``Event-by-event generation of electromagnetic fields in heavy-ion collisions,''
  Phys.\ Rev.\ C {\bf 85}, 044907 (2012),
  [arXiv:1201.5108 [nucl-th]].

\bibitem{Bloczynski:2012en} 
  J.~Bloczynski, X.~G.~Huang, X.~Zhang and J.~Liao,
  ``Azimuthally fluctuating magnetic field and its impacts on observables in heavy-ion collisions,''
  Phys.\ Lett.\ B {\bf 718}, 1529 (2013),
  [arXiv:1209.6594 [nucl-th]].

  \bibitem{KEKIntense:2010} 
  KEK. Proceedings of the international Conference on Physics in Intense Fields (PIF 2010), 
  November 24-26, 2010 Tsukuba, Japan. Editor: K.~Itakura, S.~Iso and T.~Takahashi , http://atfweb.kek.jp/pif2010/

\bibitem{Kharzeev:2013ffa} 
  D.~E.~Kharzeev,
  ``The Chiral Magnetic Effect and Anomaly-Induced Transport,''
  Prog.\ Part.\ Nucl.\ Phys.\  {\bf 75}, 133 (2014),
  [arXiv:1312.3348 [hep-ph]].

\bibitem{Bzdak:2012ia} 
  A.~Bzdak, V.~Koch and J.~Liao,
  ``Charge-Dependent Correlations in Relativistic Heavy Ion Collisions and the Chiral Magnetic Effect,''
  Lect.\ Notes Phys.\  {\bf 871}, 503 (2013),
  [arXiv:1207.7327 [nucl-th]].

\bibitem{Kharzeev:2015kna} 
  D.~E.~Kharzeev,
  ``Topology, magnetic field, and strongly interacting matter,''
  Ann.\ Rev.\ Nucl.\ Part.\ Sci.\  {\bf 65}, 193 (2015),
  [arXiv:1501.01336 [hep-ph]].

\bibitem{Tuchin:2014hza} 
  K.~Tuchin,
  ``Electromagnetic fields in high energy heavy-ion collisions,''
  Int.\ J.\ Mod.\ Phys.\ E {\bf 23}, 1430001 (2014).

\bibitem{Huang:2015oca} 
  X.~G.~Huang,
  ``Electromagnetic fields and anomalous transports in heavy-ion collisions --- A pedagogical review,''
  arXiv:1509.04073 [nucl-th].

\bibitem{Gursoy:2014aka} 
  U.~Gursoy, D.~Kharzeev and K.~Rajagopal,
  ``Magnetohydrodynamics, charged currents and directed flow in heavy ion collisions,''
  Phys.\ Rev.\ C {\bf 89}, 054905 (2014),
  [arXiv:1401.3805 [hep-ph]].

\bibitem{Hirono:2014oda} 
  Y.~Hirono, T.~Hirano and D.~E.~Kharzeev,
  ``The chiral magnetic effect in heavy-ion collisions from event-by-event anomalous hydrodynamics,''
  arXiv:1412.0311 [hep-ph].

\bibitem{Roy:2015coa} 
  V.~Roy and S.~Pu,
  ``Event-by-event distribution of magnetic field energy over initial 
  fluid energy density in $\sqrt{s_{\rm NN}}$= 200 GeV Au-Au collisions,''
  Phys.\ Rev.\ C {\bf 92}, 064902 (2015),
  [arXiv:1508.03761 [nucl-th]].

\bibitem{Tuchin:2013apa} 
  K.~Tuchin,
  ``Time and space dependence of the electromagnetic field in relativistic heavy-ion collisions,''
  Phys.\ Rev.\ C {\bf 88}, 024911 (2013),
  [arXiv:1305.5806 [hep-ph]].

\bibitem{Tuchin:2013ie} 
  K.~Tuchin,
  ``Particle production in strong electromagnetic fields in relativistic heavy-ion collisions,''
  Adv.\ High Energy Phys.\  {\bf 2013}, 490495 (2013),
  [arXiv:1301.0099].

\bibitem{Roy:2015kma} 
  V.~Roy, S.~Pu, L.~Rezzolla and D.~Rischke,
  ``Analytic Bjorken flow in one-dimensional relativistic magnetohydrodynamics,''
  Phys.\ Lett.\ B {\bf 750}, 45 (2015),
  [arXiv:1506.06620 [nucl-th]].

\bibitem{Pu:2016ayh} 
  S.~Pu, V.~Roy, L.~Rezzolla and D.~H.~Rischke,
  ``Bjorken flow in one-dimensional relativistic magnetohydrodynamics with magnetization,''
  Phys.\ Rev.\ D {\bf 93}, 074022 (2016),
  [arXiv:1602.04953 [nucl-th]].

\bibitem{Pu:2016bxy} 
  S.~Pu and D.~L.~Yang,
  ``Transverse flow induced by inhomogeneous magnetic fields in the Bjorken expansion,''
  Phys.\ Rev.\ D {\bf 93}, 054042 (2016),
  [arXiv:1602.04954 [nucl-th]].

\bibitem{Pang:2016yuh} 
  L.~G.~Pang, G.~Endrődi and H.~Petersen,
  ``Magnetic-field-induced squeezing effect at energies available at the 
  BNL Relativistic Heavy Ion Collider and at the CERN Large Hadron Collider,''
  Phys.\ Rev.\ C {\bf 93}, 044919 (2016),
  [arXiv:1602.06176 [nucl-th]].

\bibitem{Aad:2013xma} 
  G.~Aad {\it et al.} [ATLAS Collaboration],
  ``Measurement of the distributions of event-by-event flow harmonics 
  in lead-lead collisions at = 2.76 TeV with the ATLAS detector at the LHC,''
  JHEP {\bf 1311}, 183 (2013),
  [arXiv:1305.2942 [hep-ex]].

\bibitem{Bilandzic:2013kga} 
  A.~Bilandzic, C.~H.~Christensen, K.~Gulbrandsen, A.~Hansen and Y.~Zhou,
  ``Generic framework for anisotropic flow analyses with multiparticle azimuthal correlations,''
  Phys.\ Rev.\ C {\bf 89}, 064904 (2014),
  [arXiv:1312.3572 [nucl-ex]].

\bibitem{ALICE:2016kpq} 
  J.~Adam {\it et al.} [ALICE Collaboration],
  ``Correlated event-by-event fluctuations of flow harmonics in Pb-Pb collisions at $\sqrt{s_{_{\rm NN}}}=2.76$ TeV,''
  arXiv:1604.07663 [nucl-ex].

\bibitem{Chatterjee:2014nta} 
  R.~Chatterjee, D.~K.~Srivastava and T.~Renk,
  ``Triangular flow of thermal photons from an event-by-event hydrodynamic model for 2.76A TeV Pb+Pb collisions at LHC,''
  arXiv:1401.7464 [hep-ph].
  
\bibitem{Shen:2014lpa} 
  C.~Shen, J.~F.~Paquet, J.~Liu, G.~Denicol, U.~Heinz and C.~Gale,
  ``Event-by-event direct photon anisotropic flow in relativistic heavy-ion collisions,''
  Nucl.\ Phys.\ A {\bf 931}, 675 (2014),
  [arXiv:1407.8533 [nucl-th]].

\end{thebibliography}
\end{document}